\documentclass[aps,twocolumn,groupedaddress,amsmath,amssymb,prx,showkeys]{revtex4-2}
\usepackage{graphicx}
\usepackage{hyperref}
\usepackage{dcolumn}
\usepackage{bm}

\hypersetup{
    colorlinks=true, 
    linkcolor=blue,  
    citecolor=green, 
    urlcolor=magenta 
}

\begin{document}
\preprint{APS/123-QED}
\title{Understanding Flow Behaviors of Supercooled Liquids by Embodying \\ Solid-Liquid Duality at Particle Level}

\author{Dong-Xu Yu}
\author{Ke-Qi Zeng}
\author{Zhe Wang}
\email{zwang2017@mail.tsinghua.edu.cn}
\affiliation{Department of Engineering Physics and Key Laboratory of Particle and Radiation Imaging (Tsinghua University) of Ministry of Education, Tsinghua University, Beijing 100084, China}

\date{\today}

\begin{abstract}
    Understanding the flow behaviors of supercooled liquids presents a major challenge in liquid-state physics 
    due to the strong nonlinearity and rich phenomena. 
    To unravel this complexity, we introduce the concept of local configurational relaxation time $\tau_\mathrm{LC}$, 
    which allows us to embody the solid-liquid duality, 
    proposed by Maxwell for phenomenologically describing materials' response to external load, 
    at the particle level.
    The spatial distribution of $\tau_\mathrm{LC}$ in flow is heterogeneous. 
    Depending on the comparison between the local mobility measured by $\tau_\mathrm{LC}$ and the external shear rate, 
    the shear response of local regions is either solid-like or liquid-like. 
    In this way, $\tau_\mathrm{LC}$ plays a role similar to the Maxwell time.  
    By applying this microscopic solid-liquid duality to different conditions of shear flow with a wide range of shear rates, we describe the emergence of shear thinning in steady shear, 
    and predict the major characteristics of the transient response to start-up shear. 
    Furthermore, we reveal a clear structural foundation for $\tau_\mathrm{LC}$ and the solid-liquid duality associated with it by introducing an order parameter extracted from local configuration. 
    Thus, we establish a framework that connects microscopic structure, dynamics, local mechanical response, and flow behaviors for supercooled liquids. 
    Finally, we rationalize our framework in terms of activations from energy basins that are facilitated by shear. 
    This model illustrates how local structure, convection and thermal activation collectively determine $\tau_\mathrm{LC}$. 
    Notably, it predicts two distinct response groups, which well correspond to the microscopic solid-liquid duality.

\end{abstract}

\maketitle

\section{Introduction}
\label{sec:intro}
In 1867, James C. Maxwell proposed a simple model that contains the essential ingredients of viscoelasticity \cite{maxwell1867}.
Combining a perfectly elastic element with a perfectly viscous element in series, this model defines a characteristic relaxation time $\tau_\mathrm{M}$, now known as the Maxwell time, 
which is pivotal in ascertaining the response of materials to external load. 
For external loads much faster than $\tau_\mathrm{M}$, the material is unable to relax promptly and exhibits solid-like behavior. 
Conversely, for loads much slower than $\tau_\mathrm{M}$, the material responds as a viscous liquid. 
To generalize this simple picture to the vast range of viscoelastic materials, one way is to employ macroscopic laws and phenomenological construction \cite{oswald2009, larson1988}. 
The former includes laws in thermodynamics and continuum mechanics, and symmetry arguments. 
For the latter, the most typical example is serial or parallel combination of Maxwell units with certain relaxation spectrum. 
Despite the widespread success, this approach is limited by its lack of microscopic foundation. 
Consequently, the range of application and generality of such models are difficult to justify. 

To overcome the aforementioned challenge, one needs to embody Maxwell's picture of the solid-liquid duality at the microscopic level. 
Here, the key question is to identify the microstructure that determines the deformation and its relaxation time. 
A celebrated example of this microscopic approach is the tube model of entangled polymers established by Edwards, de Gennes and Doi \cite{edwards1967pps, degennes1971jcp, doi1986}. 
In this model, many-body distribution and interaction of polymer chains are abstracted to a virtual tube. 
Due to the constraint of the tube, the chain performs curvilinear diffusion along the tube axis, called reptation, 
with a microscopic relaxation time $\tau_\mathrm{d}$ that characterizes the escape process of the chain from its original tube. 
$\tau_\mathrm{d}$ plays a role as the Maxwell time, in the sense that for deformation faster than the escape, the entangled polymer melt behaves elastically as a rubber, 
while for deformation slower than the escape, the system behaves as a viscous liquid. 
The clear-cut molecular basis of $\tau_\mathrm{d}$ directly leads to the relation connecting $\tau_\mathrm{d}$ to the microstructural property: $\tau_\mathrm{d} \propto Z^3$,
where $Z$ is the number of entanglement per chain \cite{doi1986}. 
In the nonlinear rheological regime, strong convective effects, such as convective constraint release \cite{marrucci1996nonnewtonianfluid} and chain stretching \cite{doi1998mac, glamm2003}, 
are influential to chain relaxation. 
Nevertheless, the concept of reptation and $\tau_\mathrm{d}$ are still applicable, and play a crucial role in determining the chain orientation in flow \cite{glamm2003}.

For simple liquids \cite{hansen2013}, Maxwell's idea has also been introduced to describe the microscopic dynamics \cite{voigtmann2014cocs, yip1983pra,ruocco2015prl}.
For instance, by constructing a frequency-dependent viscosity with $\tau_\mathrm{M}$, one obtains the viscoelastic approximation for transverse-current correlation function \cite{hansen2013, yip1983pra}. 
It works well at the molecular level for hard-sphere fluids with moderate densities, while for supercooled states where the viscoelasticity becomes significant, this approximation breaks down. 
A pronounced challenge for applying Maxwell's picture to the supercooled state is the dynamic heterogeneity (DH) \cite{ediger2000annualreview, cavagna2009pr, weeks2012ropp, glotzer2003jcp, wangweihua2018prl}, 
which implies that different regions in a supercooled liquid exhibit different mobilities. 
Such spatially heterogeneous mobility renders the global relaxation time ineffective at the local scale \cite{tanaka2011pre, dasgupta2017prl}. 
To tackle this problem, one may assign a spatially varying relaxation time \cite{ruocco2015prl}, whereas the way to extract such local relaxation time can be tricky: 
It is found that definitions of local time from different considerations may lead to different conclusions even at qualitative level \cite{barrat2019prl, berthier2021prl}. 
This problem becomes more complicated when the external flow is applied. 
In this case, the relaxation of local configuration is a result of the combined effect of thermal activation and convection \cite{cates1997prl,lacks1999jcp,lacks2001prl}. 
Therefore, a proper definition of local characteristic time is expected to distinguish between these two effects. 

Concerning DH, it has long been argued that the local dynamics can be well predicted by local configuration 
for supercooled liquids and glasses \cite{harrowell2006prl, harrowell2008natphy, ajliu2016natphy, ajliu2019prl, ajliu2015prl, tanaka2018prx, patinet2022prl, patinet2016prl, manning2020prm, bapst2020natphy, filion2021prl, chen2023prl, han2024prl, royall2013jcp}. 
Particularly, some local structural parameters exhibit predictive ability for a wide range of temperatures and densities, 
and for both quiescent state and states under external load \cite{ajliu2016natphy, ajliu2019prl, ajliu2015prl, patinet2022prl, patinet2016prl, bapst2020natphy}. 
Considering these progresses, we expect that the aforementioned local characteristic time in supercooled liquids, which determines the local response to load, is strongly correlated to the local configuration. 
This correlation, if exists, has a further meaning. 
Unlike polymers, emulsions, foams, microgels and other soft materials composed of mesoscopic soft units \cite{oswald2009, larson1988, doi1986, cates1997prl, larson1999, cloitre2011natmat, cloitre2014cocs, fielding2022prl}, 
the constituent particle of simple liquids itself does not contribute elasticity. 
Thus, the viscoelasticity of supercooled simple liquids originates from collective behaviors of particles \cite{wang2022prx, yamamoto1998pre}, which is believed to be related to the local configuration, 
such as packing and orientation \cite{ajliu2016natphy, royall2015pr, tanaka2019natrev, sastry2016ropp}. 
Establishing this correlation is important for clarifying how the rheological response is determined by the mesoscopic structure and collective behaviors of particles. 
Moreover, the definition of such local time and its relation to local configuration should be general enough: 
they should be applicable to various flow conditions and a broad range of flow rates including the Newtonian regime and the nonlinear regime.

In this paper, we provide an understanding of the shear rheology of supercooled liquids by embodying Maxwell's picture of solid-liquid duality at the particle level. 
With non-equilibrium molecular dynamics (MD) simulations, we introduce the concept of local configurational relaxation time $\tau_\mathrm{LC}$, 
which can be measured once a microscopic configuration is given, regardless of whether the system is in an equilibrium state or under shear. 
$\tau_\mathrm{LC}$ varies in space in the same way as the commonly-understood DH. 
The competition between the local mobility represented by $\tau_\mathrm{LC}$ and the external strain rate determines how the local structure responds to shear, namely, solid-like or liquid-like. 
In this way, $\tau_\mathrm{LC}$ plays a role similar to that of $\tau_\mathrm{M}$ in the Maxwell model. 
By applying this space-dependent solid-liquid duality, we are able to quantitatively understand the rheological response of supercooled liquids to both steady shear and start-up shear in a broad range of shear rates. 
We introduce $\tau_\mathrm{LC}$ in section~\ref{sec:taucr}, and discuss the flow behaviors of supercooled liquids based on $\tau_\mathrm{LC}$ in section~\ref{sec:rheology}.

Furthermore, we reveal a robust correlation between $\tau_\mathrm{LC}$ and microscopic structural features. 
Our findings suggest a firm relation between relevant local structural parameters and $\tau_\mathrm{LC}$, irrespective of shear rates. 
Based on this correlation, we define a structural order parameter, which effectively distinguishes whether the local response to shear is liquid-like or solid-like. 
Combining these efforts and the results in sections~\ref{sec:taucr} and \ref{sec:rheology}, 
we bridge the gap between microscopic structure and macroscopic rheology---a long-standing objective in soft condensed matter physics \cite{oswald2009, doi1986, larson1999}.
These results are given in section~\ref{sec:structure}. 

In addition, we formulate the connection from the microscopic structure to flow behaviors established in sections~\ref{sec:taucr} -- \ref{sec:structure} based on a simple picture of shear-facilitated activation from basins (local configuration for particle) 
\cite{ajliu2016natphy, cates1997prl, bouchaud1996jpa}. 
With this model, we illustrate the physical significance of $\tau_\mathrm{LC}$ and how it is determined by the collective effect of local structure, convection, and thermal activation. 
Moreover, we elucidate the mechanism of the solid-liquid duality in response and its consequence in the emergence of nonlinear flow behaviors. 
The relevant discussion is given in section~\ref{sec:pel}.

\section{Local configurational relaxation time}
\label{sec:taucr} 
\subsection{Simulation method}

Three-dimensional MD simulations utilizing cubic periodic boundary conditions are employed to investigate the flow of supercooled liquids. 
Binary mixtures are adopted to prevent crystallization \cite{kob1995pre}. 
The mixture is composed of $80\%$ big particles (denoted by subscript ``b") and $20\%$ small particles (denoted by subscript ``s") with the same mass $m$. 
The inter-particle interaction is given by the Lennard-Jones (LJ) potential $V(r)=4\epsilon_{\alpha\beta}\left[ (\sigma_{\alpha\beta}/r)^{12} - (\sigma_{\alpha\beta}/r)^{6} \right]$ with $\alpha,\beta\in\{\mathrm{b},\mathrm{s}\}$.
The LJ parameters are set as follows: $\epsilon_\mathrm{bb}=1$, $\sigma_\mathrm{bb}=1$ between big particles; 
$\epsilon_\mathrm{bs}=1.5$, $\sigma_\mathrm{bs}=0.8$ between big and small particles; 
$\epsilon_\mathrm{ss}=0.5$, $\sigma_\mathrm{ss}=0.88$ between small particles. 
The units of energy, length, time, and temperature are, respectively, set by $\epsilon_\mathrm{bb}$, $\sigma_\mathrm{bb}$, $\sqrt{m\sigma_\mathrm{bb}^2/\epsilon_\mathrm{bb}}$, 
and $\epsilon_\mathrm{bb}/k_\mathrm{B}$, where $k_\mathrm{B}$ is the Boltzmann constant. 
With these basic units, the units of stress and viscosity are set by $\epsilon_\mathrm{bb}/\sigma_\mathrm{bb}^3$ and $\sqrt{m\epsilon_\mathrm{bb}}/\sigma_\mathrm{bb}^2$, respectively.
The potential interaction range is truncated and shifted at $2.5$. 
To apply shear, the SLLOD equation combined with Lees-Edwards boundary condition is used \cite{allen2017, edwards1972jpc, evans1984pra}. 
The time step is set to $0.005$. 
Constant NVT integration with a Nose-Hoover thermostat is employed to update positions and momenta of particles, with the application of the LAMMPS package \cite{lammps}.
    
\begin{figure}
    \includegraphics[width=\linewidth]{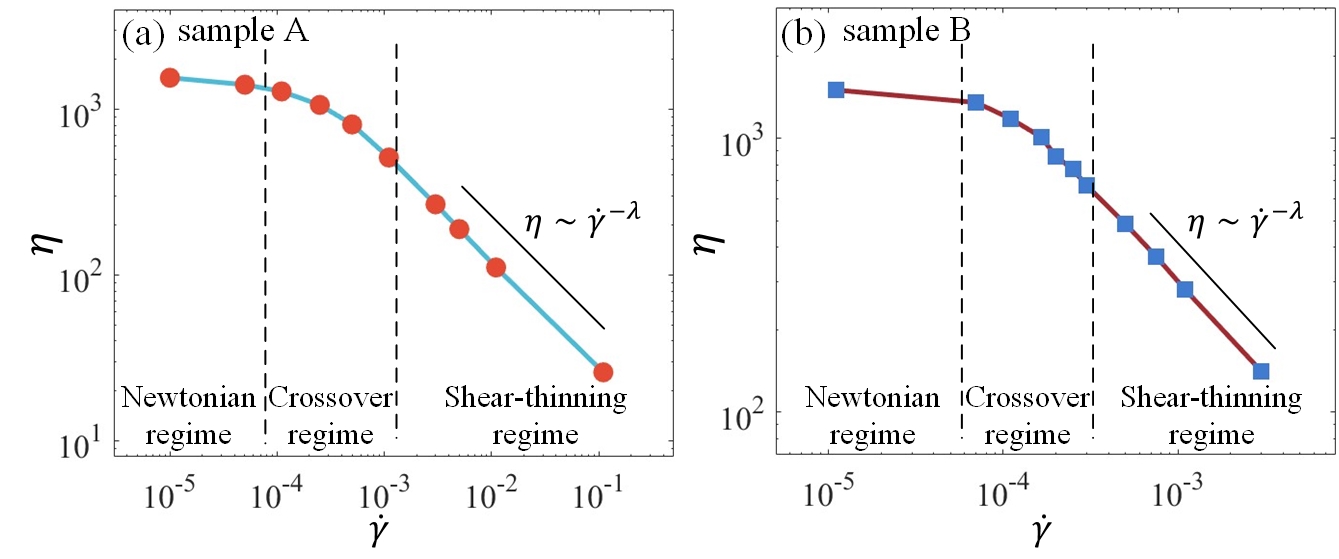}
    \caption{\label{fig:viscosity} 
    Shear viscosity $\eta$ as a function of shear rate $\dot{\gamma}$ for sample A (a) and sample B (b). 
    Symbols denote the results obtained from MD data. 
    Lines are to guide eyes. 
    Three distinct rheological regimes can be delineated, as indicated by the vertical black dashed lines. 
    In the shear-thinning regime, the relation can be described by a power law $\eta \sim \dot{\gamma}^{-\lambda}$. }
\end{figure}

Two samples are studied here. 
Sample A consists of $10976$ particles with a number density of $\rho=1.38$ and a temperature of $T=1.0$. 
Sample B consists of $9000$ particles with $\rho=1.2$ and $T=0.48$. 
The temperatures and number densities are carefully adjusted to ensure that both the Newtonian regime and the nonlinear regime can be achieved without too much computational cost. 
Under the shear with the stream velocity along the $x$ direction, the velocity gradient along the $y$ direction, and the shear rate $\dot{\gamma}$, the shear viscosity $\eta$ is calculated by:
\begin{equation}
\eta = -\frac{1}{V\dot{\gamma}} \left \langle  {\textstyle \sum_{i=1}^{N}} r_{i,x}f_{i,y} \right \rangle,
\end{equation}
where $V$ is the system volume, $N$ is the particle number, $r_{i,x}$ is the $x$ component of the position vector of particle $i$, and $f_{i,y}$ is the $y$ component of the force exerted on particle $i$. 
Figure~\ref{fig:viscosity} shows the viscosities of both samples under steady shear with various shear rates. 
At large enough $\dot{\gamma}$, both samples exhibit significant shear thinning, which can be expressed by $\eta \sim \dot{\gamma}^{-\lambda}$ \cite{larson1999}. 
The values of $\lambda$ are, respectively, $0.67$ and $0.66$ for sample A and sample B. 
For the sake of convenience, we will call the regime featured by $\eta \sim \dot{\gamma}^{-\lambda}$ as the shear-thinning regime, 
and call the regime between Newtonian regime and the shear-thinning regime as the crossover regime. 
These rheological regimes are denoted in figure~\ref{fig:viscosity}.

\subsection{\texorpdfstring{$\tau_\mathrm{LC}$ in quiescent state and local response to start-up shear}{tau\_LC in quiescent state and local response to start-up shear}}

\begin{figure}
    \includegraphics[width=\linewidth]{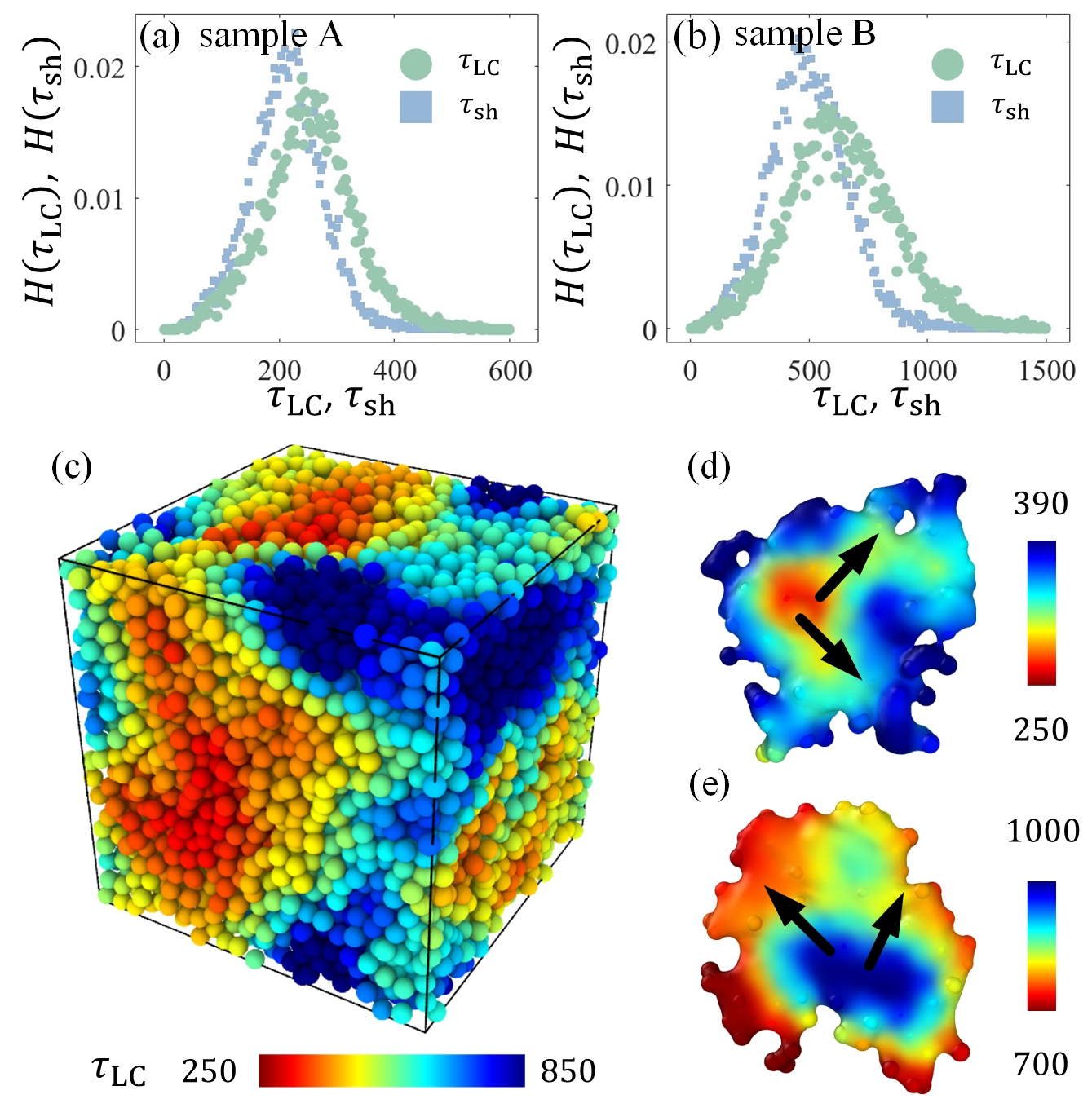}
    \caption{\label{fig:eq_spatial_taucr_dist} 
    (a) Normalized distribution of $\tau_\mathrm{LC}$ ($H(\tau_\mathrm{LC})$) and $\tau_\mathrm{sh}$ ($H(\tau_\mathrm{sh})$) of sample A. 
    (b) Normalized distribution of $\tau_\mathrm{LC}$ ($H(\tau_\mathrm{LC})$) and $\tau_\mathrm{sh}$ ($H(\tau_\mathrm{sh})$) of sample B. 
    In both (a) and (b), $\tau_\mathrm{LC}$ is measured at equilibrium, while $\tau_\mathrm{sh}$ is measured under the start-up shear condition with $\dot{\gamma}=0.00025$ for sample A and $\dot{\gamma}=0.00011$ for sample B. 
    (c) Snapshot of the spatial distribution of $\tau_\mathrm{LC}$ of sample B at equilibrium. 
    (d) and (e) illustrate the hierarchical feature of the spatial distribution of $\tau_\mathrm{LC}$. 
    (d) gives a slice of a fast cluster. It is seen that particles gradually become slower as the distance from the fast center grows. 
    (e) gives a slice of a slow cluster. It is seen that particles gradually become faster as the distance from the slow center grows. 
    Black arrows in (d) and (e) are to guide eyes, indicating that fast particles tend to ``grow" from a fast center (d), 
    and slow particles tend to ``grow" from a slow center (e).}
\end{figure}

In supercooled liquids, a particle is strongly restricted by the cage formed by its neighboring particles \cite{gotze1992ropp, tanpeng2023natphy}. 
Consequently, a particle vibrates inside its cage for most of the time, and occasionally undergoes a cage jump \cite{dyre2000jcp, heuer2008jpcm}. 
Knowing particle trajectories, cage jump events can be identified by the algorithm proposed by Candelier \textit{et al}. \cite{candelier2009prl, candelier2010prl} (Appendix \ref{sec:cage_jump_algorithm}). 
With this method, one can find the time before the first jump from a given time origin for a reference particle $\tau_\mathrm{p}$. 
Here, the subscript ``p'' denotes ``persistence'', in the sense that this definition is consistent with the persistence time in continuous time random walk framework \cite{chandler2005jcp, pastore2021jcp}. 
To enhance the statistics, we employ the isoconfigurational ensemble (ICE) \cite{harrowell2004prl}, 
which consists of multiple trajectories from the same configuration but with random initial momenta sampled from the Maxwell-Boltzmann distribution. 
Note that, by averaging across ICE, one suppresses the uncertainty related to thermal fluctuations. 
Thus, the ICE-averaged result $\left \langle \tau_\mathrm{p} \right \rangle_\mathrm{ICE}$ is mainly determined by the configuration at the given time origin, and is irrelevant to the instantaneous momenta. 
For this reason, we name it local configurational relaxation time, and denote it as $\tau_\mathrm{LC}$. 
Figure~\ref{fig:eq_spatial_taucr_dist} (a) and (b) display the normalized equilibrium distribution of $\tau_\mathrm{LC}$ of sample A and sample B, respectively. 
The value of $\tau_\mathrm{LC}$ varies from particle to particle, and exhibits significant spatial heterogeneity, as shown in figure~\ref{fig:eq_spatial_taucr_dist} (c) for sample B at the equilibrium state. 

In principle, $\tau_\mathrm{LC}$, which represents the local relaxation rate, is relevant to the local response to external load. 
To explore the relation between $\tau_\mathrm{LC}$ and local response, we apply the start-up shear to the equilibrium state, 
and record the first-jump time from the starting time of this shear for each particle $\tau_\mathrm{p,sh}$, where the subscript ``sh" represents ``shear". 
Then, we still perform the ICE average on $\tau_\mathrm{p,sh}$ with respect to the configuration at the start of shear, and denote the result as $\tau_\mathrm{sh}$ ($\tau_\mathrm{sh} = \left \langle \tau_\mathrm{p,sh} \right \rangle_\mathrm{ICE}$). 
In figure~\ref{fig:eq_spatial_taucr_dist} (a) and (b), we plot the distribution of $\tau_\mathrm{sh}$ at a given $\dot{\gamma}$ for sample A and sample B, respectively. 
Compared with $\tau_\mathrm{LC}$, the distribution of $\tau_\mathrm{sh}$ shifts to lower values, manifesting the facilitation of dynamics by the external shear. 
For a specific particle, the difference between $\tau_\mathrm{sh}$ and $\tau_\mathrm{LC}$ reflects the impact of shear on its mobility. 
Intuitively, slow particles, i.e., particles with large $\tau_\mathrm{LC}$, are more susceptible to imposed shear. 
On the other hand, particles with very small $\tau_\mathrm{LC}$ cannot perceive the shear effect, because their spontaneous relaxations are faster than the shear rate.

\begin{figure}
    \includegraphics[width=\linewidth]{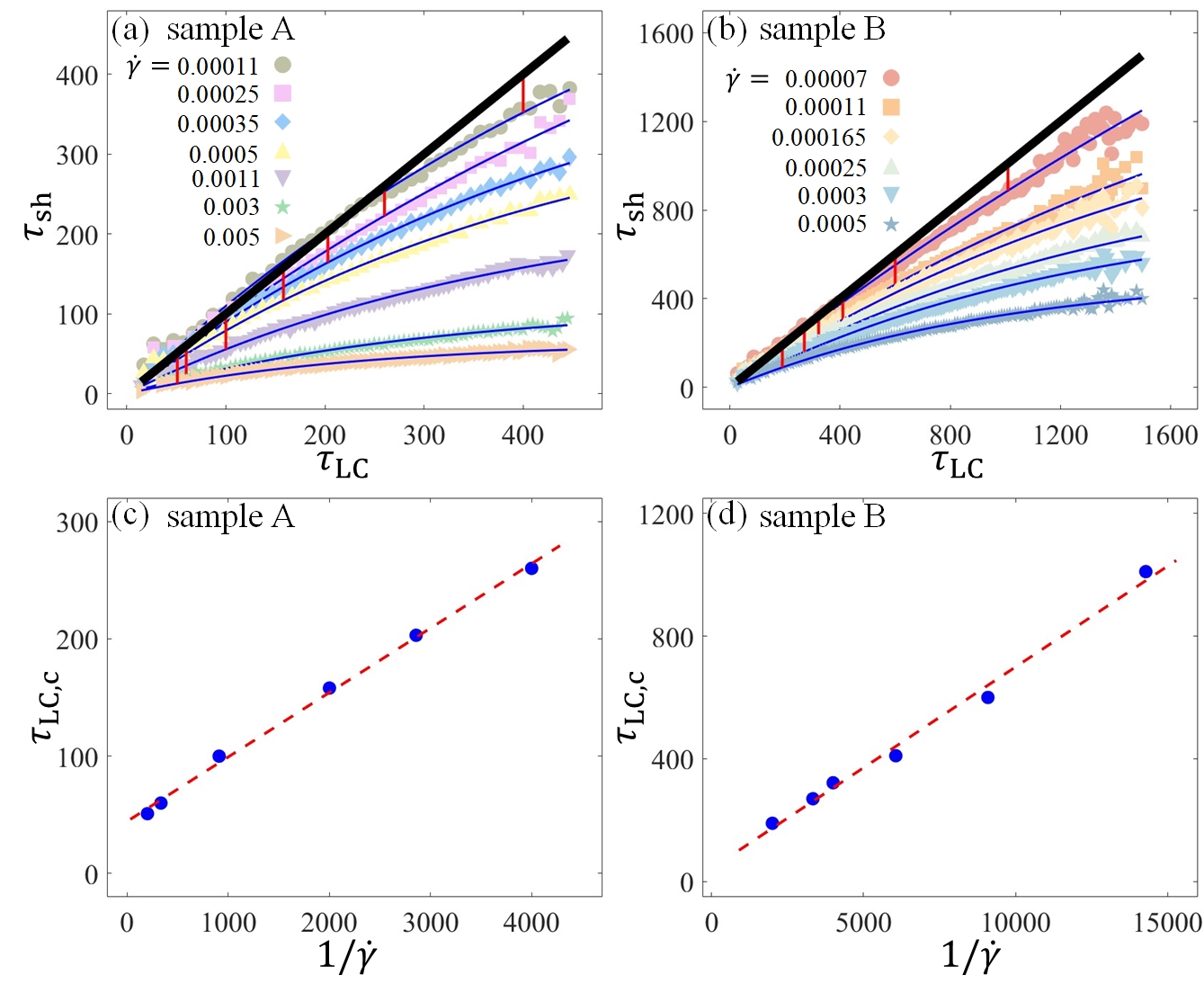}
    \caption{\label{fig:eq_taucr_sh} 
    (a) and (b) display the relations between $\tau_\mathrm{LC}$ and $\tau_\mathrm{sh}$, denoted as $\tau_\mathrm{sh}(\tau_\mathrm{LC})$, 
    obtained from start-up shear with various $\dot{\gamma}$ for sample A and sample B, respectively. 
    In both (a) and (b), symbols denote the results, thick black lines denote the reference condition of $\tau_\mathrm{sh}=\tau_\mathrm{LC}$, thin blue lines denote the exponential fits for data, 
    the positions of red vertical bars denote the boundaries between fast and slow groups $\tau_\mathrm{LC,c}$, and the length of red vertical bars denotes the value of $\tau_2$. 
    (c) and (d) display the boundaries between the fast and slow groups $\tau_\mathrm{LC,c}$ as a function of $\dot{\gamma}^{-1}$ for sample A and sample B, respectively. 
    In both (c) and (d), symbols denote the results, dashed lines denote the linear fits to data points. }
\end{figure}

To quantitatively evaluate the difference between $\tau_\mathrm{sh}$ and $\tau_\mathrm{LC}$, we perform the following two-step operation for each sample: 
First, we divide all particles into many bins according to the ascending sequence of $\tau_\mathrm{LC}$. 
In each bin, particles are approximately equal in the value of $\tau_\mathrm{LC}$ with a difference up to the bin width. 
Subsequently, for each bin, we calculate the averaged $\tau_\mathrm{sh}$ for particles within that bin. 
With this operation, we can compare the bin-averaged values of $\tau_\mathrm{sh}$ and $\tau_\mathrm{LC}$. 
The results, shown in figure~\ref{fig:eq_taucr_sh} (a) and (b) for both samples and various $\dot{\gamma}$, can be summarized as follows: 
(1) For small $\dot{\gamma}$, $\tau_\mathrm{sh}$ is approximately equal to $\tau_\mathrm{LC}$ for the majority of $\tau_\mathrm{LC}$ values, 
as indicated by the close alignment with the black thick line denoting $\tau_\mathrm{sh}=\tau_\mathrm{LC}$. 
(2) As $\dot{\gamma}$ increases, deviation from this alignment grows at large values of $\tau_\mathrm{LC}$, suggesting that more and more particles are accelerated by shear. 
This deviation is pronounced in the shear-thinning regime, manifested by the flattening of the $\tau_\mathrm{sh}(\tau_\mathrm{LC})$ curve. 
These results are consistent with our intuition.

Knowing the relation between $\tau_\mathrm{sh}$ and $\tau_\mathrm{LC}$ at a specific $\dot{\gamma}$,
we can categorize all particles into a \textit{fast} group and a \textit{slow} group, depending on the extent to which the particle is facilitated by shear. 
The aim of this dichotomy is to draw a correspondence to the solid-liquid duality in the Maxwell model, as we will discuss later. 
Here, we identify a particle as a slow one if its first-jump time in shear ($\tau_\mathrm{sh}$) is significantly accelerated compared with its equilibrium counterpart ($\tau_\mathrm{LC}$):
\begin{equation}
    \tau_\mathrm{LC} - \tau_\mathrm{sh} \ge \tau_2,
    \label{eq:divide_fast_slow}
\end{equation}
where $\tau_2$ is the non-Gaussian time at equilibrium, which measures the waiting time for the system to exhibit significant DH \cite{weitz2000science}.  
According to the preceding criterion, slow particles exhibit distinct relaxation heterogeneity compared with the equilibrium situation, thereby being considered as being strongly influenced by shear. 
Then, the fast particles are identified by satisfying $\tau_\mathrm{LC} - \tau_\mathrm{sh} < \tau_2$. 
The boundary between fast and slow groups $\tau_\mathrm{LC,c}$ ($\tau_\mathrm{LC,c}=\tau_\mathrm{sh}+\tau_2$) is denoted by red vertical bar for each $\dot{\gamma}$ in figure~\ref{fig:eq_taucr_sh} (a) and (b), 
where the length of the bar is set to $\tau_2$. 
$\tau_\mathrm{LC,c}$ reflects the time scale of external shear, so that it is expected to linearly depend on $\dot{\gamma}^{-1} $ \cite{petekids2015prl}.  
This linearity is shown in figure~\ref{fig:eq_taucr_sh} (c) and (d), implying that our way of determining $\tau_\mathrm{LC,c}$ is reasonable. 

\begin{figure*}
    \includegraphics[width=\linewidth]{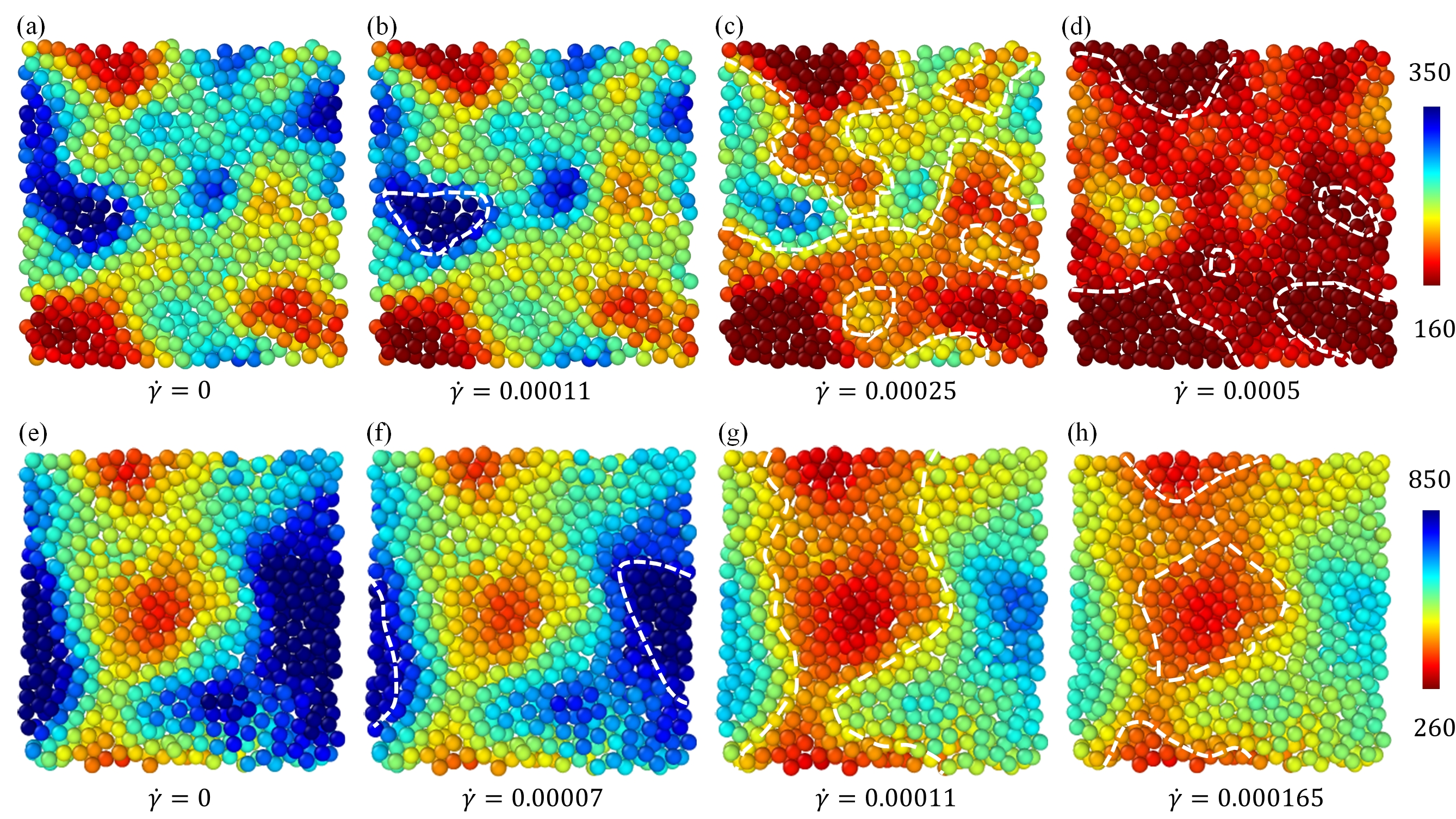}
    \caption{\label{fig:eq_spatial_expand} 
    (a) Slice of the spatial distribution of $\tau_\mathrm{LC}$ of sample A at equilibrium. 
    (b) -- (d) Slices of the spatial distribution of $\tau_\mathrm{sh}$ of sample A obtained under start-up shear with various $\dot{\gamma}$ denoted at each panel. 
    In (b) -- (d), white dashed lines denote the boundaries between fast and slow groups of particles found by equation~\ref{eq:divide_fast_slow}. 
    (e) -- (h) show the results for sample B. }
\end{figure*}

The spatial distribution of fast or slow particles and its variation with $\dot{\gamma}$ are of particular importance for clarifying the local response. 
To understand this point, we first notice that the distribution of $\tau_\mathrm{LC}$ is hierarchical. 
Taking fast particles as an example: Fast particles tend to aggregate, as a manifestation of DH. 
The center of an aggregation is formed by a few very fast particles. 
As the distance from the center increases, the particles gradually become slower, as shown in figure~\ref{fig:eq_spatial_taucr_dist} (d). 
This is consistent with the dynamic facilitation mechanism, which suggests that local relaxation events propagate to neighboring regions \cite{chandler2011prx, candelier2009prl, candelier2010prl, berthier2022prx}. 
Similarly, slow particles exhibit clustering with centers formed by a few very slow particles. 
Within a slow cluster, particles gradually become faster as the distance from center increases, as shown in figure~\ref{fig:eq_spatial_taucr_dist} (e). 
Analogous pattern is also observed in a recent study \cite{berthier2024prl}.
When a start-up shear is applied, the local response is reflected by the distribution of $\tau_\mathrm{sh}$. 
According to the previous analysis, as $\dot{\gamma}$ increases, more particles will be involved as slow ones, manifested by the expansion of slow clusters. 
This trend is clearly seen in figure~\ref{fig:eq_spatial_expand}. 
Figure~\ref{fig:eq_spatial_expand} (a) shows the spatial distribution of $\tau_\mathrm{LC}$ for sample A at equilibrium. 
Figure~\ref{fig:eq_spatial_expand} (b) -- (d) show the spatial distributions of $\tau_\mathrm{sh}$ of sample A under start-up shear with various $\dot{\gamma}$. 
The boundaries of slow clusters, found by $\tau_\mathrm{sh,b} = \tau_\mathrm{LC,c} - \tau_2$, are denoted by dashed lines (the detailed method for circling these clusters is given in Appendix A). 
It is seen that $\tau_\mathrm{LC}$ and $\tau_\mathrm{sh}$ exhibit similar heterogeneous features. 
Moreover, the boundaries of slow clusters expand as $\dot{\gamma}$ increases. 
Same observation is found for sample B, as shown in figure~\ref{fig:eq_spatial_expand} (e) -- (h). 

For the meaning of ``slow" and ``fast", we would like to discuss the following two points. 
(i) As $\dot{\gamma}$ increases, the mean value of $\tau_\mathrm{sh}$ decreases, suggesting faster dynamics. 
On the other hand, more particles are involved in slow regions. 
These two observations are not contradictory. 
When we identify a group of particles as slow, it means that their mobility is slower than external shear. 
Thus, as $\dot{\gamma}$ increases, the dynamics is facilitated by shear, meanwhile, more particles become slower than the external shear and are identified as slow ones. 
(ii) Considering that the distributions of $\tau_\mathrm{sh}$ and $\tau_\mathrm{LC}$ are not bimodal, one may feel that the dichotomy used here is too rough. 
Especially, for particles with mediate mobility, whether they are fast or slow is obscure. 
This problem can be clarified by the clustering feature of slow/fast regions. 
As discussed in the preceding paragraph, typically, a slow (or fast) cluster has a center formed by a few very slow (or fast) particles. 
As the distance from the center increases, particles gradually become faster (or slower). 
Consequently, the particles with mediate mobility are mainly at the boundaries between fast and slow clusters. 
Whether they are identified as slow or fast may result in small fluctuation of the boundaries, but will not influence the identification of a cluster to be slow or fast. 

\begin{figure}
    \includegraphics[width=\linewidth]{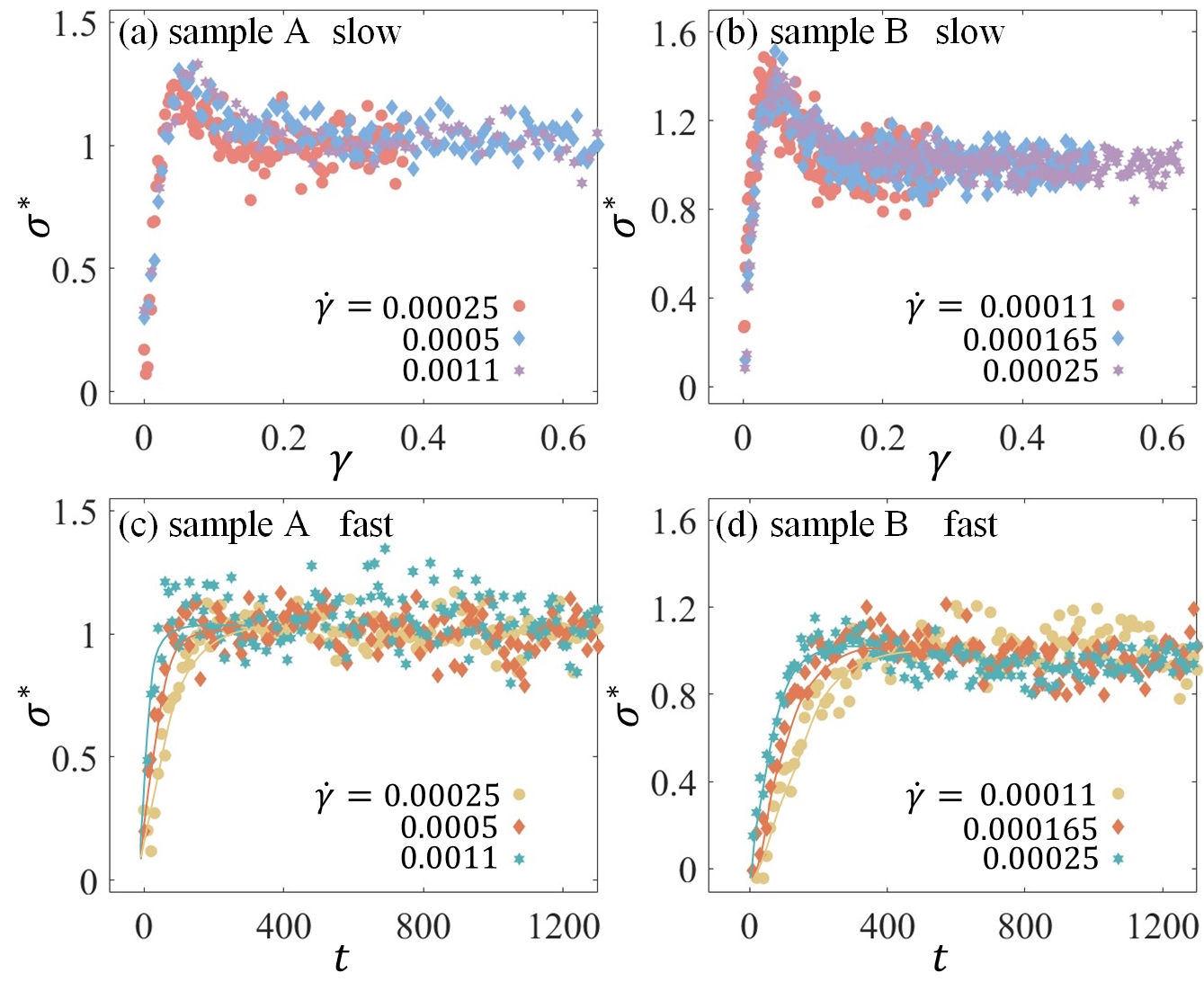}
    \caption{\label{fig:startup_stress} 
    (a) and (b) show the transient stresses of the slow part, normalized by the corresponding stable-state stresses, under start-up shear with various $\dot{\gamma}$ for sample A and sample B, respectively. 
    In both (a) and (b), we adopt $\gamma=\dot{\gamma}t$ as the time variable. 
    (c) and (d) show the transient stresses of the fast part, normalized by the corresponding stable-state stresses, under start-up shear with different $\dot{\gamma}$ for sample A and sample B, respectively. 
    In both (c) and (d), lines denote the fits with $\sigma^*(t) = 1 - \exp(-t/\tau_\mathrm{f})$. }
\end{figure}

The emergence of slow and fast clusters allows us to explore the heterogeneity in mechanical response. 
For each cluster, we calculate its shear stress by summing the atomic level stresses \cite{egami2011pms} of its constituent particles during the start-up process. 
Then, by averaging the results of all slow (or fast) clusters, we obtain the response of the slow (or fast) part. 
To mitigate the effect of thermal fluctuations, stresses are further averaged over ICE. 
In figure~\ref{fig:startup_stress}, we present the start-up shear stresses of slow part ((a) and (b)) and fast part ((c) and (d)) for both samples at different $\dot{\gamma}$. 
For clarity, the transient stress $\sigma(t)$ is scaled using the corresponding stable-state stress, ensuring that the resulting dimensionless stress $\sigma^*(t)$ is equal to $1$ at long time. 
We find that the stress of slow part exhibits an overshoot, a feature typically observed in glasses \cite{petekidis2012prl, zaccone2017prl}. 
In contrast, the stress of fast part exhibits a smooth transition to stable state without a distinct overshoot. 
This behavior can be described by the Maxwell model as $\sigma^*(t) = 1 - \exp(-t/\tau_\mathrm{f})$, where $\tau_\mathrm{f}$ is the Maxwell time of fast part. 
As $\dot{\gamma}$ increases, $\tau_\mathrm{f}$ is seen to decrease, meaning that the fast part becomes faster and approaches ideal viscous liquid, 
corresponding to the shrinkage of fast regions with increasing $\dot{\gamma}$. 
This observation is consistent with our previous analysis given by figures~\ref{fig:eq_taucr_sh} and \ref{fig:eq_spatial_expand}. 
It should be pointed out that whether there is an overshoot is not a necessary and sufficient condition to distinguish between solid-like and liquid-like responses \cite{petekidis2012prl, schall2016sofmat}. 
Even so, the difference between the responses of slow and fast parts is unambiguous. 
In the next subsection, the coexistence of solid-like and liquid-like responses will be clearly revealed in the flow under steady shear.

\subsection{\texorpdfstring{$\tau_\mathrm{LC}$ in flow and local response to steady shear}{tau\_LC in flow and local response to steady shear}}
\label{sec:taucr_steady_shear}
In the preceding subsection, we explore the connection between DH and local mechanical response to start-up shear through the concept of $\tau_\mathrm{LC}$. 
A key to this connection is the functional form of $\tau_\mathrm{sh}(\tau_\mathrm{LC})$, which determines whether a particle behaves slowly or fast in response to shear. 
Then, a natural question is raised: How can the insight derived from the start-up shear guides our understanding of other flow conditions? 
In this subsection, we will discuss the steady shear, the most common nonequilibrium condition in fluid study. 

\begin{figure}
    \includegraphics[width=\linewidth]{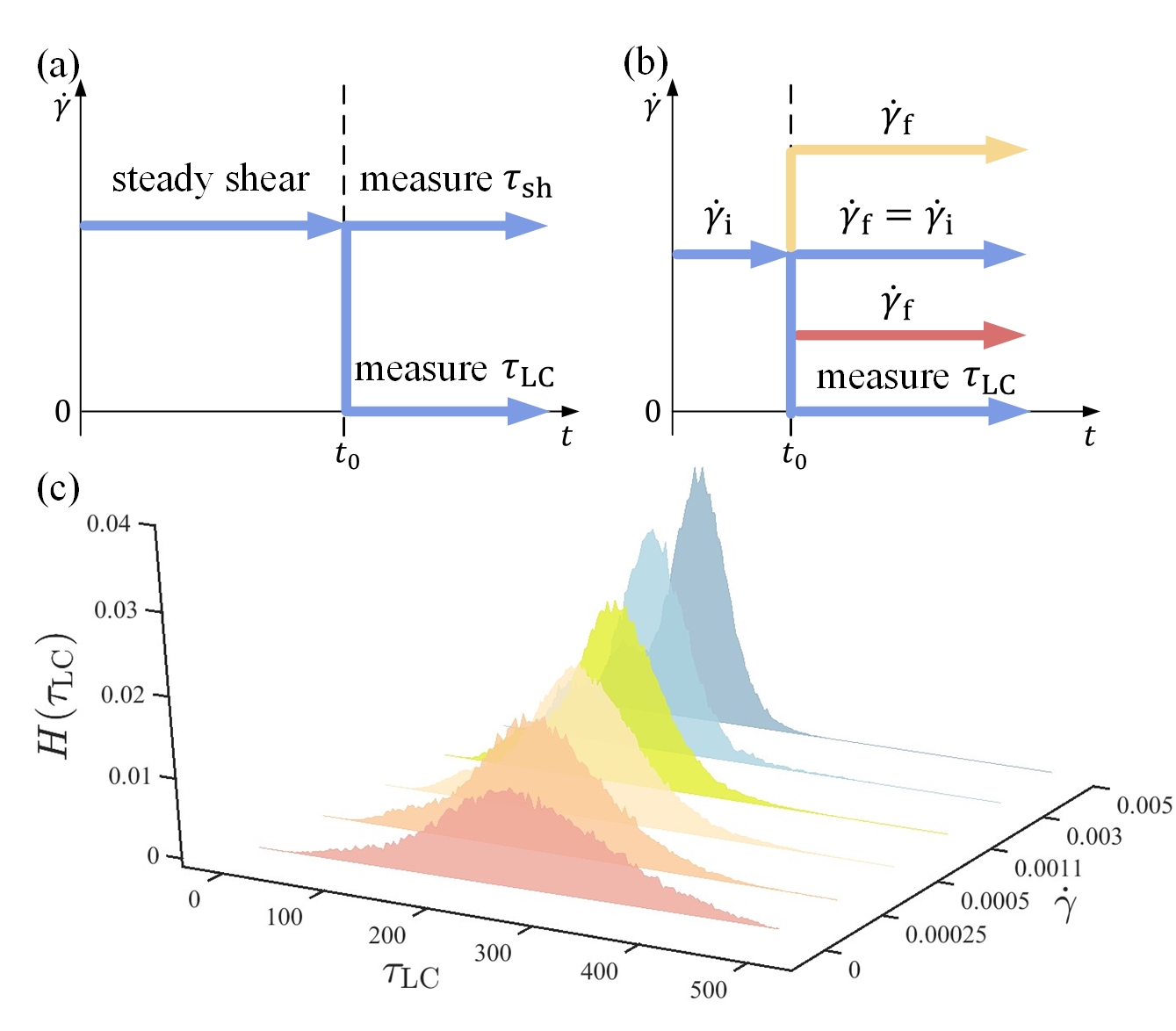}
    \caption{\label{fig:measure_taucrsh_diag} 
    (a) Schematic diagram for illustrating the flow conditions for measuring $\tau_\mathrm{LC}$ and $\tau_\mathrm{sh}$ under steady shear with respect to a configuration at time $t_0$. 
    (b) Schematic diagram for illustrating the flow conditions for testing the universality of $\tau_\mathrm{sh}(\tau_\mathrm{LC})$. 
    Here, the shear rate is abruptly altered from the initial value $\dot{\gamma}_\mathrm{i}$ to a final value $\dot{\gamma}_\mathrm{f}$ at $t=t_0$. 
    (c) Distributions of $\tau_\mathrm{LC}$ ($H(\tau_\mathrm{LC})$) at equilibrium and under steady shear with various $\dot{\gamma}$.  }
\end{figure}

For a steady shear with $\dot{\gamma}$, we calculate its $\tau_\mathrm{LC}$ with respect to a given time $t_0$ as follows: 
First, we switch off the shear flow at $t=t_0$, and measure the first-jump time from $t_0$ for every particle. 
Then, we perform the ICE average according to the configuration at $t=t_0$ with keeping the shear off from $t=t_0$. 
Besides $\tau_\mathrm{LC}$, we also calculate the first-jump time from $t=t_0$ in shear flow $\tau_\mathrm{sh}$. 
In calculating $\tau_\mathrm{sh}$, the steady shear is kept on, and the ICE average is still employed. 
The flow conditions for calculating $\tau_\mathrm{LC}$ and $\tau_\mathrm{sh}$ are illustrated in figure~\ref{fig:measure_taucrsh_diag} (a). 
In the study of polymer dynamics, the abrupt cessation of flow is widely used, and its relaxation time is intimately linked to the constitutive model \cite{doi1986, wang2017prx}. 
However, for most studies of flow behaviors of particulate liquids and glasses, characteristic relaxation times are measured in the continuous flow \cite{yamamoto1998pre, poon2007prl, schall2012pre, tanaka2009prl}. 
The times characterizing the process after the cessation of flow, though appearing in a few studies \cite{schall2016sofmat, horbach2009epl}, have not attracted enough attention. 
Note that $\tau_\mathrm{sh}$ is determined by the combined effect of the steady convection and the thermal activation from the configuration at $t=t_0$. 
In contrast, by switching off the shear, the steady convection is eliminated and, thus, $\tau_\mathrm{LC}$ is mainly determined by the thermal activation from the distorted configuration at $t=t_0$. 
In figure~\ref{fig:measure_taucrsh_diag} (c), we plot the distributions of $\tau_\mathrm{LC}$ at equilibrium and under steady shear with various $\dot{\gamma}$ for sample A. 
The distribution of $\tau_\mathrm{LC}$ shifts to lower values as $\dot{\gamma}$ increases, corresponding to more distorted configurations at larger $\dot{\gamma}$.

\begin{figure}
    \includegraphics[width=\linewidth]{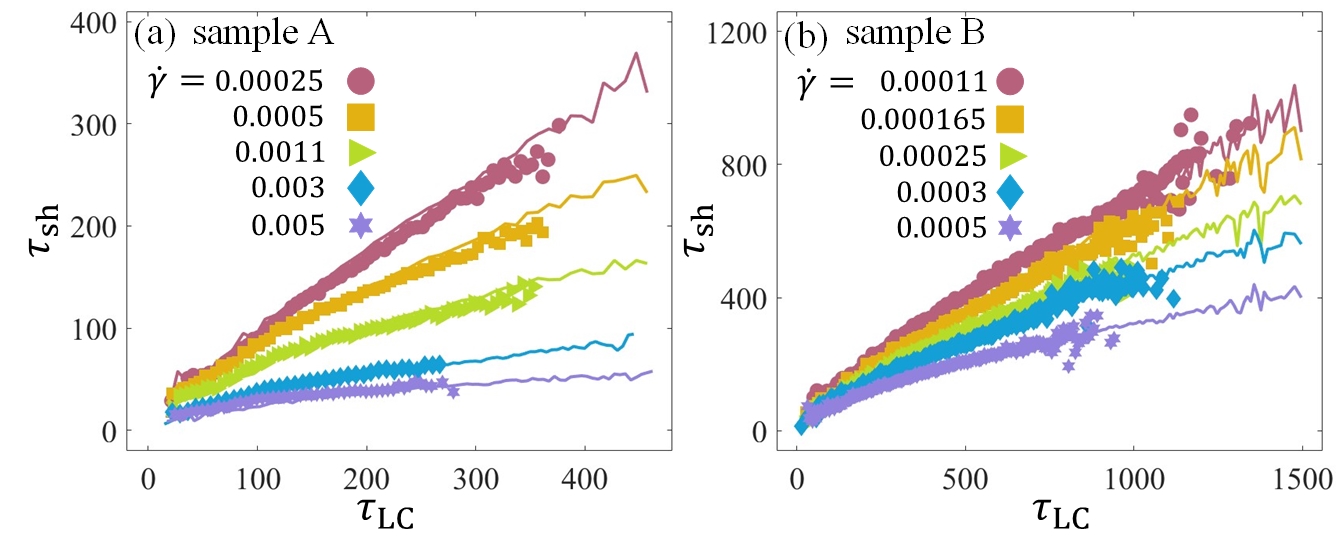}
    \caption{\label{fig:compare_eqsh_stopcont} 
    (a) and (b) show $\tau_\mathrm{sh}(\tau_\mathrm{LC})$ obtained under steady shear (symbols) with various $\dot{\gamma}$ for sample A and sample B, respectively. 
    In addition, the results shown in figure~\ref{fig:eq_taucr_sh} (a) and (b), which are obtained by applying start-up shear, are replotted here by lines. 
    Here, lines and symbols corresponding to the same $\dot{\gamma}$ are plotted in the same color.  }
\end{figure}

Figure~\ref{fig:compare_eqsh_stopcont} gives the $\tau_\mathrm{sh}(\tau_\mathrm{LC})$ relations in steady shear (symbols) for both samples at various $\dot{\gamma}$ 
covering both the crossover regime and the shear-thinning regime. 
In addition, we replot the results for the start-up shear that have been shown in figure~\ref{fig:eq_taucr_sh} (a) and (b) (lines). 
It is noteworthy that the symbols well align with the line for each $\dot{\gamma}$. 
This nontrivial agreement suggests that the relation between $\tau_\mathrm{sh}$ and $\tau_\mathrm{LC}$ is universal for a given $\dot{\gamma}$, irrespective of whether the flow is steady or start-up. 

\begin{figure}
    \includegraphics[width=\linewidth]{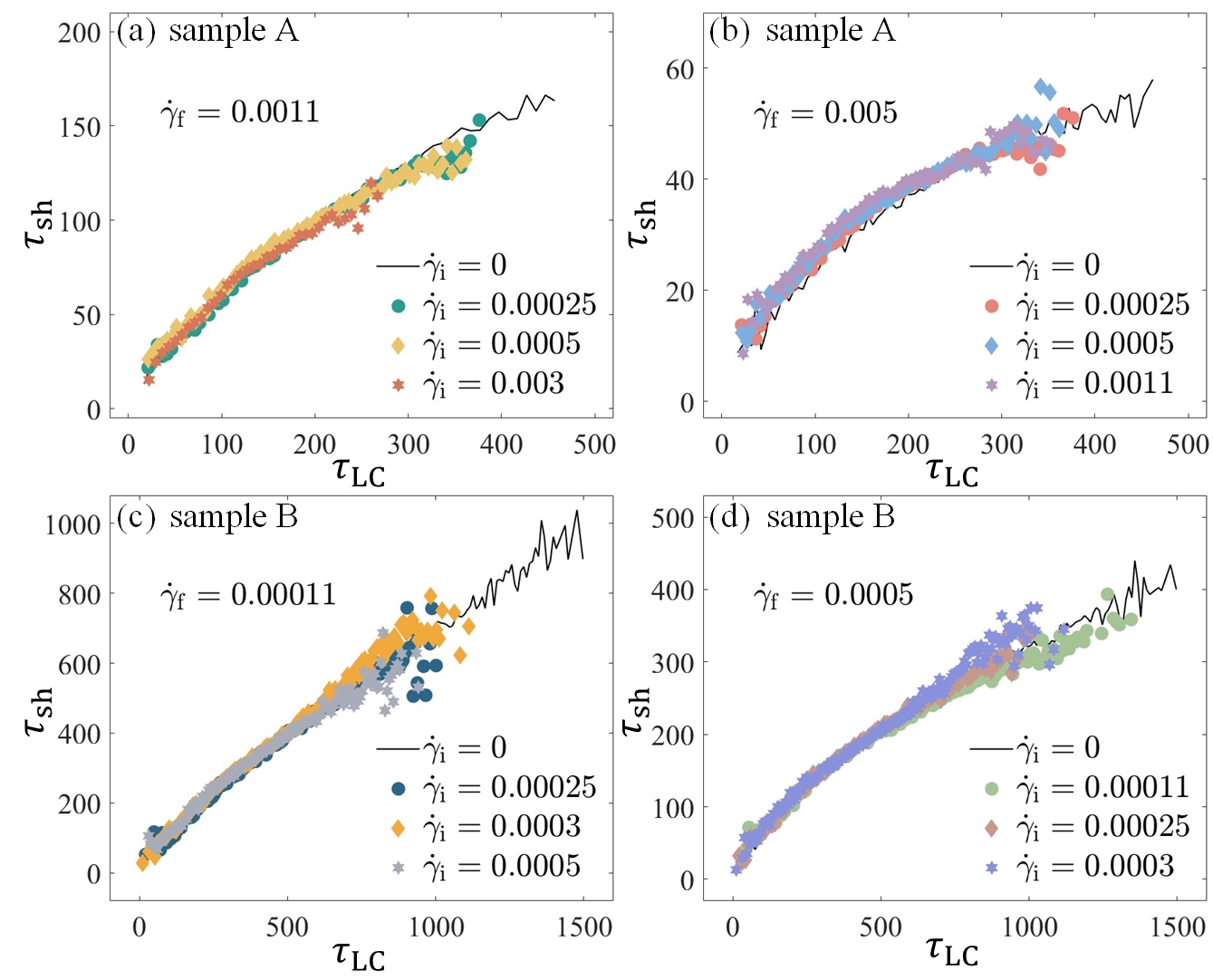}
    \caption{\label{fig:compare_multigammai} 
    Demonstration of the universality of the $\tau_\mathrm{sh}(\tau_\mathrm{LC})$ relation with respect to $\dot{\gamma}_\mathrm{f}$. 
    (a) shows the $\tau_\mathrm{sh}(\tau_\mathrm{LC})$ relations obtained with $\dot{\gamma}_\mathrm{f}=0.0011$ and various $\dot{\gamma}_\mathrm{i}$ for sample A. 
    (b) shows the results for another given final shear rate $\dot{\gamma}_\mathrm{f}=0.005$ for sample A. 
    (c) and (d) display the results for sample B. 
    In all panels, the results of $\tau_\mathrm{sh}(\tau_\mathrm{LC})$ with different $\dot{\gamma}_\mathrm{i}$ collapse for a given $\dot{\gamma}_\mathrm{f}$.}
\end{figure}

To examine the likely universality of the $\tau_\mathrm{sh}(\tau_\mathrm{LC})$ relation mentioned above, we conduct additional tests as follows. 
As illustrated in figure~\ref{fig:measure_taucrsh_diag} (b), we first let the system be in a steady state with a shear rate $\dot{\gamma}_\mathrm{i}$. 
Then, we perform an abrupt change of shear rate from $\dot{\gamma}_\mathrm{i}$ to $\dot{\gamma}_\mathrm{f}$ at $t=t_0$. 
Steady shear is just a special case where $\dot{\gamma}_\mathrm{f}=\dot{\gamma}_\mathrm{i}$. 
Both $\tau_\mathrm{LC}$ and $\tau_\mathrm{sh}$ with respect to $t=t_0$ are calculated. 
In figure~\ref{fig:compare_multigammai} (a), we show the results of $\tau_\mathrm{sh}(\tau_\mathrm{LC})$ with $\dot{\gamma}_\mathrm{f}=0.0011$ and several $\dot{\gamma}_\mathrm{i}$ 
ranging from $\dot{\gamma}_\mathrm{i}=0$ to $\dot{\gamma}_\mathrm{i}=0.003$ for sample A. 
Notice that, the conditions here include both $\dot{\gamma}_\mathrm{i}<\dot{\gamma}_\mathrm{f}$ and $\dot{\gamma}_\mathrm{i}\geq\dot{\gamma}_\mathrm{f}$. 
It is seen that the $\tau_\mathrm{sh}(\tau_\mathrm{LC})$ relations at different $\dot{\gamma}_\mathrm{i}$ remarkably coincide for the same $\dot{\gamma}_\mathrm{f}$. 
Such coincidence is also found for different $\dot{\gamma}_\mathrm{f}$ (figure~\ref{fig:compare_multigammai} (b)) and different samples (figure~\ref{fig:compare_multigammai} (c) and (d)). 
The results shown in figures~\ref{fig:compare_eqsh_stopcont} and \ref{fig:compare_multigammai} suggest that the functional form of $\tau_\mathrm{sh}(\tau_\mathrm{LC})$ is universal 
once the final shear rate $\dot{\gamma}_\mathrm{f}$ is given. 
This observation implies that the transient configuration (i.e., the configuration at $t=t_0$, determined by the initial shear rate $\dot{\gamma}_\mathrm{i}$) and 
the convective effect (determined by $\dot{\gamma}_\mathrm{f}$) independently contribute to the local relaxation. 

\begin{figure}
    \includegraphics[width=\linewidth]{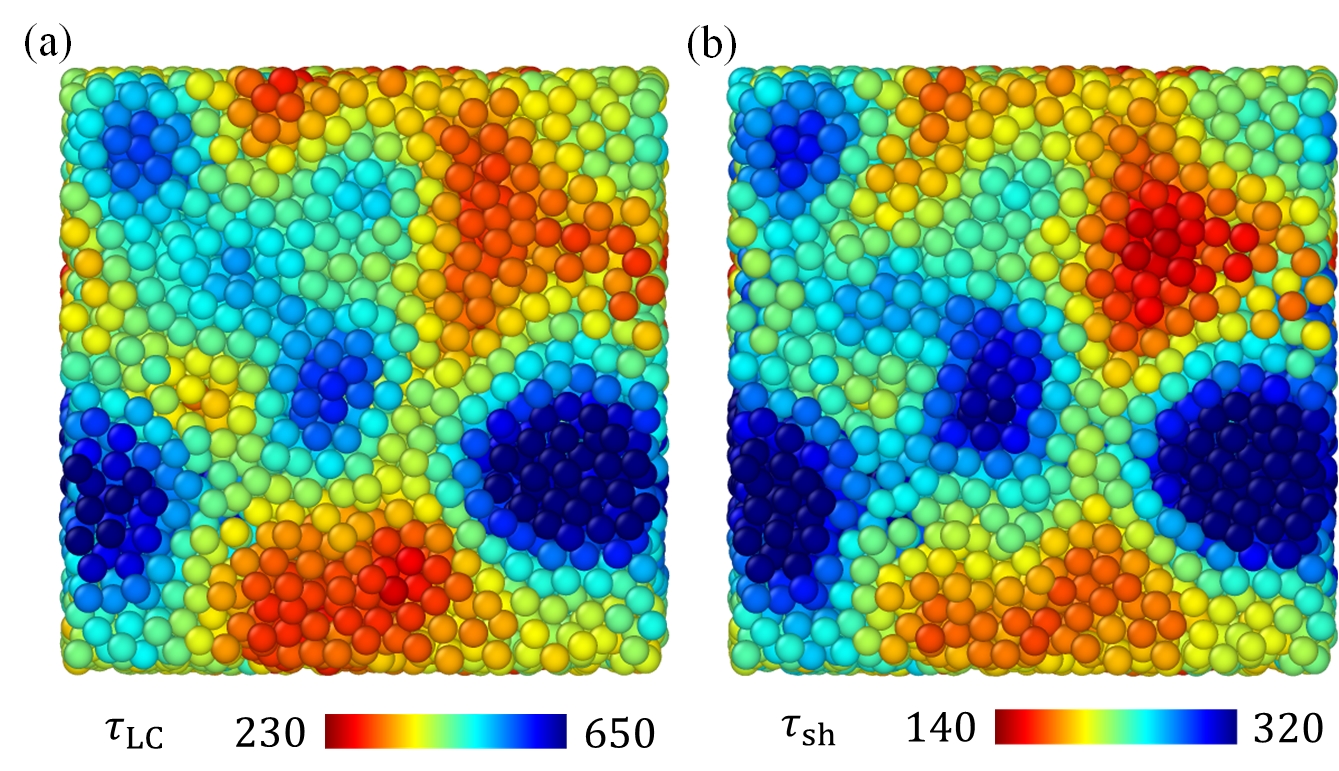}
    \caption{\label{fig:compare_cr_sh_spatial_dist} 
    Snapshots of the slice of the spatial distribution of $\tau_\mathrm{LC}$ (a) and $\tau_\mathrm{sh}$ (b) for sample B under steady shear with $\dot{\gamma}=0.0003$. }
\end{figure}

In figure~\ref{fig:eq_spatial_expand}, we show that the heterogeneous patterns of $\tau_\mathrm{LC}$ and $\tau_\mathrm{sh}$ are highly alike for the start-up shear. 
This likeness is also found in all other flow conditions. 
An example for steady shear is shown in figure~\ref{fig:compare_cr_sh_spatial_dist}. 
The similarity between the DH patterns of $\tau_\mathrm{LC}$ and $\tau_\mathrm{sh}$, together with the universal form of $\tau_\mathrm{sh}(\tau_\mathrm{LC})$ relation 
shown in figures~\ref{fig:compare_eqsh_stopcont} and \ref{fig:compare_multigammai}, suggest an approximate one-to-one correspondence between $\tau_\mathrm{LC}$ and $\tau_\mathrm{sh}$. 
It is interesting to note that in steady-shear cases, $\tau_\mathrm{LC}$ is a hidden parameter because it is not directly measurable in the continuous flow. 
To find $\tau_\mathrm{LC}$, one has to switch off the shear, which is incompatible with the condition of steady shear. 
On the other hand, $\tau_\mathrm{sh}$ can be measured directly in the flow. 
With the one-to-one correspondence between $\tau_\mathrm{LC}$ and $\tau_\mathrm{sh}$, we can obtain the hidden $\tau_\mathrm{LC}$ once the distribution of $\tau_\mathrm{sh}$ is known.

\begin{figure*}
    \includegraphics[width=\linewidth]{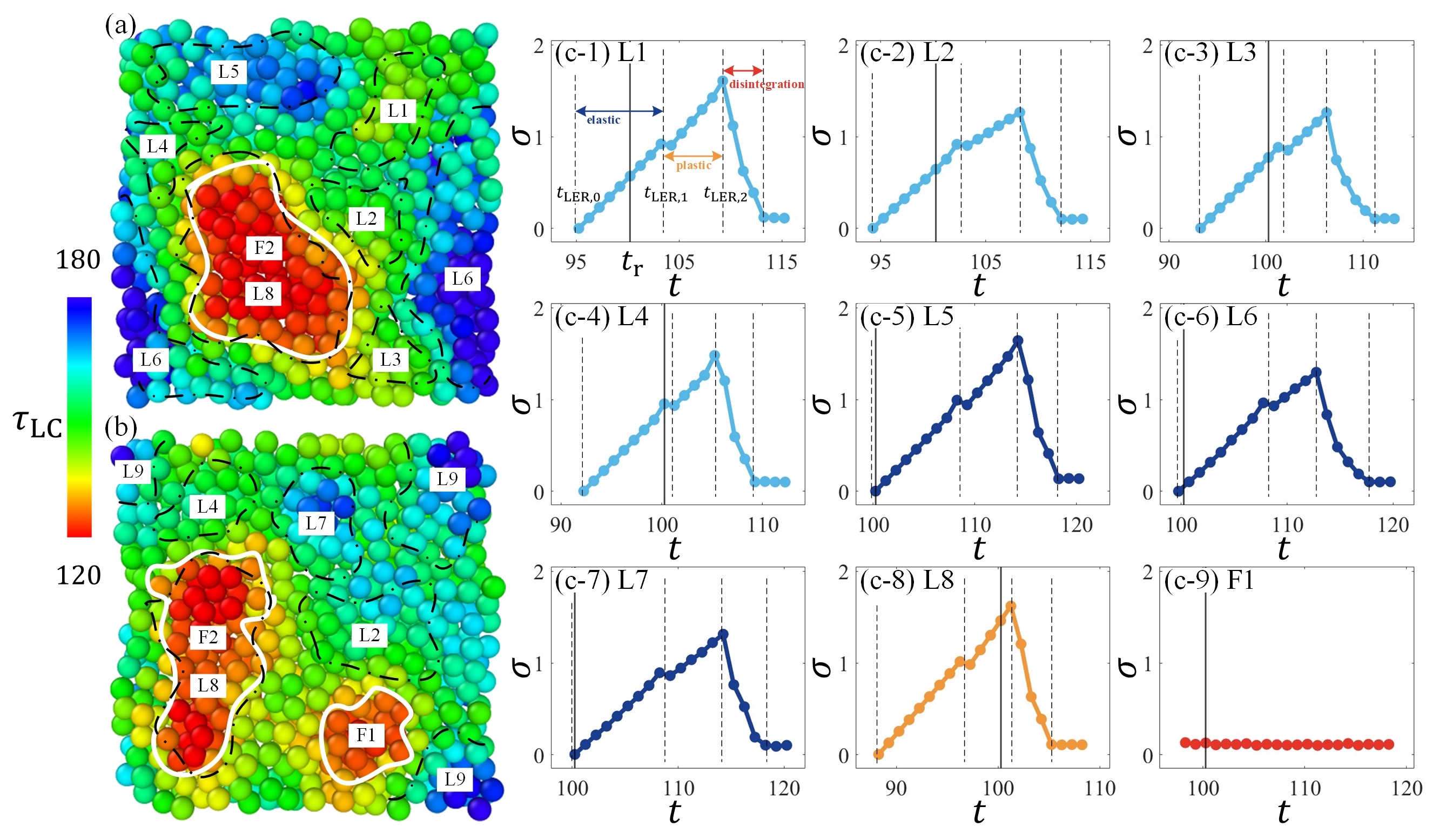}
    \caption{\label{fig:compare_to_ler} 
    DH and heterogeneity in local response under steady shear. 
    (a) and (b) show two slices of the spatial distribution of $\tau_\mathrm{LC}$ of sample A under steady state with $\dot{\gamma}=0.003$ at a reference time $t_\mathrm{r}$. 
    Nine LERs are found at $t=t_\mathrm{r}$. 
    We label them with L1, L2, \dots, L9, and denote their boundaries by black dash-dot lines. 
    The boundaries of slow and fast groups are denoted by white solid lines. 
    There are two fast clusters within these two slices, and are labeled with F1 and F2. 
    In (c-1) -- (c-8), we respectively show the evolution of the stress sustained by the LER for L1 -- L8. 
    In each panel, we use vertical dashed lines to demarcate the three stages, namely, elastic deformation, plastic yielding, and disintegration, as illustrated in (c-1). 
    In (c-9), we show the evolution of the stress of the fast cluster F1. 
    In (c-1) -- (c-9), vertical solid lines are used to denote the reference time $t_\mathrm{r}$.}
\end{figure*}

Knowing the approximate one-to-one correspondence between $\tau_\mathrm{LC}$ and $\tau_\mathrm{sh}$ in flow, 
we conjecture that the local mechanical response, which should be closely related to $\tau_\mathrm{sh}$, exhibits spatial heterogeneity with a pattern similar to that of $\tau_\mathrm{LC}$. 
As to the heterogeneity in response, we proposed a concept of localized elastic region (LER) \cite{wang2022prx} recently, 
inspired by the idea of ``solidity" in supercooled liquids \cite{egami2012prl, dyre1999pre, dyre2006rmp, dyre2024jpc,lemaitre2013prl}. 
An LER, typically composed of hundreds of particles, is a transient, solid-like cluster capable of accumulating elastic stress in the flow of supercooled liquid. 
Under the steady shear, especially in the nonlinear regime, the generation, deformation and relaxation of LER are ubiquitous and successive \cite{wang2022prx, wang2024arxiv}, 
contributing to the viscoelasticity of supercooled liquids. 
The existence of LER signifies the heterogeneity in the local response to shear: 
some regions behave elastically, while other regions behave as viscous liquid. 
Then a question arises naturally: Does LER correlate with the heterogeneous distribution of $\tau_\mathrm{LC}$ and the grouping of fast and slow particles? 
Figure~\ref{fig:compare_to_ler} (a) and (b) show two slices of system A under the steady shear with $\dot{\gamma}=0.003$ at a reference time denoted as $t_\mathrm{r}$. 
The color of each particle denotes its value of $\tau_\mathrm{LC}$ measured at $t=t_\mathrm{r}$. 
With the method given in Ref.~\cite{wang2024arxiv}, we identify nine LERs in the system at $t=t_\mathrm{r}$. 
In figure~\ref{fig:compare_to_ler} (a) and (b), we denote their boundaries with black dash-dot lines, and label them with L1, L2, \dots, L9. 
Meanwhile, we also plot the boundaries between the fast and slow groups determined by equation~\ref{eq:divide_fast_slow} with white solid lines. 
Two distinct fast clusters are found within these two slices, and are labeled with F1 and F2 in figure~\ref{fig:compare_to_ler} (a) and (b). 
In figure~\ref{fig:compare_to_ler} (c-1) -- (c-8), we plot the evolution of the stresses sustained by LERs L1 -- L8 (the stress of L9 is not shown due to the limited space of figure~\ref{fig:compare_to_ler}). 
It is seen that an LER sequentially undergoes elastic deformation, plastic yielding, and sharp disintegration that eventually transforms it into a liquid-like state \cite{wang2024arxiv}. 
An example of these three stages is denoted in figure~\ref{fig:compare_to_ler} (c-1): 
Here, the beginnings of elastic deformation, plastic yielding, and disintegration are denoted as $t_\mathrm{LER,0}$, $t_\mathrm{LER,1}$, and $t_\mathrm{LER,2}$, respectively. 
In our recent work \cite{wang2024arxiv}, we have shown that the appearance of plasticity at $t=t_\mathrm{LER,1}$ is accompanied by the emergence of cage-jump spots, 
where cage jumps highly concentrate within a few diameters of particle, inside the LER. 
By calculating the pseudoharmonic modes \cite{lerner2021prl}, we show that \cite{wang2024arxiv} these soft spots 
constitute the manifestation of the shear transformation zones (STZ) \cite{argon1979acta, langer1998pre, egami2014nat, weitz2007science}. 
This is very similar to amorphous solids, where STZs play as the precursor of the mechanical failure. 
In figure~\ref{fig:compare_to_ler} (c-9), we show the stress evolution of the fast cluster labeled with F1. 
No stress accumulation is seen here. 
The stress is steady, similar to the macroscopic behavior of viscous liquids under steady shear.

In the system, LERs are in different stages at a given time. 
To clarify the stages of the LERs at $t=t_\mathrm{r}$, we denote the reference time $t_\mathrm{r}$ by vertical solid lines in figure~\ref{fig:compare_to_ler} (c-1) -- (c-8). 
As seen in figure~\ref{fig:compare_to_ler} (c-1) -- (c-7), for LERs L1 -- L7 (and L9, not shown in figure~\ref{fig:compare_to_ler}), 
the corresponding LER is in the elastic deformation stage (i.e., within the period of $\left[ t_\mathrm{LER,0}, t_\mathrm{LER,1} \right]$). 
In contrast, for LER L8, as seen from figure~\ref{fig:compare_to_ler} (c-8), it is in the plastic yielding stage (i.e., within the period of $\left[ t_\mathrm{LER,1}, t_\mathrm{LER,2} \right]$), 
and is very close to the disintegration. 
By examining figure~\ref{fig:compare_to_ler}, we find following correlation between LER and the spatial distribution of $\tau_\mathrm{LC}$:

(i) For slow particles, most of them are involved in the elastic deformation stage of LER. 
Their atomic level stresses will continue to build up until the yielding of LER. 
This point is manifested by the observation that the slow region highly overlaps with the regions occupied by LERs L1 -- L7 and L9, 
for which the reference time $t_\mathrm{r}$ is within the elastic stages of these LERs. 
By checking several flow conditions, we find that there are nearly $80\%$ of slow particles are involved in the elastic deformation stage of LER. 

(ii) For fast particles, there are two situations. 
One is exemplified by the fast cluster F1. It is out of all LERs. 
As shown in figure~\ref{fig:compare_to_ler} (c-9), it behaves as a typical viscous liquid. 
The other one is exemplified by the fast cluster F2. 
In space, it nearly coincides with the LER L8, which is very close to its disintegration at $t=t_\mathrm{r}$ as shown in figure~\ref{fig:compare_to_ler} (c-8). 
This coincidence is understandable, since the imminent disintegration of LER involves strong displacements of most of its constituent particles \cite{wang2024arxiv}. 
In either case, the stress of the fast cluster cannot accumulate. 

Summarizing the results shown in figure~\ref{fig:compare_to_ler}, we conclude that under steady shear, the regions composed of slow particles exhibit elastic response to imposed shear, 
while the fast regions behave as viscous liquid. 
In this way, we establish a correlation between DH and local mechanical response. 

\subsection{\texorpdfstring{Functional form of $\tau_\mathrm{sh}(\tau_\mathrm{LC})$}{Functional form of tau\_sh(tau\_LC)}}
\label{sec:exp_para}

\begin{figure}
    \includegraphics[width=\linewidth]{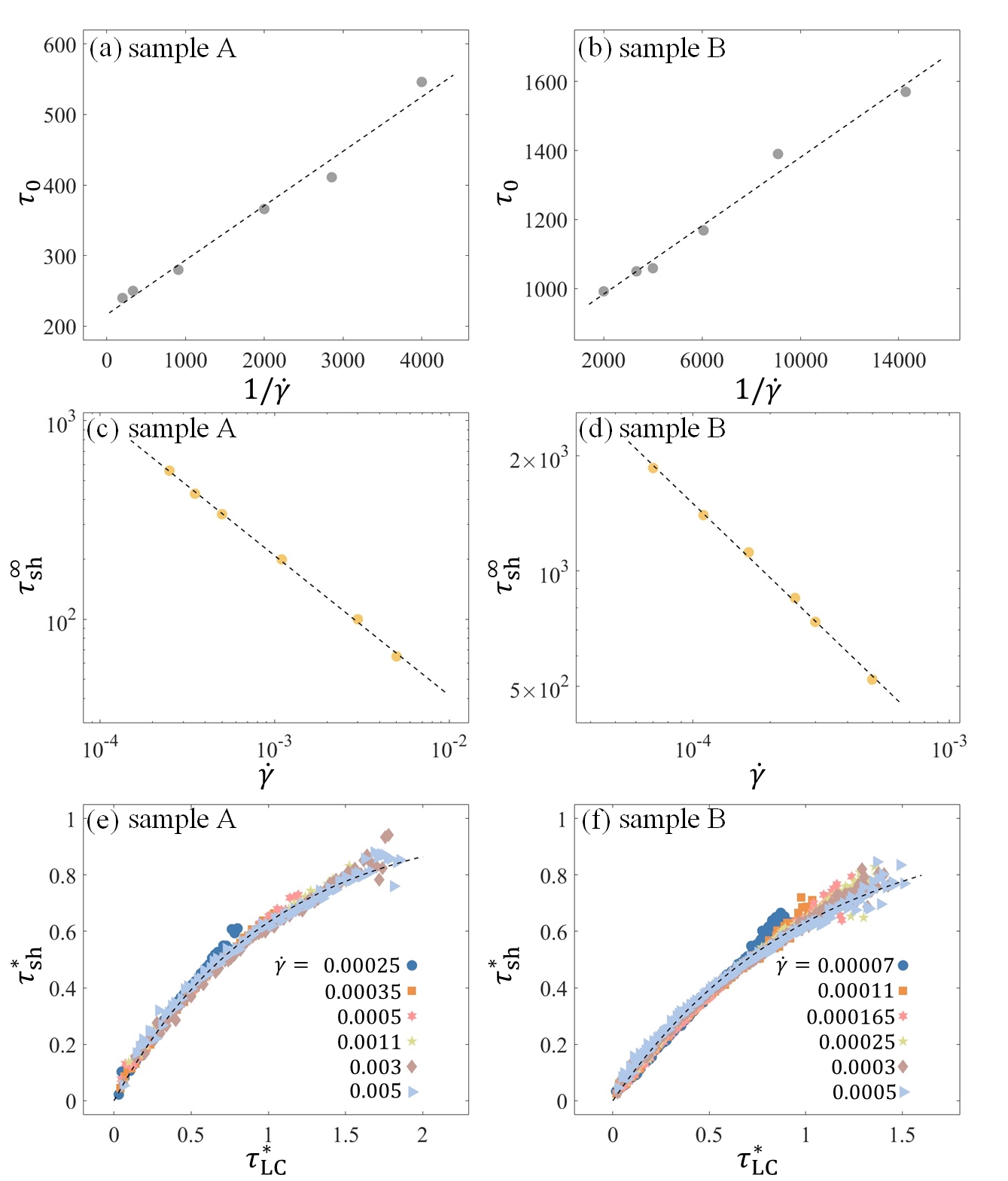}
    \caption{\label{fig:exp_fit_para} 
    (a) and (b) show the fitted $\tau_0$ as a function of $\dot{\gamma}^{-1}$ for sample A and sample B, respectively. 
    Dashed lines in (a) and (b) denote the linear relation between $\tau_0$ and $\dot{\gamma}^{-1}$. 
    (c) and (d) show the fitted $\tau_{\text{sh}}^\infty$ as a function of $\dot{\gamma}$ for sample A and sample B, respectively. 
    Dashed lines in (c) and (d) denote the relation of $\tau_\mathrm{sh}^\infty \sim \dot{\gamma}^{-\lambda_\tau}$. 
    (e) and (f) give the rescaled relations of $\tau_\mathrm{sh}(\tau_\mathrm{LC})$, denoted as $\tau_\mathrm{sh}^*(\tau_\mathrm{LC}^*)$, for sample A and sample B, respectively. 
    Dashed lines in (e) and (f) represent the master curves given by equation~\ref{eq:scaled_exp_fit_transformation_function}.}
\end{figure}

The functional form of $\tau_\mathrm{sh}(\tau_\mathrm{LC})$ plays a critical role in the above discussion. 
In fact, for all cases studied in this work, $\tau_\mathrm{sh}(\tau_\mathrm{LC})$ can be nicely fitted by an exponential form:
\begin{equation}
    \tau_\mathrm{sh}= \tau_\mathrm{sh}^\infty \left [ 1-\exp(-\frac{\tau_\mathrm{LC}}{\tau_0} ) \right ],
    \label{eq:exp_fit_transformation_function}
\end{equation}
where $\tau_\mathrm{sh}^\infty$ and $\tau_0$ are fitting parameters. 
$\tau_0$ governs the position at which $\tau_\mathrm{sh}(\tau_\mathrm{LC})$ significantly deviates from $\tau_\mathrm{sh}=\tau_\mathrm{LC}$. 
It reflects the time scale of the external shear. 
Thus, we expect that $\tau_0$ depends on $\dot{\gamma}^{-1}$ linearly. 
Figure~\ref{fig:exp_fit_para} (a) and (b) give the fitted $\tau_0$ as a function of $\dot{\gamma}^{-1}$ for sample A and sample B, respectively, from where the linearity is clearly seen. 
$\tau_\mathrm{sh}^\infty$ represents the dynamics of the slowest particles in the flow. 
These particles, according to our above analysis, are mostly involved in LERs. 
We can define the characteristic strain for these particles by $\gamma_\mathrm{sh}^\infty = \dot{\gamma} \tau_\mathrm{sh}^\infty$. 
In Ref.~\cite{wang2022prx}, it is demonstrated that in the shear-thinning regime, the strain of LER $\gamma_\mathrm{LER}$ varies 
with $\dot{\gamma}$ by $\gamma_\mathrm{LER} \sim \dot{\gamma}^\epsilon$ with $\epsilon = 1 - \lambda$. 
Therefore, we expect $\gamma_\mathrm{sh}^\infty \sim \dot{\gamma}^\epsilon$, or equivalently, $\tau_\mathrm{sh}^\infty \sim \dot{\gamma}^{-\lambda}$. 
The fitted $\tau_\mathrm{sh}^\infty$ as a function of $\dot{\gamma}$ are plotted in log-log scale for both samples in figure~\ref{fig:exp_fit_para} (c) and (d). 
The relation of $\tau_\mathrm{sh}^\infty(\dot{\gamma})$ indeed exhibits a power-law dependence, denoted as $\tau_\mathrm{sh}^\infty \sim \dot{\gamma}^{-\lambda_\tau}$. 
The fitted value of the power $\lambda_\tau$ is $0.70$ for sample A, and $0.65$ for sample B. 
As we expect, these values are very close to the corresponding $\lambda$. 
Note that the power law of $\tau_\mathrm{sh}^\infty \sim \dot{\gamma}^{-\lambda_\tau}$ covers the whole range of $\dot{\gamma}$ studied in this work, and is not limited to the shear-thinning regime. 
According to these considerations, we can rescale $\tau_\mathrm{sh}$ as $\tau_\mathrm{sh}^*=\tau_\mathrm{sh}/(c \, \dot{\gamma}^{-\lambda})$, 
and rescale $\tau_\mathrm{LC}$ as $\tau_\mathrm{LC}^*=\tau_\mathrm{LC}/(a\dot{\gamma}^{-1}+b)$, where $a$, $b$ and $c$ are material-dependent parameters. 
Then, equation~\ref{eq:exp_fit_transformation_function} is rewritten as
\begin{equation}
    \tau_\mathrm{sh}^*=1-\exp(-\tau_\mathrm{LC}^*).
    \label{eq:scaled_exp_fit_transformation_function}
\end{equation}
We rescale our data found from different conditions with the above method for both samples, and show the results in figure~\ref{fig:exp_fit_para} (e) and (f). 
All data well collapse onto the master curve given by equation~\ref{eq:scaled_exp_fit_transformation_function}. 

Both the values of $\lambda$ of two samples are close to $2/3$. 
Similar results are also observed in other LJ liquids in a wide range of supercooled temperatures \cite{berthier2002jcp}.  
In fact, $2/3$ is the prediction of a mean-field theory \cite{berthier2000pre}. 
Recently, we furtherly show that the value of $\lambda$ is determined by the behaviors of localized elasticity and slightly alters among different samples \cite{wang2022prx,wang2024arxiv}.
Nevertheless, by assuming $\tau_\mathrm{sh}^\infty = c \, \dot{\gamma}^{-2/3}$, the obtained $\tau_\mathrm{sh}^\infty$ still effectively 
collapse the data of $\tau_\mathrm{sh}(\tau_\mathrm{LC})$ with equation~\ref{eq:scaled_exp_fit_transformation_function}. 

We will provide a model analysis on the form of $\tau_\mathrm{sh}(\tau_\mathrm{LC})$ in section~\ref{sec:pel}. 

\section{SOLID-LIQUID DUALITY AND RHEOLOGY}
\label{sec:rheology}
In the preceding section, we establish the correlation between DH and local mechanical response to imposed shear in supercooled liquids. 
It is found that the way of local response highly depends on the comparison between the local mobility, represented by $\tau_\mathrm{LC}$, 
and the time scale of external shear, which linearly depends on $\dot{\gamma}^{-1}$. 
Though the distribution of $\tau_\mathrm{LC}$ value is smooth, the local response exhibits two major ways: solid-like for slow particles, and liquid-like for fast particles. 
This solid-liquid duality in response can be viewed as a specific implementation of Maxwell's insight into viscoelasticity at the microscopic level. 
Additionally, it is crucial to note that slow/fast particles do not randomly distribute through the system. 
Instead, they tend to aggregate into clusters. 
With this picture, we will explore the microscopic mechanisms of two important rheological behaviors for supercooled liquids, namely, steady shear and start-up shear, in this section.

\subsection{Rearrangements of solid-like and liquid-like regions}
\label{sec:rearrange_solid_liquid_region}
\begin{figure}
    \includegraphics[width=\linewidth]{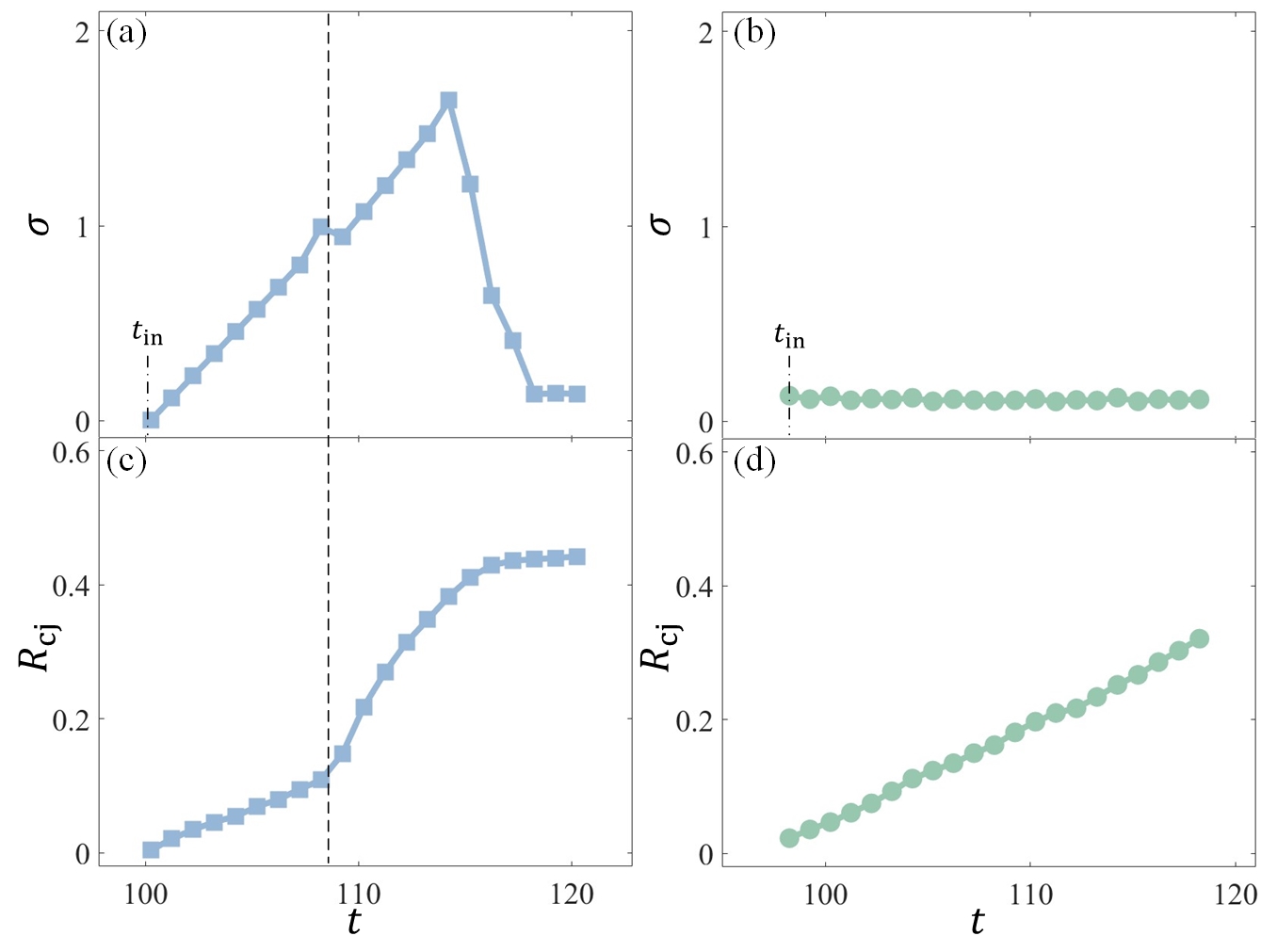}
    \caption{\label{fig:cj_ratio} 
    (a) and (b) reproduce the stress evolutions of the LER L5 (representing a solid-like region) and the fast cluster F1 (representing a liquid-like region) in figure~\ref{fig:compare_to_ler}. 
    $t_\mathrm{in}$ is the time at which the LER L5 starts to accumulate elastic stress. 
    (c) and (d) give the jump ratio $R_\mathrm{cj}(t)$ for the above two regions. 
    In (a) and (c), we denote the end of elastic deformation stage by a vertical dashed line.}
\end{figure}

To employ the microscopic solid-liquid duality, it is important to clarify the difference between the rearrangements of solid-like and liquid-like regions first. 
Figure~\ref{fig:cj_ratio} (a) and (b) display the stress evolutions of a solid-like region (the LER L5 in figure~\ref{fig:compare_to_ler}) 
and a liquid-like region (the fast cluster F1 in figure~\ref{fig:compare_to_ler}) in simple A, respectively. 
For each region, we calculate the ratio of constituent particles that have undergone cage jump since $t=t_\mathrm{in}$, denoted as $R_\mathrm{cj}(t)$. 
Here, $t_\mathrm{in}$, marked in figure~\ref{fig:cj_ratio} (a) and (b), is the beginning time of the elastic deformation of LER L5. 
The results of $R_\mathrm{cj}(t)$ are shown in figure~\ref{fig:cj_ratio} (c) and (d). 
As seen from figure~\ref{fig:cj_ratio} (c), for the solid-like region, $R_\mathrm{cj}(t)$ increases slowly in the elastic deformation stage, and then grows much more rapidly as the cluster enters the plastic yielding process. 
In contrast, $R_\mathrm{cj}(t)$ for the liquid-like region increases in a nearly linear form, as shown in figure~\ref{fig:cj_ratio} (d). 
The difference between figure~\ref{fig:cj_ratio} (c) and (d) reflect that solid-like region and liquid-like region rearrange in fundamentally different ways, as we will discuss in following two paragraphs. 

For liquid-like regions, the linear form of $R_\mathrm{cj}(t)$ suggests that cage jumps take place independently. 
This result is consistent with Ref.~\cite{egami2013prl}, which demonstrates that the phonon propagation induced by a cage jump in normal liquids is highly localized and short-lived. 
In this case, due to the absence of the long-ranged elasticity-mediated interaction, most cage jumps cannot be correlated and, thus, are activated independently. 

For solid-like regions, $R_\mathrm{cj}(t)$ exhibits an evident crossover at the end of the elastic deformation stage. 
To understand this nontrivial phenomenon, we recall that the yielding process of amorphous solids is collective and hierarchical \cite{candelier2009prl}, 
where the facilitation due to the long-ranged anisotropic elasticity-mediated interaction plays a crucial role \cite{barrat2018rmp, biroli2023prl,lemaitre2014prl}. 
Particularly, when only a small fraction of particles undergo rearrangements, the system remarkably loses its ability to further accumulate elastic stress: 
As suggested in Ref.~\cite{ajliu2011prl}, the failure of amorphous solids occurs when only about $10\%$ of particles are involved in STZs. 
As for LER, its response is very similar to that of amorphous solids, as shown in figure~\ref{fig:compare_to_ler} and Ref.~\cite{wang2024arxiv}. 
Especially, the end of its elastic deformation is accompanied by the emergence of cage-jump spots that manifest STZs \cite{wang2024arxiv}. 
With these considerations, the observation shown in figure~\ref{fig:cj_ratio} (c) is understandable: 
During the elastic deformation stage of LER, cage jumps are relatively rare, and $R_\mathrm{cj}(t)$ grows slowly with $t$. 
As $R_\mathrm{cj}(t)$ exceeds about $10\%$, the elasticity cannot be well sustained any longer. 
Then, plastic yielding and disintegration successively occur, accompanied by overwhelming cage jumps that result in the rapid growth of $R_\mathrm{cj}(t)$. 

\subsection{Steady shear}
Under steady shear, both samples exhibit shear thinning, as shown in figure~\ref{fig:viscosity}.
It can be described by a phenomenological expression $\eta(\dot{\gamma}) \approx \eta_\mathrm{eq} \big/ \left [ 1 + (\tau_\eta\dot{\gamma})^\lambda \right]$, 
where $\eta_\mathrm{eq}$ is the zero-shear viscosity, and $\tau_\eta$ is a parameter with the dimension of time \cite{oswald2009, yamamoto1998pre, cross1965jcs}. 
As $\dot{\gamma}$ increases from null, $\eta$ gradually deviates from $\eta_\mathrm{eq}$ and finally behaves with a power law $\eta(\dot{\gamma}) \sim \dot{\gamma}^{-\lambda}$.  
To understand this complicated nonlinear behavior, we employ the solid-liquid duality by expressing the shear stress $\sigma$ as: 
\begin{equation}
    \sigma(\dot{\gamma}) = \frac{1}{2} \, p_\mathrm{slow} \, G_\mathrm{eff} \, \dot{\gamma} \, \tau_\mathrm{el} + (1 - p_\mathrm{slow}) \, \eta_\mathrm{eq} \, \frac{\bar{\tau}_\mathrm{sh, fast}}{\bar{\tau}_\mathrm{LC, eq}} \, \dot{\gamma},  
    \label{eq:predict_stress}
\end{equation}
where the first term in the right-hand side of equation~\ref{eq:predict_stress} represents the contribution from slow particles, 
and the second term represents the contribution from fast particles. 
We discuss the former first. $p_\mathrm{slow}$ is the fraction of slow particles. 
As demonstrated in section~\ref{sec:taucr_steady_shear}, most slow particles are involved in the elastic stage of LERs. 
Thus, this contribution is given by the average elastic stress of LER. 
Here, $G_\mathrm{eff}$ is the effective shear modulus of LER, and $\tau_\mathrm{el}$ is the average time of duration of the elastic deformation of LER. 
In flow, an LER undergoes elastic loading, and then a sharp disintegration to unload. 
The factor $1/2$ accounts for the averaging effect of this saw-tooth-like loading-unloading process. 

$\tau_\mathrm{el}$ and $G_\mathrm{eff}$ are needed to quantify the contribution from slow particles. 
$\tau_\mathrm{el}$ is evaluated as follows. 
As we discuss in the preceding subsection, the end of the elastic deformation stage of an LER is accompanied by the cage jumps of about $10\%$ of its constituent particles. 
Therefore, we determine $\tau_\mathrm{el}$ as the waiting time that $10\%$ of slow particles have relaxed according to the distribution of $\tau_\mathrm{sh}$ value. 

We now turn our attention to the effective modulus of LER $G_\mathrm{eff}$. 
By ``effective", we mean that it encapsulates both elastic and plastic effects \cite{zaccone2017prl, zaccone2014prb}. 
To be specific, even in the elastic stage of LER, small amount of nonaffine displacements of constituent particles take place, as shown in figure~\ref{fig:cj_ratio} (c) and Ref.~\cite{wang2024arxiv}. 
This plasticity slightly weakens the modulus of LER. 
Therefore, the modulus during the elastic stage can be written as $G_\mathrm{eff}=w G_\mathrm{ini}$, 
where $G_\mathrm{ini}$ is the initial shear modulus of LER, and $w$ ($w<1$) is a weakening factor reflecting this plasticity. 
To estimate $w$, we adopt a strategy that ties the variation in shear modulus to the neighbor loss process \cite{zaccone2017prl, zaccone2014prb}. 
The detail for calculating $w$ and $G_\mathrm{eff}$ is given in Appendix \ref{sec:effective_modulus}. 
For all studied conditions, we find that $G_\mathrm{eff} \approx 0.8 G_\mathrm{ini}$.

For fast particles, as we show in figure~\ref{fig:compare_to_ler}, they do not exhibit elastic deformation. 
Instead, such regions behave like viscous liquids. 
Considering that the average persistence time is found to be proportional to viscosity \cite{pastore2021jcp}, 
we can estimate the viscosity of fast particles $\eta_\mathrm{fast}$ by the relation:  
\begin{equation}
    \frac{\eta_\mathrm{fast}}{\eta_\mathrm{eq}} =\frac{\bar{\tau}_\mathrm{sh, fast}}{\bar{\tau}_\mathrm{LC, eq}},
    \label{eq:viscosity_for_fast_particles}
\end{equation}
where $\bar{\tau}_\mathrm{LC, eq}$ is the average of $\tau_\mathrm{LC}$ at equilibrium, 
and $\bar{\tau}_\mathrm{sh, fast}$ is the average of $\tau_\mathrm{sh}$ of fast particles at the given shear rate. 
Here, $\bar{\tau}_\mathrm{sh}$ can be approximated by $\bar{\tau}_\mathrm{LC}$ for fast particles. 
With equation~\ref{eq:viscosity_for_fast_particles}, we obtain the second term in the right-hand side of equation~\ref{eq:predict_stress}.

\begin{figure}
    \includegraphics[width=\linewidth]{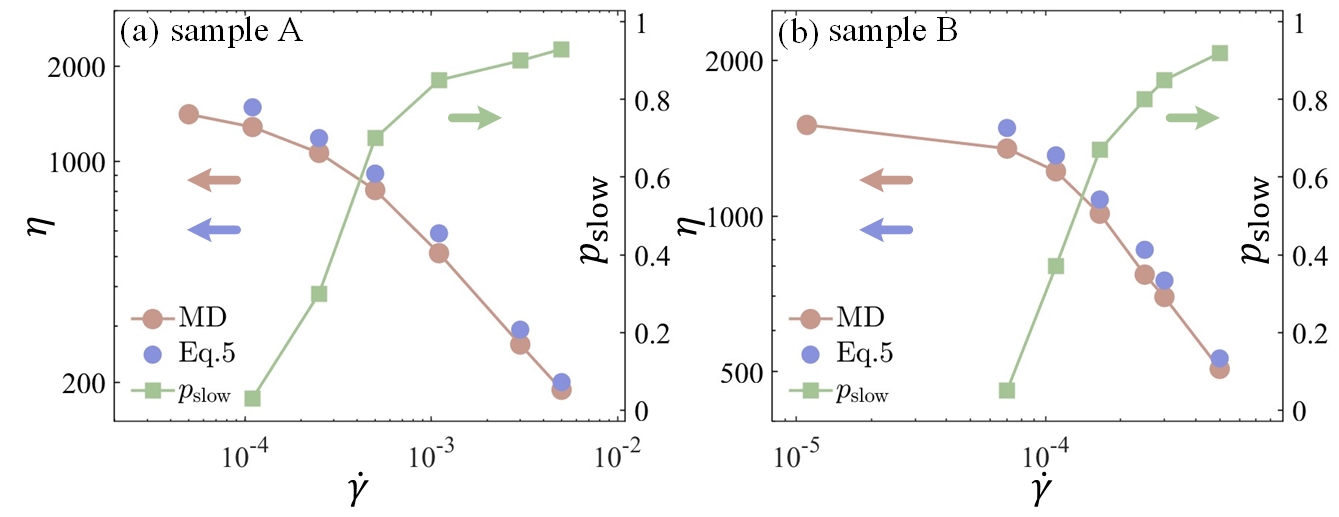}
    \caption{\label{fig:predict_viscosity_and_slowratio} 
    In both (a) (representing sample A) and (b) (representing sample B), the viscosity estimated by equation~\ref{eq:predict_stress}, 
    the viscosity directly measured from MD data, and the proportion of slow particles $p_\mathrm{slow}$ are given. }
\end{figure}

Figure~\ref{fig:predict_viscosity_and_slowratio} shows the results of viscosity calculated by equation~\ref{eq:predict_stress} for two samples. 
The accurate results, shown in figure~\ref{fig:viscosity}, are replotted here for comparison. 
The agreement between two results is remarkable. 
We also show the results of $p_\mathrm{slow}$ in figure~\ref{fig:predict_viscosity_and_slowratio}.
It is seen that $p_\mathrm{slow}$ is negligible at the Newtonian regime. 
As $\dot{\gamma}$ increases, $p_\mathrm{slow}$ becomes finite, meanwhile, noticeable shear thinning appears. 
As $\dot{\gamma}$ further increases, $p_\mathrm{slow}$ grows rapidly, and then saturates in the shear thinning regime. 
These observations clearly reveal the relation between DH and the onset of shear thinning: 
Due to the heterogeneous dynamics, different regions in a supercooled liquid respond to shear in different ways. 
Slow particles tend to aggregate into LERs. 
The accumulation of elastic stress and strain of LER, which is limited by thermal activation and shear-driven effects, 
cannot catch up with the growth of $\dot{\gamma}$. 
Thus, the viscosity contributed by slow particles shear-thins. 
As $\dot{\gamma}$ becomes larger, more particles are involved as slow particles, results in progressively prominent shear thinning. 
When $\dot{\gamma}$ is very large, most particles are involved in LERs. 
Consequently, the nonlinear rheology of the system, described by $\eta \propto \dot{\gamma}^{-\lambda}$, is dominated by the behavior of LER. 
As we demonstrate in Ref.~\cite{wang2022prx,wang2024arxiv}, in the shear-thinning regime, the characteristic length $\xi_\mathrm{LER}$ and strain $\gamma_\mathrm{LER}$ of LER, respectively, 
depend on $\dot{\gamma}$ by $\xi_\mathrm{LER} \propto \dot{\gamma}^{-v}$ and $\gamma_\mathrm{LER} \propto \dot{\gamma}^{\epsilon}$. 
Then, the exponent $\lambda$ is determined by LER through the scaling relation $\lambda =1-\epsilon = (D + 1)v$, where $D$ is the dimension of the system. 

It is worth noting that the onset of noticeable thinning takes place at a shear rate much smaller than $\tau_\alpha^{-1}$ \cite{yamamoto1998pre, mizuno2024comphy, webb1990},
where $\tau_\alpha$ is the equilibrium $\alpha$ relaxation time \cite{hansen2013}. 
The origin of this observation has been much debated. 
Particularly, the mode-coupling-theory (MCT) approach predicts that this onset shear rate is close to $\tau_\alpha^{-1}$ \cite{cates2002prl, miyazaki2002pre, cates2009pnas}. 
It should be noted that the MCT approach highlights the caging effect and advection, while it does not explicitly involve DH. 
Within our framework, the aforementioned observation is easy to understand. 
Owing to DH, there are regions with relaxation times much longer than $\tau_\alpha$. 
Such slow regions can be accelerated by shear at a shear rate much smaller than $\tau_\alpha^{-1}$, which leads to the emergence of shear thinning.

\subsection{Start-up shear}
The success shown in figure~\ref{fig:predict_viscosity_and_slowratio} implies that the particle-level solid-liquid duality provides a potential scenario for understanding the flow of supercooled liquids. 
With above considerations, now we explore the response of supercooled liquids to the start-up shear. 
Our objective is to predict the stress evolution during the start-up process with the knowledge of the equilibrium state, such as the equilibrium distribution of $\tau_\mathrm{LC}$. 
The form of $\tau_\mathrm{sh}(\tau_\mathrm{LC})$ is crucial here. 
As we discussed in section~\ref{sec:exp_para}, for a specific sample, the form of $\tau_\mathrm{sh}(\tau_\mathrm{LC})$ is given by 
equation~\ref{eq:scaled_exp_fit_transformation_function} once a few material-dependent parameters are known. 
Note that for the study of amorphous solids, the deformation and plastic yielding in the start-up process are of central importance \cite{langer1998pre, bonn2017rmp}.
However, studies focusing on the start-up process of supercooled liquids are much less. 
While some research \cite{ilg2017sofmat} has identified the shear banding phenomenon in supercooled liquids under high shear rates, a more general understanding remains elusive.   

Recently, the progress in the elastoplastic model (EPM) significantly advances the quantitative understanding of the start-up behavior of glass \cite{barrat2018rmp,martens2021prl,biroli2022prl,fielding2020prl}. 
In this approach, a glass is discretized into mesoscopic grid cells in space. 
Each grid cell is assigned an initial stress. 
Under external shear, these grid cells undergo stress variations over time, which can be described in terms of cyclic behavior. 
The elementary cycle of a grid cell includes three stages: elastic deformation, plastic yielding, and restructuring for the next loading cycle. 
During the restructuring stage, the grid cell redistributes its accumulated stress through the 
long-ranged anisotropic elasticity-mediated interaction crossing the system \cite{barrat2018rmp, bocquet2004epje}. 

\begin{figure}
    \includegraphics[width=\linewidth]{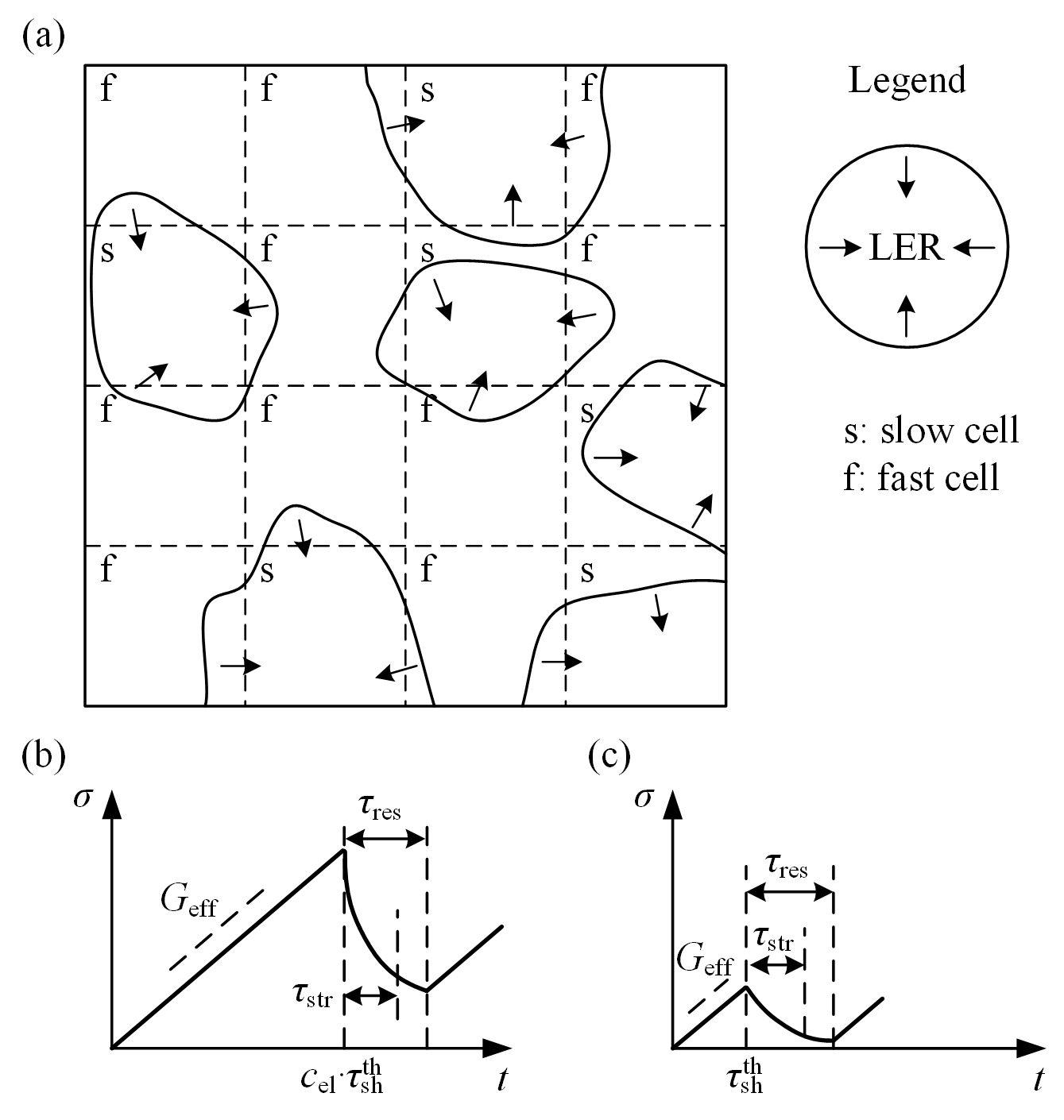}
    \caption{\label{fig:statrup_model_diag} 
    (a) Schematic plot for illustrating the grid cell model of the flow of supercooled liquid. 
    Conceptually, the system is divided into many cells. 
    Since the system contains both slow and fast particles, each cell is classified as either slow (labeled as ``s'') or fast (labeled as ``f''). 
    According to our analysis, LERs, represented by the curves, are expected to be predominantly composed of slow particles. 
    In slow cells, phonons can propagate (as indicated by the arrows); otherwise, their propagation is damped. 
    (b) One stress cycle of a slow cell. 
    (c) One stress cycle of a fast cell.  }
\end{figure}

Borrowing some ideas from EPM, we construct a simple model for supercooled liquids under shear. 
We still assume that the system is separated into many cells. 
Nevertheless, there is a key difference between our model and EPM: 
In our model, the inter-cell elasticity-mediated interaction is not present, and cells are independent of each other. 
To understand this setting, we recall that in supercooled liquids, LERs are localized, and are separated by liquid-like regions \cite{wang2022prx, wang2024arxiv}. 
In this case, phonons can only propagate within LER, while are quickly damped in liquid-like regions \cite{egami2013prl}. 
Therefore, if we assume that a cell conceptually corresponds to an LER or a liquid-like region (see figure~\ref{fig:statrup_model_diag} (a) for an illustration), 
it will be unnecessary to consider the elasticity-mediated interaction between cells. 
Consequently, we assume that each cell evolves independently. 
Notice that in supercooled liquids, it is suggested that the relaxations of different dynamic regions are also coupled \cite{bouchaud1996jpa, heuer2005pre}. 
To describe this effect, a coupling factor can be introduced \cite{heuer2005pre}. 
At the current stage, we ignore this effect to keep the model as simple as possible.

As shown in figure~\ref{fig:eq_spatial_taucr_dist}, the spatial distribution of $\tau_\mathrm{LC}$ is smooth and tends to cluster in space. 
Thus, we can assign an average $\tau_\mathrm{LC}$ value to each cell. 
In the beginning of the shear, we sample the average $\tau_\mathrm{LC}$ value for each cell according to the equilibrium distribution of $\tau_\mathrm{LC}$ value ($H_\mathrm{eq}(\tau_\mathrm{LC})$). 
Then, cells can be identified as fast or slow according to equation~\ref{eq:divide_fast_slow}.

Recall that $\tau_\mathrm{LC}$ is obtained with ICE, where the thermal fluctuation is eliminated. 
In other words, $\tau_\mathrm{LC}$ should be regarded as the thermal-average jump time for a particle. 
In the presence of thermal fluctuation, we suggest that the actual jump time, $\tau_\mathrm{LC}^\mathrm{th}$, 
follows a distribution with $\tau_\mathrm{LC}$ being the mean, which is given by: 
\begin{equation}
    p_\mathrm{th}(\tau_\mathrm{LC}^\mathrm{th})=\frac{1}{\tau_\mathrm{LC}}\exp(-\frac{\tau_\mathrm{LC}^\mathrm{th}}{\tau_\mathrm{LC}}),
    \label{eq:taucrth_dist}
\end{equation}
To account for this thermal effect, after assigning each cell a $\tau_\mathrm{LC}$ sampled from $H_\mathrm{eq}(\tau_\mathrm{LC})$, 
we use the preceding distribution to sample $\tau_\mathrm{LC}^\mathrm{th}$ for each cell. 
With $\tau_\mathrm{LC}^\mathrm{th}$, the local dynamics under shear, represented by $\tau_\mathrm{sh}^\mathrm{th}$, can be found through equation~\ref{eq:scaled_exp_fit_transformation_function}. 

\begin{figure}
    \includegraphics[width=\linewidth]{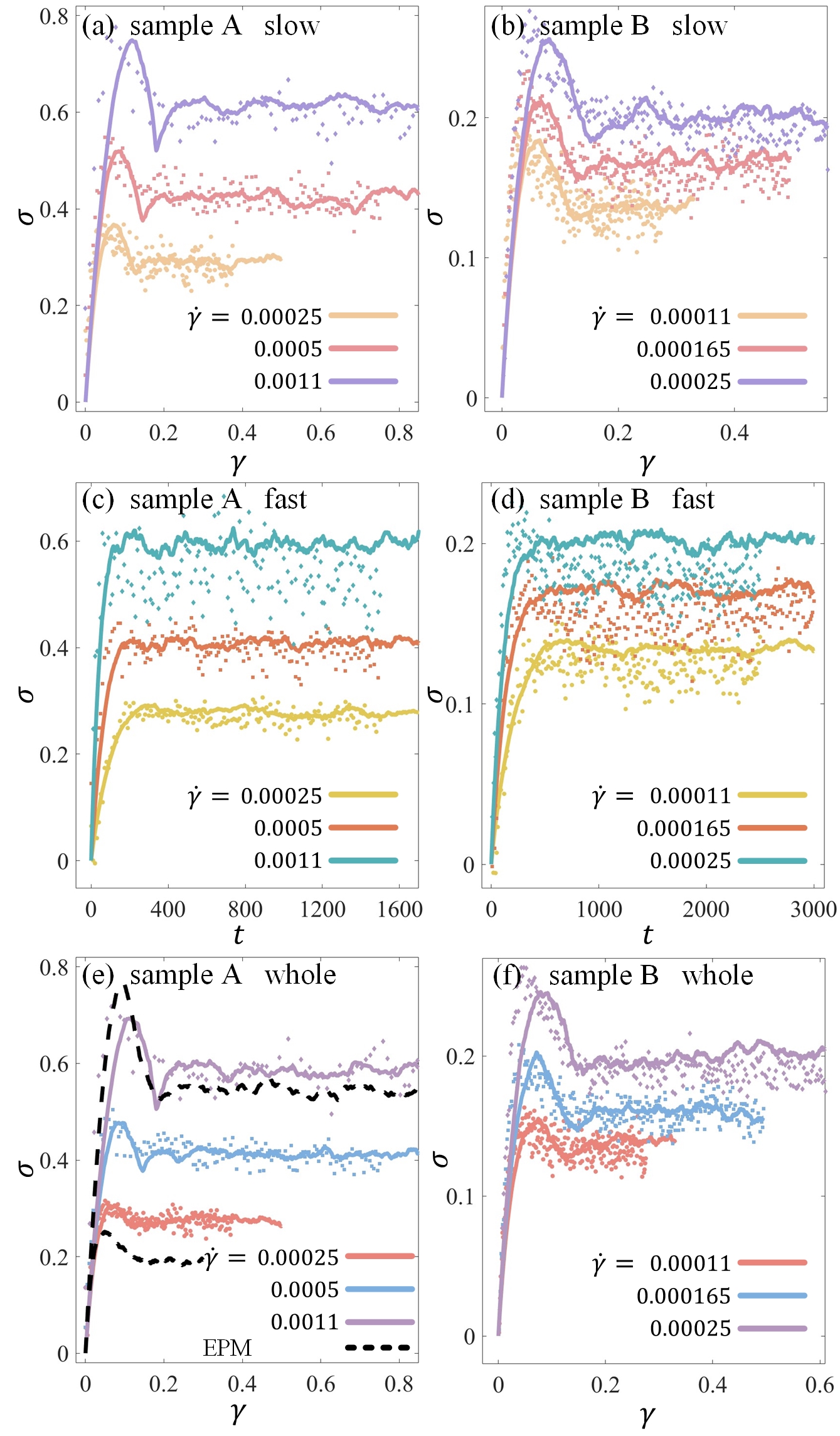}
    \caption{\label{fig:predict_startup_stress} 
    Stress responses of both samples during the start-up process with several $\dot{\gamma}$.
    (a) and (b) respectively show the results of slow particles in sample A and sample B. 
    (c) and (d) respectively show the results of fast particles in sample A and sample B. 
    Note that, ``slow" and ``fast" denote the classification at the beginning of shear. 
    (e) and (f) respectively show the results of the whole system for sample A and sample B. 
    In all panels, symbols denote the MD results, and solid lines denote the results predicted by our model. 
    In (e), black dashed lines denote the results calculated by the EPM approach. 
    For the results of slow particles and the whole system, we adopt $\gamma=\dot{\gamma}t$ as the time variable.}
\end{figure}

Now we discuss the evolution of cell. 
Slow cells behave as solid and can be viewed as LERs. 
As we discuss in section~\ref{sec:rearrange_solid_liquid_region}, the yielding and rearrangement of LER are collective due to the existence of the elasticity-mediated interaction. 
To embody this consideration in modeling, we recall our calculation in the preceding subsection: 
We use $\tau_\mathrm{el}$ as the duration time of the elastic stage of LER, rather than the average of $\tau_\mathrm{sh}$ of slow particles $\bar{\tau}_\mathrm{sh, slow}$. 
$\tau_\mathrm{el}$ is noticeably shorter than $\bar{\tau}_\mathrm{sh, slow}$. 
$\tau_\mathrm{el}$ corresponds to the waiting time for about $10\%$ of slow particles to relax. 
In contrast, $\bar{\tau}_\mathrm{sh, slow}$ represents the time at which nearly half of slow particles have relaxed. 
For such a long time, the slow region has long lost its elasticity. 
To capture this effect, we introduce the ratio $c_\mathrm{el}$:
\begin{equation}
    c_\mathrm{el} = \tau_\mathrm{el} / \bar{\tau}_\mathrm{sh, slow}.
\end{equation}
Then, for a slow cell whose dynamics under shear is characterized by $\tau_\mathrm{sh}^\mathrm{th}$, its duration of elasticity can be written as $c_\mathrm{el} \, \tau_\mathrm{sh}^\mathrm{th}$.

The slow cell starts to yield after the elastic stage. 
Drawing upon the EPM framework \cite{martens2021prl}, we employ an exponential form for the stress decrease during the unloading as $\sigma \sim \exp(-t/t_\mathrm{str})$, 
where $t_\mathrm{str}$ represents a ballistic timescale. 
$t_\mathrm{str}$ is estimated by particle's mean-square displacement (MSD) data prior to the onset of plateau behavior. 
For the sake of simplicity, we assume a constant $t_\mathrm{str}$ for all shear rates. 
Throughout this unloading stage, the cell is restructuring and preparing for the ensuing loading. 
Following Ref.~\cite{martens2016prl}, we introduce a restructuring strain $\gamma_\mathrm{res}$. 
Given a shear rate $\dot{\gamma}$, the restructuring time $\tau_\mathrm{res}$ is modeled as an exponentially distributed random variable with the mean given by $\gamma_\mathrm{res}/\dot{\gamma}$. 
Figure~\ref{fig:statrup_model_diag} (b) illustrates a cycle of a slow cell. 

In fast cells, the elasticity spanning an extended spatial region is not present. 
Cage jump is highly independent, in the sense that the phonon propagation induced by such a flow event is very short-lived and localized \cite{egami2013prl}.
Therefore, we can put our attention on the cage. 
Under shear, the cage of a fast particle slightly deforms \cite{hess1986pra,hess1983pla}
and rearranges quickly after a jump time $\tau_\mathrm{sh}$. 
The massive and successive cage jumps in a fast region result in a stress evolution that is relatively stationary, without an overall stress accumulation process. 
To model the cycle of fast cells, however, we still assume a deformation--yielding process, as illustrated in figure~\ref{fig:statrup_model_diag} (c). 
This form should not be viewed as the evolution of a solid-like cluster. 
On the contrary, it represents the averaged \textit{cage distortion--cage jump} process in a fast cell. 
Notice that, here we still use $G_\mathrm{eff}$ as the slope of stress, as shown in figure~\ref{fig:statrup_model_diag} (c). 
In principle, the modulus here should be smaller than that of solid-like regions. 
However, considering that modulus is not very sensitive to relaxation time \cite{dyre2006rmp}, we adopt $G_\mathrm{eff}$ for fast cells as an acceptable approximation. 
By assuming a similar apparent form of cycle for both fast and slow cells, we simplify the calculation of our model.

After the first cycle, a cell will sample a new $\tau_\mathrm{LC}$, which corresponds to new $\tau_\mathrm{LC}^\mathrm{th}$ and $\tau_\mathrm{sh}^\mathrm{th}$. 
Then, the next cycle starts. 
The treatment is the same: The new cycle will be identified as slow or fast. 
Then, it evolves following the corresponding form of cycle. 
This process will be carried out repeatedly. 
Here, we still use $H_\mathrm{eq}(\tau_\mathrm{LC})$ for sampling $\tau_\mathrm{LC}$ of subsequent cycles. 
The stress of the system is given by the average stress of all cells. 
In this model, for a specific sample, the only free adjustable parameter is the restructuring strain $\gamma_\mathrm{res}$.

In figure~\ref{fig:predict_startup_stress}, we compare the start-up stress predicted by our model with the accurate results for both samples. 
In each column, we display the stress evolutions of slow particles, fast particles, and the whole system in sequence for a specific sample. 
Here, ``slow" and ``fast" denote the classification at the beginning of shear. 
It is seen that this simple model captures some essential features: 

(i) For slow particles, stress exhibits a pronounced overshoot before yielding. 
Both the magnitude of the stress overshoot and the yielding strain increase with $\dot{\gamma}$. 

(ii) For fast particles, the stress evolves in a smoother way, which can be described by $\sigma(t) \propto \left [ 1 - \exp(-t/\tau_\mathrm{f}) \right ]$.
The waiting time prior to steady state $\tau_\mathrm{f}$ decreases as $\dot{\gamma}$ increases. 

There are some aspects that our model does not well align with the MD results, particularly at large shear rates. 
First, our model displays an undershoot following the overshoot. 
This undershoot arises from our assumption that $\tau_\mathrm{sh}^\mathrm{th}$ is solely determined from $\tau_\mathrm{LC}^\mathrm{th}$ through equation~\ref{eq:scaled_exp_fit_transformation_function}. 
In reality, there could be other uncertainties in this relation. 
By incorporating additional parameter, such as the yielding rate \cite{martens2021prl}, this undershoot can be mitigated. 
More notably, the model overestimates the yielding strain for shear rates within the shear-thinning regime. 
This problem should be related to the fact that at high shear rates, the majority of the system behaves as solid-like regions. 
In this case, the behaviors of different cells are inevitably correlated by the elasticity-mediated interaction. 
As discussed above, the key difference between our model and EPM is that our model does not incorporate the correlation between different cells, while EPM does. 
Following this line of thought, we perform an EPM analysis \cite{barrat2018rmp, martens2021prl, martens2016prl, fielding2024prl} on our samples. 
The details of the EPM we use will be given in a separated paper. 
In figure~\ref{fig:predict_startup_stress} (e), we plot the EPM predictions for sample A at $\dot{\gamma}=0.00025$ and $0.0011$. 
It is seen that for $\dot{\gamma}=0.0011$ (within the shear-thinning regime), the EPM gives a better prediction on the yielding strain than our model. 
However, at $\dot{\gamma}=0.00025$ (within the crossover regime), for which the solid-like behavior is not dominant, the performance of EPM is worse than our model. 
The reason could be that the inter-cell correlation mediated by elasticity, which is essential in EPM, is weak here due to the presence of massive liquid-like regions. 
In principle, EPM treats the mechanical behavior from the viewpoint of solids. 
On the contrary, the starting point of our model is the viewpoint of supercooled liquids---the heterogeneities in dynamics and response. 
Considering these results, a possible improvement of our model is to introduce the inter-cell correlation between adjacent solid-like cells. 
This work is in progress.

\section{STRUCTURAL BASIS}
\label{sec:structure}
The above analysis shows that the intricate flow behaviors of supercooled liquids can be understood by focusing on $\tau_\mathrm{sh}$ and $\tau_\mathrm{LC}$, 
which are connected by the relation $\tau_\mathrm{sh}(\tau_\mathrm{LC})$ given by equation~\ref{eq:scaled_exp_fit_transformation_function}. 
There are several key observations related to $\tau_\mathrm{sh}$ and $\tau_\mathrm{LC}$: 
First, $\tau_\mathrm{LC}$ exhibits DH features, in the sense that $\tau_\mathrm{LC}$ exhibits a broad distribution and significant clustering in space. 
Second, the functional form of $\tau_\mathrm{sh}(\tau_\mathrm{LC})$ of a liquid is universal once the final shear rate is given. 
To be specific, it converts a $\tau_\mathrm{LC}$ value to the corresponding $\tau_\mathrm{sh}$, regardless of the configuration from which $\tau_\mathrm{LC}$ is extracted. 
The degeneracy of $\tau_\mathrm{LC}$ implies that there exists an underlying configurational parameter (or a set of parameters) that 
can be applied to both equilibrium and shear states to determine the $\tau_\mathrm{LC}$ value through the same functional form. 
Third, as the final shear rate increases, $\tau_\mathrm{sh}(\tau_\mathrm{LC})$ bends at smaller $\tau_\mathrm{LC}$ with more remarkable flattening. 
This change directly leads to the increase of the solid-like component in flow, which critically enhances shear thinning. 
Here, a question arises: How can we understand these dynamical behaviors within a unified framework? 

As we mentioned in section~\ref{sec:intro}, in the last two decades, substantial efforts have been made to explore the role of structure in determining dynamics of supercooled liquids and glasses. 
Inspired by this thinking, in this section, we aim to seek the structural basis of the dynamical behaviors mentioned in the preceding paragraph. 
The key is to construct a parameter or a set of parameters using the structural properties of equilibrium state and shear states with various $\dot{\gamma}$. 
By employing this structural parameter, we reveal a connection between microscopic structure and dynamics for flowing supercooled liquids. 
This connection, together with the established relation between $\tau_\mathrm{LC}$ and rheology, creates a pathway from microscopic structure to macroscopic flow behavior.  

\subsection{\texorpdfstring{Structural parameters for $\tau_\mathrm{LC}$}{Structural parameters for tau\_LC}}
\label{sec:linear_regression}
The discussion in this section will be based on the condition of steady shear. 
First, we will establish the connection between $\tau_\mathrm{LC}$ and configuration. 
In literature of supercooled liquids and glasses, numerous structural parameters have been proposed to predict dynamics, ranging from simple, physically informed metrics to more complex ones. 
Some examples include free volume \cite{han2024prl, enma2016nat}, local favored structures \cite{royall2013jcp, tanaka2019natrev}, soft modes \cite{harrowell2008natphy}, structural entropy \cite{chen2023prl}, etc. 
In recent years, approaches based on statistical learning have gained prominence \cite{ajliu2015prl, manning2020prm, bapst2020natphy, filion2021prl}.  
Especially, some sophisticated models, such as graph neural network \cite{bapst2020natphy, yuhaibin2024scienceadvance} and multilayer perceptron \cite{berthier2023prl}, 
have been employed and achieved unprecedentedly high prediction accuracies compared with the single, simpler parameters, albeit at the cost of being order-agnostic. 
Notably, linear regression method offers a balance between high prediction accuracy and interpretability. 
Another advantage of linear regress method is that it does not need a very large data set. 
In previous studies \cite{filion2021prl, filion2023jcp}, Filion \textit{et al}. have connected the local structure of reference particle to its dynamic propensity using linear regression, 
which achieves an accuracy comparable to that of more complex models. 
In this subsection, linear regression will be used to connect $\tau_\mathrm{LC}$ and local configuration. 

The first step is to construct a data set to train the model. 
Here, we include configurations under steady shear with all studied shear rates in our data set. 
Specifically, the shear rates span from the Newtonian regime to the shear-thinning regime. 
For each $\dot{\gamma}$, we extract $10$ configurations in the steady state. 
The time interval between adjacent configurations is greater than $4\tau_\alpha$, where $\tau_\alpha$ is the $\alpha$ relaxation time at the corresponding state. 
We then measure $\tau_\mathrm{LC}$ for every particle in each configuration in the data set. 
Additionally, we have verified that including more configurations does not alter our main findings. 

Next, for a reference particle $i$ in a given configuration, we use the following quantity to describe its local configuration, as suggested by Ref.~\cite{filion2021prl}:
\begin{equation}
g_{l,i}^m(s,r)=\frac{1}{r^2}  {\textstyle \sum_{j \in s} \exp \left [ -\frac{(r-r_{ij})^2}{2\delta^2} \right ] Y_l^m(\boldsymbol{r}_{ij})},
\label{eq:glm_for_particle}
\end{equation}
where $r$ denotes the distance to particle $i$, $r_{ij}$ is the displacement between particle $j$ and particle $i$, 
$s$ denotes the species of particles (big or small) whose distribution we want to probe, and $Y_l^m (\boldsymbol{r}_{ij})$ is the $(l,m)$-order spherical harmonic function. 
It can be seen that $g_{l,i}^m(s,r)$ gives the spherical harmonic expansion coefficient of the transient density of neighbors belonging to species $s$ at distance $r$ from the reference particle $i$. 
To ensure that the density varies smoothly in space, we use a Gaussian density with a width $\delta$ in the sum. 
We set $\delta=0.1$, and consider $48$ values of $r$ including $0.3, 0.4, 0.5, \cdots, 5.0$. 

\begin{figure}
    \includegraphics[width=\linewidth]{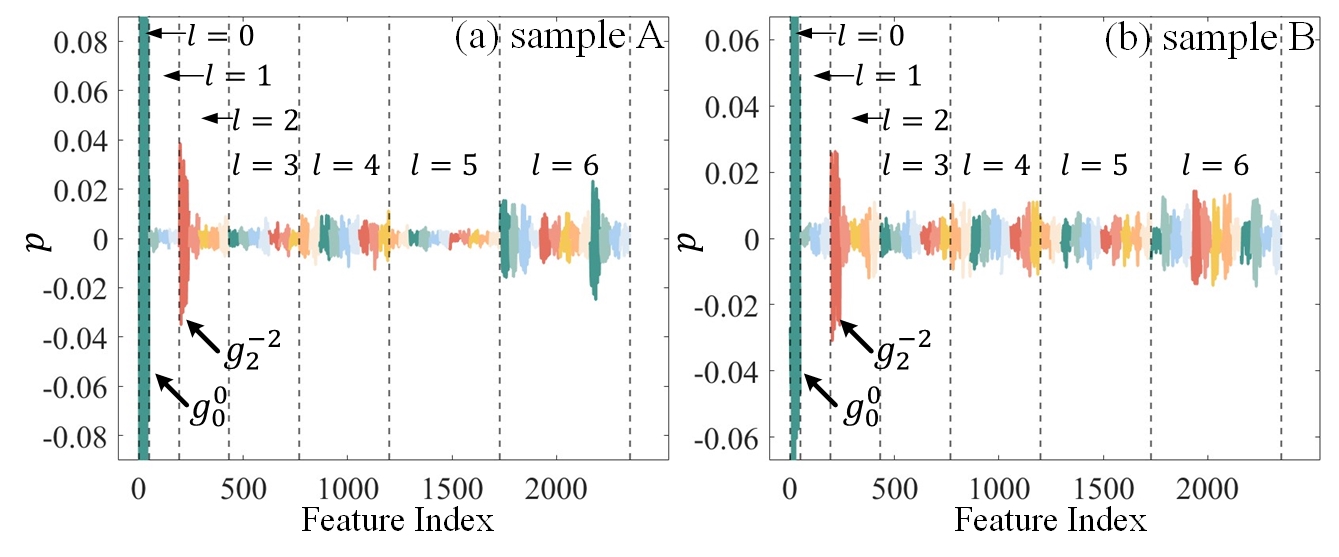}
    \caption{\label{fig:lm_correlation} 
    Correlation between $\tau_\mathrm{LC}$ and the local structural feature defined by equation~\ref{eq:glm_for_particle}.
    (a) and (b) give the results of sample A and sample B, respectively. 
    The correlation is characterized by the Pearson correlation coefficient $p$. 
    The horizontal axis represents the element in the feature vector. 
    Each element is characterized by $l$, $m$ and $r$ (only big particles are considered here). 
    $l$ ranges from $0$ to $6$, dividing the horizontal coordinates into $7$ regimes separated by vertical dashed lines. 
    For each $l$, the values of $m$ are arranged from $-l$ to $l$. 
    Different $m$ features with the same $l$ are distinguished by different colors. 
    For each $(l,m)$ combination, there are $48$ values of $r$. 
    In total, there are $2352$ elements are shown. 
    Two prominent peaks are highlighted by black arrows, representing the features constructed using $g_0^0$ and $g_2^{-2}$. }
\end{figure}

For a given combination of $(l,m)$, the feature vector $g_l^m$ comprises $96$ elements, involving $48$ $r$ values and $2$ species. 
In principle, one may include numerous $(l,m)$ combinations to formulate the total feature vector for a particle. 
However, too many $(l,m)$ combinations would yield a feature vector that is cumbersome and computationally intensive. 
To make our feature representation as simple as possible, it is necessary to judiciously select a subset of $(l,m)$ combinations to serve as the major features. 
For this purpose, we examine the importance of every $(l,m)$ combination in affecting dynamics up to $l=6$ by 
computing the Pearson correlation coefficient $p$ between each element in the feature vector with respect to the given $(l,m)$ and the target property $\tau_\mathrm{LC}$. 
For the sake of simplicity, only big particles are considered here. 
Thus, $48$ elements, covering all $r$ values, are examined for a $(l,m)$ combination. 
Considering that the number of $(l,m)$ combinations up to $l=6$ is $49$, we examine $48 \times 49=2352$ elements in total for each sample. 
Figure~\ref{fig:lm_correlation} shows the results. 
The horizontal axis represents the feature index ranging from $1$ to $2352$. 
These indexes are divided into $7$ regimes according to the value of $l$. 
For each $l$, the values of $m$ are arranged from $-l$ to $l$; and for each $(l,m)$ combination, the values of $r$ are arranged from $0.3$ to $5.0$ in figure~\ref{fig:lm_correlation}. 
Two prominent peaks in figure~\ref{fig:lm_correlation} are seen, respectively corresponding to $(l=0,m=0)$ and $(l=2,m=-2)$. 
The $(l=0,m=0)$ features reflect the packing efficiency around the reference particle, and it has been reported that local packing capacity is highly correlated with dynamics \cite{ajliu2016natphy, tanaka2018prx}.
The $(l=2,m=-2)$ features have a symmetry consistent with the shear geometry \cite{wang2022prx, hess1986pra, hess1983pla}, 
and thus reflect the local shear strain \cite{wang2022prx, schall2016sofmat, egami2012prl, wang2019pccp}.  
As discussed in the previous sections, different $\tau_\mathrm{LC}$ values correspond to different deformation phases in solid-like regions. 
A larger local strain indicates that it is closer to the local yielding point, leading to greater instability. 
Thus, the high correlation of $(l=2,m=-2)$ features is expected. 
In passing, the importance of the local packing and configurational distortion in the dynamics and rheology of supercooled liquids has been highlighted in recent studies \cite{furukawa2023prr,furukawa2017pre}.
It is also seen that features corresponding to $l=6$ show relatively high correlation. 
This is consistent with the studies reporting the existence of icosahedral pattern in supercooled liquids, which is strongly correlated with dynamics \cite{kob2020pnas, kob2023pnas}. 

According to the observation shown in figure~\ref{fig:lm_correlation}, we use the features with $(l=0,m=0)$ and $(l=2,m=-2)$ to construct the feature vector. 
For particle $i$, the feature vector is given by:
\begin{equation}
    X_i^{(0)}=\left ( g_{0,i}^0(s,r), g_{2,i}^{-2}(s,r) \right ).
    \label{eq:feature_vector_generation_0}
\end{equation}
Incorporating both species and $48$ $r$ values, the elements of $X_i^{(0)}$ amount to $192$. 
The superscript ``$(0)$" denotes that this is the zeroth generation of a series of feature vectors. 

It is seen that both solid-like and liquid-like regions are spatially extended, indicating the existence of medium-ranged order. 
In literature, medium-ranged orders have also been emphasized on glass transition \cite{kob2023pnas, sastry2018prl, egami2021pre, enma2006nat}.
To incorporate this medium-ranged structural effect, we adopt a coarse-graining method \cite{tanaka2018prx, filion2021prl}. 
Specifically, the next generation of the feature vector is produced by coarse-graining the last generation by:
\begin{equation}
    X_i^{(n+1)}=\frac{1}{Z} {\textstyle \sum_{j, r_{ij}<r_\mathrm{c}}\exp(-\frac{r_{ij}}{r_\mathrm{c}})X_j^{(n)}},
    \label{eq:feature_coarse_grain}
\end{equation}
where $r_\mathrm{c}$ is a parameter characterizing the range of coarse-graining and is set to $4$, and $Z$ is written as:
\begin{equation}
    Z={\textstyle \sum_{j, r_{ij}<r_\mathrm{c}}\exp(-\frac{r_{ij}}{r_\mathrm{c}})}.
\end{equation}
Another advantage of the coarse-graining is that it averages out irrelevant thermal noise. 
With equations~\ref{eq:feature_vector_generation_0} and \ref{eq:feature_coarse_grain}, we construct the final feature vector for particle $i$ by combining the feature vectors with generations $0$, $1$ and $2$:
\begin{equation}
    X_i=\left ( X_i^{(0)}, X_i^{(1)}, X_i^{(2)} \right ).
    \label{eq:feature_vector}
\end{equation}
There are $576$ elements in $X_i$.

\begin{figure}
\includegraphics[width=\linewidth]{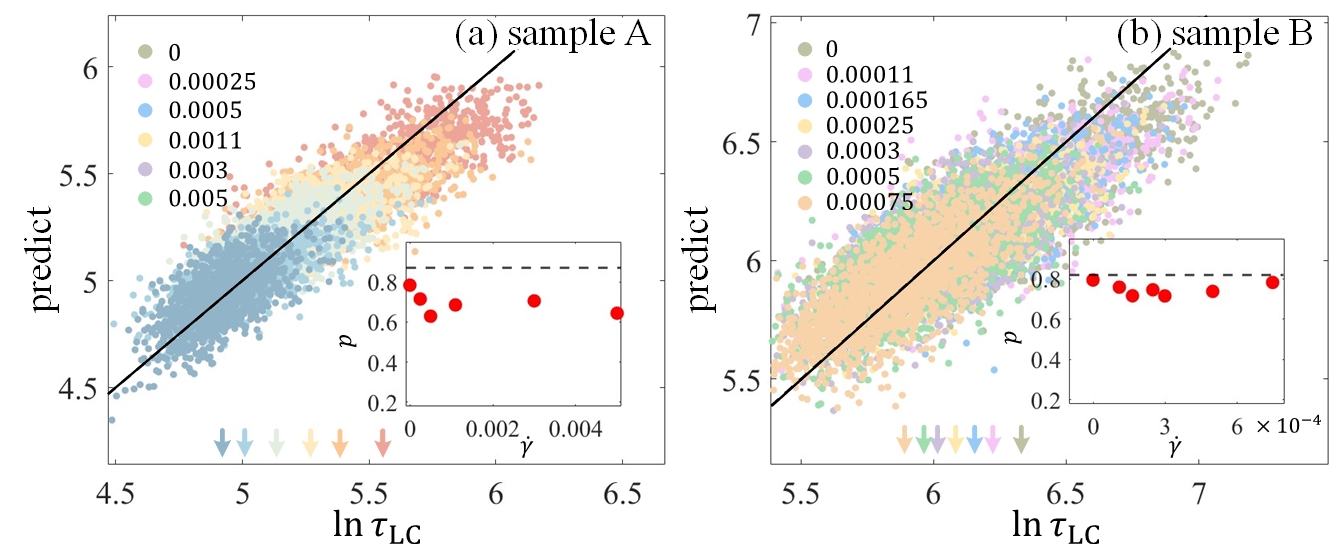}
    \caption{\label{fig:linearregression_predict_true} 
    Comparison between the true values of $\ln(\tau_\text{cr})$ and the values predicted by the linear regression model for particles extracted from configurations at various $\dot{\gamma}$. 
    (a) and (b) show the results of sample A and sample B, respectively. 
    Each point corresponds to a particle. 
    Points belonging to the same $\dot{\gamma}$ are in the same color. 
    Black solid lines denote the situation that the predicted value equals to the true value. 
    Arrows positioned at the bottom of panels indicate the average $\ln(\tau_\text{cr})$ for each $\dot{\gamma}$. 
    Inset: Pearson correlation coefficient between the true and predicted values of $\ln(\tau_\text{cr})$ for each $\dot{\gamma}$.
    Horizontal dashed line represents the Pearson correlation coefficient calculated using the data across all $\dot{\gamma}$. }
\end{figure}

Now, we try to establish the connection between $\tau_\mathrm{LC}$ and the local configuration represented by $X_i$. 
Note that, in many studies on glass transition, researchers explore the connection between $\tau_\alpha$ and structural parameters through a Vogel-Fulcher-Tammann form \cite{tanaka2019nat, smallenburg2020prl}. 
Inspired by this approach, we transform $\tau_\mathrm{LC}$ to $\ln \tau_\mathrm{LC}$ and employ the Ridge linear regression model \cite{sklearn} to
establish the correlation between feature vectors and $\ln \tau_\mathrm{LC}$. 
Such a transformation will be further rationalized in the next section.

\begin{figure}
\includegraphics[width=\linewidth]{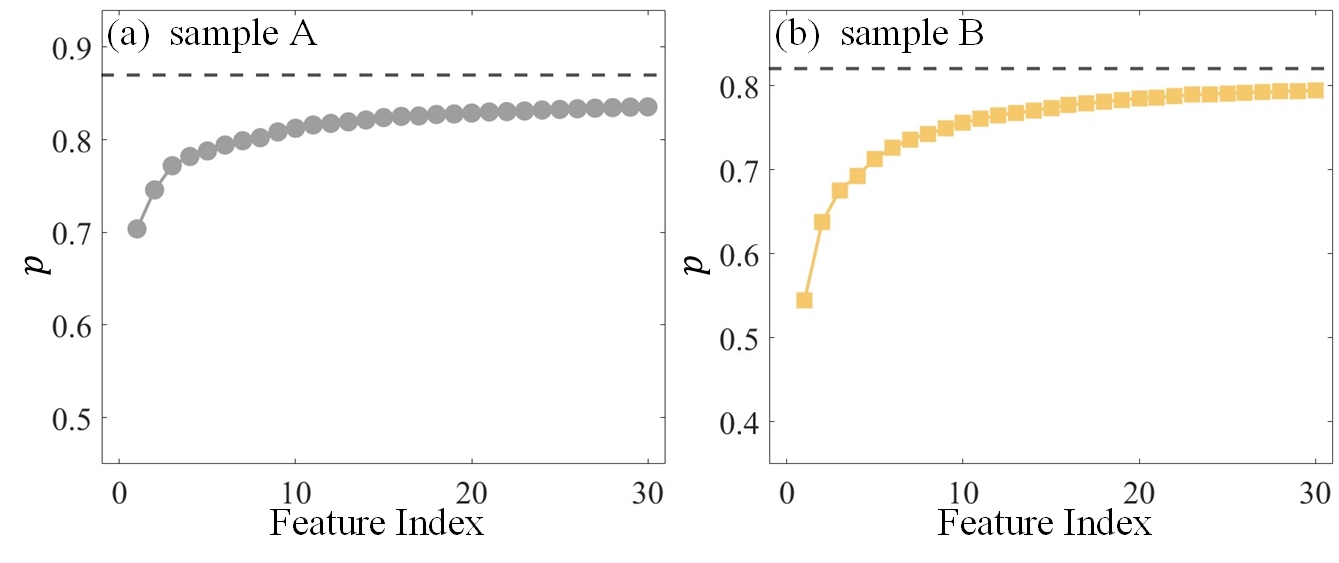}
    \caption{\label{fig:maxcorr_features} 
    Cumulative Pearson correlation coefficients of the top $30$ most important elements in the feature vector defined by equation~\ref{eq:feature_vector}.
    (a) and (b) show the results of sample A and sample B, respectively. 
    The dashed line in each panel denotes the Pearson correlation coefficient obtained with the entire feature vector. }
\end{figure}

We randomly split our data set with a ratio of $8:2$, with $80\%$ of the data used for training and the remaining $20\%$ used for testing. 
Figure~\ref{fig:linearregression_predict_true} compares the true values of $\ln \tau_\mathrm{LC}$ and the values predicted by the regression model for the test set, where each point represents a particle. 
Here, results under steady shear with various $\dot{\gamma}$, including equilibrium, Newtonian, crossover, and shear-thinning regimes, are plotted together. 
These results suggest a positive correlation between the predicted and true values of $\ln \tau_\mathrm{LC}$, 
quantified by the Pearson correlation coefficient $p$ of $0.87$ for sample A, and $0.82$ for sample B. 
We also calculate the Pearson correlation coefficient for each $\dot{\gamma}$. 
The results, given in the insets of figure~\ref{fig:linearregression_predict_true}, are around $0.7$. 
It is seen that the correlation for single $\dot{\gamma}$ remains robust. 
The good correlations shown in figure~\ref{fig:linearregression_predict_true} suggest that the structural parameters used here are applicable to both equilibrium and a wide range of $\dot{\gamma}$.  

An important observation from figure~\ref{fig:linearregression_predict_true} is that the data points corresponding to different $\dot{\gamma}$ approximately align along the same truth line. 
This consistency suggests the existence of a common structural basis for $\tau_\mathrm{LC}$ that is invariant to the applied shear. 
As a result, we unify the $\tau_\mathrm{LC}$ found from equilibrium and shear states by leveraging these structural parameters. 

One may argue that the $576$ elements contained in one feature vector given by equation~\ref{eq:feature_vector} are still too many, which calls for further interpretation and simplification. 
We sort these elements according to their contributions to the cumulative Pearson correlation coefficient. 
Figure~\ref{fig:maxcorr_features} gives the cumulative Pearson correlation coefficients of the top $30$ most important elements in sorting sequence. 
It is seen that the first several elements contribute the majority of the total correlation, while the remaining ones' contribution is marginal. 
The detailed sequences of important elements for both samples are given in Appendix \ref{sec:algorithm_select_features}. 

\begin{figure*}
\includegraphics[width=\linewidth]{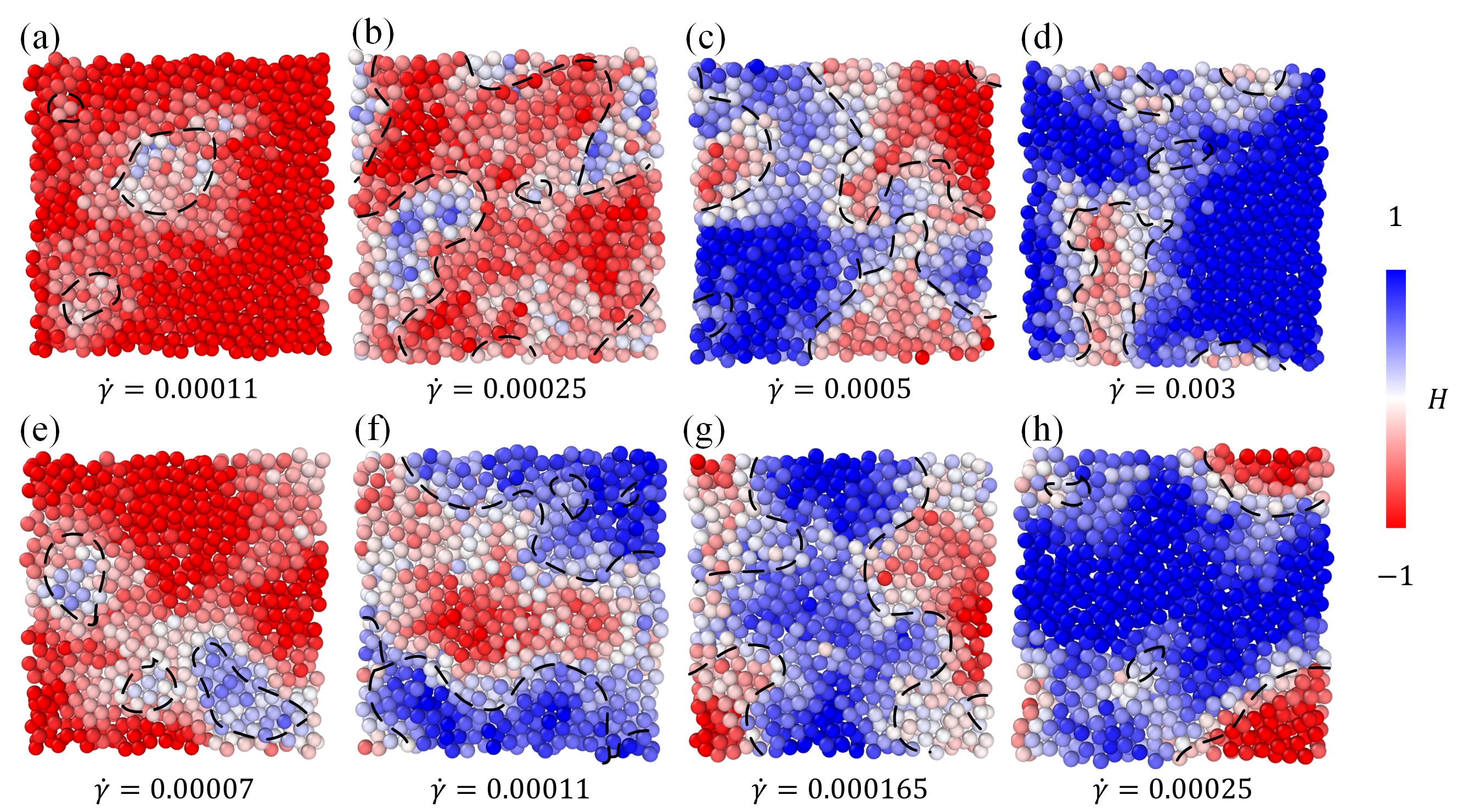}
    \caption{\label{fig:spatial_h_dist} 
    Spatial distributions of the hardness $H$ at various $\dot{\gamma}$ for both samples. 
    (a) -- (d) show the configuration slices selected at different $\dot{\gamma}$ for sample A. 
    (e) -- (h) show the configuration slices selected at different $\dot{\gamma}$ for sample B. 
    In all panels, the color of each particle denotes its value of $H$ with a common color bar. 
    The boundaries between liquid-like (fast) and solid-like (slow) groups are denoted by black dashed lines. 
    A nice agreement between the boundaries and the particles with $H$ close to $0$ is seen in all panels. }
\end{figure*}

\subsection{Structural basis for solid-liquid duality}
\label{sec:classification}

In the preceding subsection, $\tau_\mathrm{LC}$ extracted from different $\dot{\gamma}$ have been quantified through a unified viewpoint of structure. 
Moreover, in section~\ref{sec:taucr}, we reveal a firm relation between the local response and the local dynamics. 
Therefore, one may expect a structural basis for the local response. 
Particularly, we have demonstrated that the local response has two forms: solid-like and liquid-like. 
As $\dot{\gamma}$ increases, the ratio of these two components varies, corresponding to the flattening of the form of $\tau_\mathrm{sh}(\tau_\mathrm{LC})$. 
Here, we aim to comprehend such variation with the assistance of structural parameters. 
The key is to find a structural criterion, which is universal for all shear rates, for classifying particles into solid-like and liquid-like groups.

In the study of amorphous solids, research focusing on identifying soft spots that are more susceptible and likely to transform into STZs based on structural information is highly active. 
Ref.~\cite{manning2020prm} gives a comprehensive review on this topic. 
Recently, such research was advanced by a pioneering study \cite{ajliu2015prl}, in which the soft spots in amorphous solids are identified by employing support vector machine (SVM) \cite{sklearn}. 
To be specific, the authors firstly define a structural feature vector for particles. 
By viewing each particle as a point in the high-dimensional space defined by the feature vector, 
the SVM can identify a hyperplane in this space that optimally separates particles into two categories according to their nonaffine displacements under deformation. 
Particles located at the side of the hyperplane featured by large nonaffine displacements are considered as soft spots. 
Furthermore, the authors define the \textit{softness} parameter for each particle by its distance to the hyperplane. 
Particles with higher softness are more likely to undergo rearrangement under shear. 

Inspired by Ref.~\cite{ajliu2015prl}, we use SVM to explore the structural basis of the solid-liquid duality of the local response. 
It is worth noting that there is a fundamental difference between the considerations of Ref.~\cite{ajliu2015prl} and our work. 
In the study of the deformation of amorphous solids such as Ref.~\cite{ajliu2015prl}, the key question is to figure out \textit{why and how a solid flows}. 
A soft spot is a liquid-like region in the solid background, which plays as the precursor of the bulk yielding and flow. 
For supercooled liquids, on the contrary, the key question is shifted to \textit{why a liquid exhibits strong viscoelasticity} \cite{wang2022prx}. 
Consequently, the focus of our work is shifted to the hard regions, 
i.e., the solid-like regions in the liquid background \cite{wang2022prx, egami2012prl, dyre1999pre, dyre2006rmp}.

Here, we still use the feature vector introduced in equation~\ref{eq:feature_vector} to characterize the local structure of each particle. 
To construct the data set for the SVM model, we select $10$ configurations from each $\dot{\gamma}$, ranging from the equilibrium state to the shear-thinning regime. 
For each configuration, particles are classified as slow and fast using the method given in section~\ref{sec:taucr}. 
Slow particles are assigned to the solid-like group $G_\mathrm{s}$, while fast particles are assigned to the liquid-like group $G_\mathrm{l}$. 
The number of elements in $G_\mathrm{s}$ and that in $G_\mathrm{l}$ are set to be equal. 
The data set is randomly partitioned into a training set and a test set with a ratio of $8:2$. 
Particles in $G_\mathrm{s}$ are labeled as $5$, while those in $G_\mathrm{l}$ are labeled as $-5$. 
Then, a SVM algorithm with a linear kernel is employed on the training set. 
The penalty factor is finely tuned to achieve the optimal classification. 
With this operation, we construct a hyperplane in the high dimensional space defined by the feature vector to separate the two categories of particles. 
We examine the accuracy of the classification with the test set, and the result is $82\%$. 

Moreover, we calculate the distance to the hyperplane for each particle. 
Positive and negative values of the distance represent solid-like and liquid-like particles, respectively. 
This distance is expected to reflect the extent to which the particle exhibits solid-like response. 
Thus, we term it as the \textit{hardness} of particle, denoted as $H$. 
Considering that the classification of response is related to $\tau_\mathrm{LC}$, we expect that $H$ and $\tau_\mathrm{LC}$ is positively correlated. 
We test their correlation with the test set. 
The Pearson correlation coefficients of $H$ and $\tau_\mathrm{LC}$ are about $0.5$ for all shear rates and both samples. 

To validate our SVM model, it is crucial to check if it can correctly capture the solid-like and liquid-like regions. 
In figure~\ref{fig:spatial_h_dist}, we plot several slices of both systems at different $\dot{\gamma}$. 
The upper and lower panels represent sample A and sample B, respectively. 
Particles are colored based on their values of $H$. 
We also plot the boundary between fast and slow regions for all slices according to the method given in section~\ref{sec:taucr} with black dashed lines. 
It is seen that the two methods give highly consistent spatial features of the solid-liquid duality in local response.  

\begin{figure}
\includegraphics[width=\linewidth]{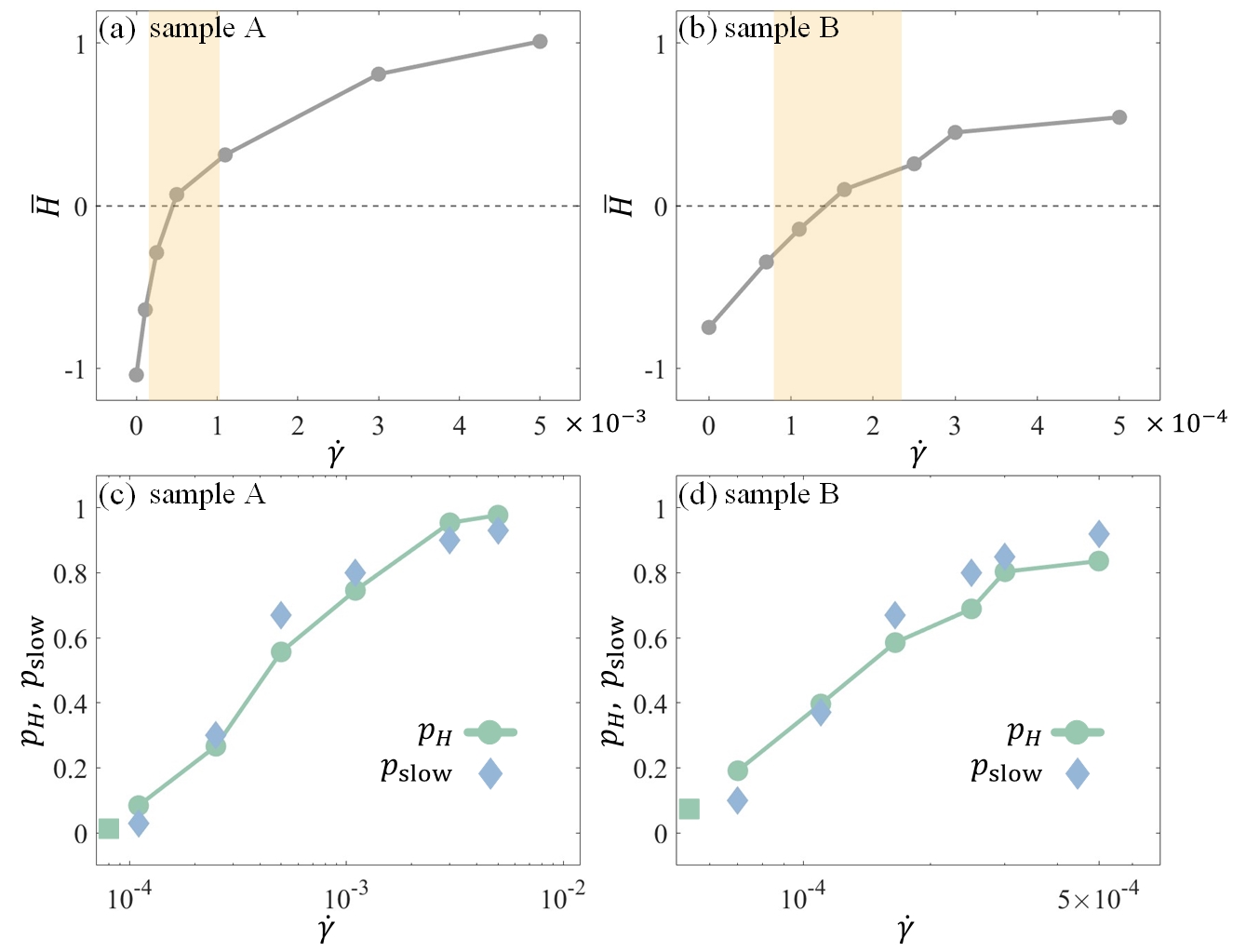}
    \caption{\label{fig:h_vary} 
    (a) and (b) show the variation of the configuration-averaged $H$, denoted as $\bar{H}$, with respect to $\dot{\gamma}$ for sample A and sample B, respectively. 
    In both (a) and (b), vertical yellow bands denote the crossover regime. 
    (c) and (d) show the proportion $p_H$ of particles with $H>0$ (circles and squares) for sample A and sample B, respectively. 
    Squares denote the results at the equilibrium state, and circles denote the results at other shear rates. 
    In both (c) and (d), the results of $p_\mathrm{slow}$ (diamonds), which have been shown in figure~\ref{fig:predict_viscosity_and_slowratio}, are replotted.}
\end{figure}

There is another important observation from figure~\ref{fig:spatial_h_dist}. 
Firstly, we recall that the model for determining $H$ is universal and valid for all shear rates. 
Then, as shown in figure~\ref{fig:spatial_h_dist}, we find that more particles migrate to the $H>0$ side with the growth of $\dot{\gamma}$. 
To quantify this observation, we calculate the average value of $H$ over all particles $\bar{H}$ for each $\dot{\gamma}$ and both samples. 
The results are given in figure~\ref{fig:h_vary} (a) and (b). 
As $\dot{\gamma}$ increases, $\bar{H}$ starts to increase from a negative value, passes through zero, and finally becomes positive. 
It is interesting to see that the shear rate at which $\bar{H}$ crosses zero just coincides with the crossover regime. 
We can evaluate the percentage of slow particles by calculating the proportion of particles with $H>0$. 
The results, denoted as $p_H$, are shown in figure~\ref{fig:h_vary} (c) and (d) for sample A and sample B, respectively. 
Meanwhile, we replot the results of $p_\mathrm{slow}$ shown in figure~\ref{fig:predict_viscosity_and_slowratio}.
It is seen that $p_H$ and $p_\mathrm{slow}$ are consistent with each other. 
This agreement provides a structural view for the shear-induced changes in dynamics and local response. 

The result discussed in this subsection makes an advance to the preceding subsection. 
In the preceding subsection, we find the correlation between structure and $\tau_\mathrm{LC}$. 
This correlation is somewhat natural: By definition, $\tau_\mathrm{LC}$ is calculated by using the ICE method and stopping external shear. 
In this way, the convection and thermal noise are removed, which highlights the role of the transient configuration in determining $\tau_\mathrm{LC}$. 
This correlation is irrelevant to local response to shear and the form of $\tau_\mathrm{sh}(\tau_\mathrm{LC})$. 
In this subsection, by utilizing the same feature vector, we build the connection between local response and local structure. 
Note that the form of local response involves the interplay among external shear, thermal activation and structure. 
The effectiveness of $H$ in classification and the variations of $\bar{H}$ and $p_H$ with $\dot{\gamma}$ further emphasize the importance of local structure in determining the rheology of supercooled liquids. 
Summarizing all results in this section, we establish the structural basis for the dynamical and rheological behaviors discussed in sections~\ref{sec:taucr} and \ref{sec:rheology}. 

\section{DISCUSSION: A MODEL ANALYSIS}
\label{sec:pel}
In sections~\ref{sec:taucr} -- \ref{sec:structure}, we find the connection from the local configuration to the local dynamics and response, and finally to the flow behaviors. 
Following this line of thought, one may expect to formulate the behaviors of $\tau_\mathrm{LC}$, the form of $\tau_\mathrm{sh}(\tau_\mathrm{LC})$, and the duality in local response, 
which compose the foundation for understanding the flow of supercooled liquids in our framework, with theoretical tools involving local configuration and dynamics. 
The key issue is to elucidate how the structure, flow and thermal effect collectively affect the local dynamics. 
In this section, this problem will be discussed with a simple model employing the picture of shear-facilitated activation at the semiquantitative level.  

\subsection{\texorpdfstring{Shear-facilitated-activation picture of $\tau_\mathrm{LC}$ and $\tau_\mathrm{sh}$}{Shear-facilitated-activation picture of tau\_LC and tau\_sh}}

Studies of dynamical processes in supercooled liquids typically conceptualize the structure as a basin (or trap) 
that can be overcome by thermal activation to realize structural relaxation \cite{heuer2008jpcm, sciortino2005jsm,stillinger1998nat}.
Such an activation picture has been extensively applied at the local level, resulting in the trap model \cite{bouchaud1996jpa, bouchaud2003prl}, 
and even at the particle level, leading to a series of studies by Liu \textit{et al}. wherein a ``softness" parameter is introduced \cite{ajliu2016natphy, ajliu2023epl}. 
As previously noted, particles in supercooled liquids are confined within cages for the majority of time, 
exhibiting intermittent hoppings to new environments. 
Thus, an energy basin that temporarily confines an individual particle can be unambiguously defined, 
whereby the rearrangement corresponds to the thermal activations out of the basin.

In the presence of external shear, the configuration is distorted, implying that the energy basin is tilted \cite{lacks1999jcp}, 
which facilitates the out-of-basin motion on average \cite{barrat2015epje,lacks2004prl}. 
An intriguing interplay between thermal activation and shear deformation emerges under such conditions. 
The incorporation of both effects was pioneered in the soft glassy rheology (SGR) model \cite{cates1997prl} and certain EPMs \cite{fielding2024prl}, 
where mesoscopic regions are considered. 
Since such shear-induced distortion also manifests at the cage level, 
the interplay between thermal activation and shear deformation should be applicable at this scale.

\label{sec:sgr_startup}
\begin{figure}
\includegraphics[width=0.8\linewidth]{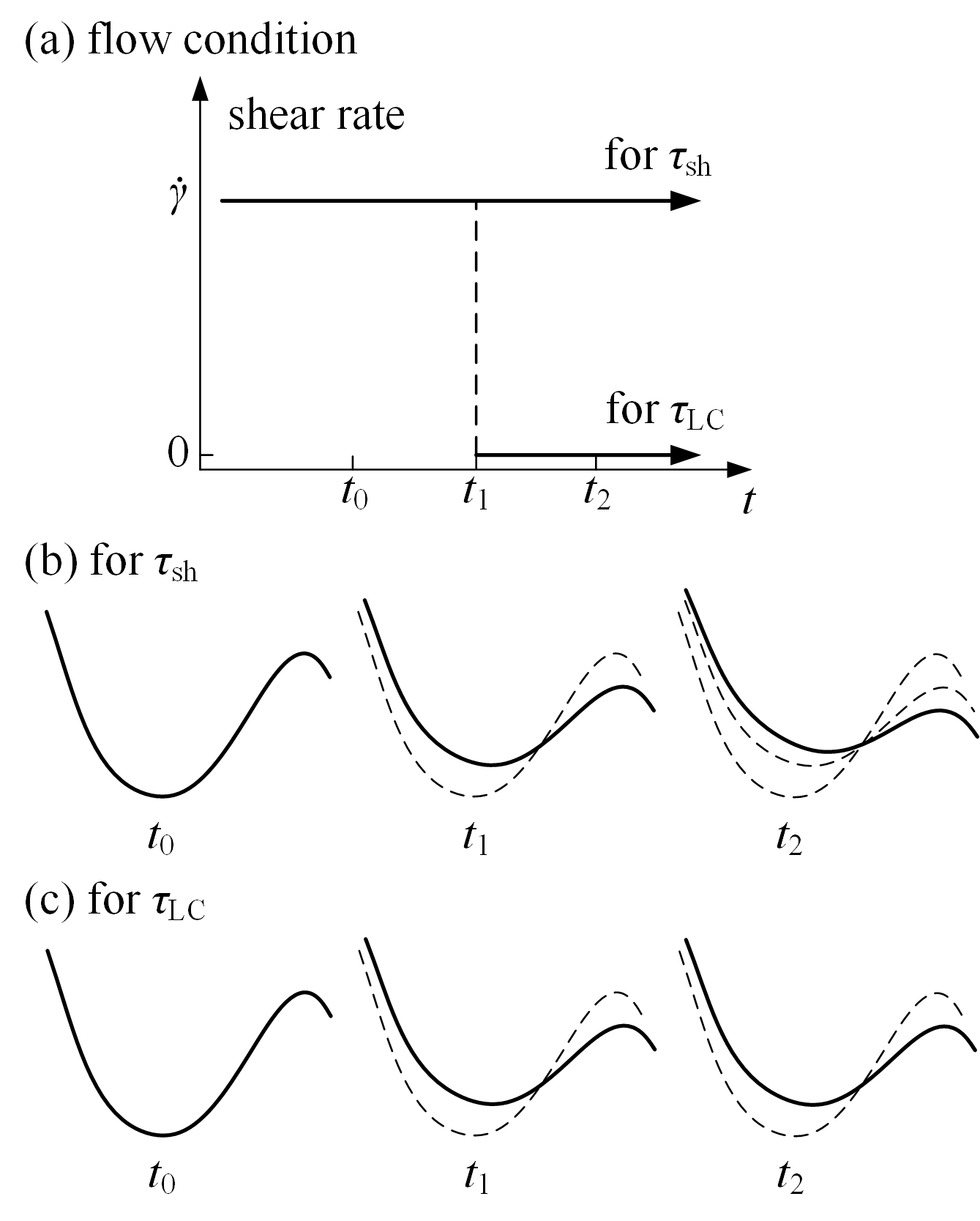}
    \caption{\label{fig:pel_tilt} 
    (a) A review of the flow conditions for measuring $\tau_\mathrm{LC}$ and $\tau_\mathrm{sh}$ under steady shear with respect to a configuration at time $t_1$. 
    (b) Illustration of the evolution of the basin for measuring $\tau_\mathrm{sh}$ under steady shear. 
    Here, the basin shows an increasing tilt as time goes, represented by the change in shape from dashed lines to solid lines. 
    (c) Illustration of the evolution of the basin for measuring $\tau_\mathrm{LC}$. 
    Here, the tilting of basin stops at $t=t_1$ due to the abrupt cession of flow.  }
\end{figure}

In this subsection, we will model the behaviors of $\tau_\mathrm{LC}$ and $\tau_\mathrm{sh}$ by combining the ideas discussed in the above two paragraphs. 
Considering that both $\tau_\mathrm{LC}$ and $\tau_\mathrm{sh}$ are defined as the time for a reference particle to jump out, 
we can assume a basin for the particle to represent the potential energy barrier it experiences \cite{ajliu2016natphy}. 
The shape of the basin is determined by the local configuration of the reference particle. 
When shear is applied, the basin is tilted. 
In figure~\ref{fig:pel_tilt} (a), we review the flow conditions for extracting $\tau_\mathrm{LC}$ and $\tau_\mathrm{sh}$ with respect to the configuration at $t=t_1$. 
Figure~\ref{fig:pel_tilt} (b) illustrates the evolution of basin for measuring $\tau_\mathrm{sh}$. 
In this case, the shear is steady, and the basin tilts more and more with time before the activation or the mechanical instability occurs. 
Figure~\ref{fig:pel_tilt} (c) illustrates the evolution of basin for measuring $\tau_\mathrm{LC}$. 
Here, the external shear stops at $t=t_1$. 
Consequently, the basin also stops tilting at $t=t_1$. 
Figure~\ref{fig:pel_tilt} elucidates the difference between $\tau_\mathrm{LC}$ and $\tau_\mathrm{sh}$. 
$\tau_\mathrm{sh}$ describes the local relaxation in flow. 
It is determined by the combined effect of thermal activation and convection. 
For $\tau_\mathrm{LC}$, the convection is eliminated by stopping external shear and, thus, 
it is determined by the thermal activation from the distorted configuration at $t=t_1$. 
From the above discussion, it is seen that $\tau_\mathrm{LC}$, rather than $\tau_\mathrm{sh}$, directly reflects the local configuration and the particle-level energy barrier. 

Now we put our attention on the particle-level basin. 
We denote the energy barrier of the basin at zero strain as $E_\mathrm{b}$.
Without external shear, the probability of thermal activation occurring within a small time interval $\mathrm{d}t$ is proportional to $\Gamma \exp \left( -E_\mathrm{b} /k_\mathrm{B}T \right) \mathrm{d}t$, 
where $\Gamma$ is the attempt frequency. 
The value of $E_\mathrm{b}$ varies in space \cite{wyart2024pnas, wyart2025pnas}.  
According to the activation picture, $E_\mathrm{b}$ is related to $\tau_\mathrm{LC}$ at equilibrium by:
\begin{equation}
    \tau_\mathrm{LC,eq} \sim \frac{1}{\Gamma} \exp \left( \frac{E_\mathrm{b}}{k_\mathrm{B}T} \right),
    \label{eq:sgr_eqtaucr}
\end{equation}
where $\tau_\mathrm{LC,eq}$ denotes the $\tau_\mathrm{LC}$ at equilibrium. 
With equation~\ref{eq:sgr_eqtaucr} and the distribution of $\tau_\mathrm{LC,eq}$, the distribution of $E_\mathrm{b}$ can be obtained. 
As evident from the spatial distribution of $\tau_\mathrm{LC}$ (figure \ref{fig:eq_spatial_taucr_dist}), 
while energy barriers $E_\mathrm{b}$ are defined for individual particles, they exhibit a smooth spatial variation with clustering features. 

\begin{figure}
\includegraphics[width=0.8\linewidth]{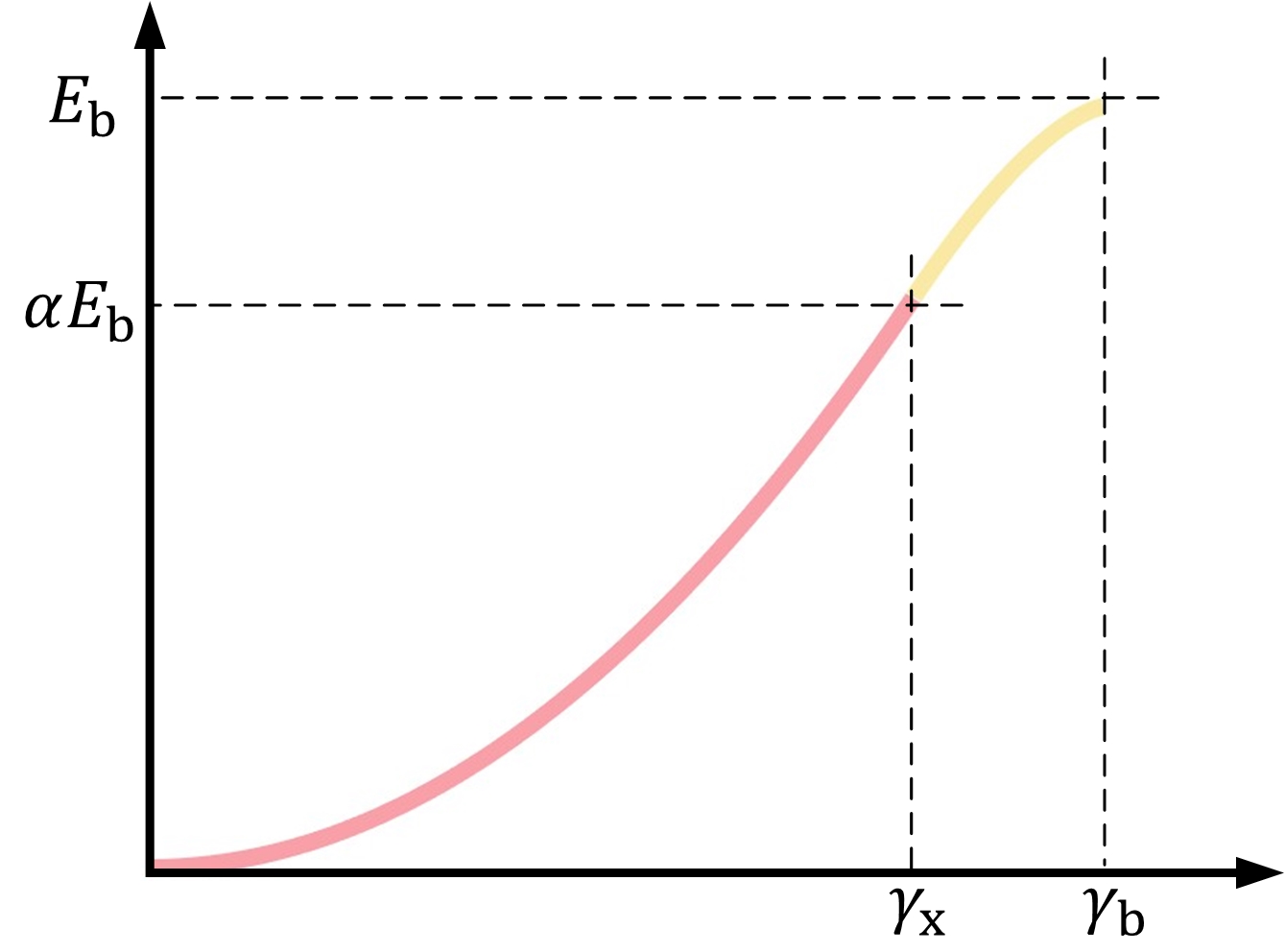}
    \caption{\label{fig:pel_function} 
    Illustration of the deformation energy $E(\gamma)$ given by equation~\ref{eq:pel_function}. 
    The harmonic part is colored by light red, and the catastrophe part is colored by yellow. 
    $\gamma_\mathrm{x}$ denotes the boundary between these two parts. 
    $\alpha E_\mathrm{b}$ ($0<\alpha<1$) gives the value of $E(\gamma=\gamma_\mathrm{x})$. }
\end{figure}

In the presence of external shear, the local structure undergoes deformation at a strain rate of $\dot{\gamma}$. 
During the deformation, the local structure will store potential energy until the activation takes place. 
This deformation energy is denoted as $E(\gamma)$, where $\gamma$ is the local strain. 
Taking into account both thermal activation and shear effect, 
the probability for a particle to escape the basin within $\mathrm{d}t$ is given by:
\begin{equation}
    \Gamma \exp \left [ -\frac{E_\mathrm{b}-E(\dot{\gamma} t)}{k_\mathrm{B}T}  \right ] \mathrm{d}t.
    \label{eq:sgr_outbarrier}
\end{equation}
Here, the form of $E(\gamma)$ is important. 
$E(\gamma)$ is harmonic at small strains. 
As $\gamma$ increases, $E(\gamma)$ gradually deviates from the harmonic form. 
As predicted by the catastrophe theory, when $\gamma$ approaches the edge of basin $\gamma_\mathrm{b}$, 
the energy increment $\Delta E$ scales with the residual load $\gamma_\mathrm{b} - \gamma$ as $\Delta E \propto (\gamma_\mathrm{b} - \gamma)^{3/2}$ \cite{wales2001science}.
Moreover, Maloney and Lacks demonstrate that this scaling remains valid well beyond the vanishing regime of $\gamma_\mathrm{b} - \gamma$, even at finite temperatures \cite{lacks2006pre}. 
Considering the behaviors of $E(\gamma)$ at small and large strains, we model its form as:
\begin{equation}
    E(\gamma)=
    \begin{cases}
    \frac{1}{2} G \Omega \gamma^2  &  \;\,0 \le \gamma \le \gamma_\mathrm{x}  \\
    E_\mathrm{b} - A(\gamma_\mathrm{b} - \gamma)^{3/2}    & \gamma_\mathrm{x} < \gamma \le \gamma_\mathrm{b} 
    \end{cases},
    \label{eq:pel_function}
\end{equation}
where $G$ is the local shear modulus, $\Omega$ is the activation volume, $\gamma_\mathrm{x}$ is the crossover strain, and $A$ is a numerical factor. 
Figure~\ref{fig:pel_function} illustrates the form of $E(\gamma)$.  
By requiring that both $E(\gamma)$ and its first-order derivative are continuous at $\gamma=\gamma_\mathrm{x}$, 
the three parameters, $\gamma_\mathrm{x}$, $\gamma_\mathrm{b}$ and $A$, can be reduced to a single parameter. 
As illustrated in figure~\ref{fig:pel_function}, we can denote the value of $E(\gamma_\mathrm{x})$ by $\alpha E_\mathrm{b}$, 
where $\alpha \, \left( \alpha \in (0,1) \right)$ is employed as the free adjustable parameter that represents $\gamma_\mathrm{x}$, $\gamma_\mathrm{b}$ and $A$. 

Besides the free parameter $\alpha$, other parameters in equation~\ref{eq:sgr_outbarrier} are set as follows: 
$\Gamma^{-1}$ is determined from the MSD data at equilibrium as the time at which the particle enters the plateau regime. 
$G$ is set to the effective modulus $G_\mathrm{eff}$ defined in section~\ref{sec:rheology}. 
$\Omega$ represents the size that will be involved for the reference particle to jump out. 
At the current stage, we estimate $\Omega$ by the volume of the cage, i.e., the sphere whose radius is given by the distance of the first minimum of $g(r)$. 
For the sake of simplicity, we assume that all these parameters do not vary with $\dot{\gamma}$. 
Note that our assumption on $\Omega$ is oversimplified. 
As demonstrated in section~\ref{sec:rheology}, the component of solid-like parts increases with $\dot{\gamma}$, which may result in the enhancement of the correlation between jump events. 
We will revisit this point in section~\ref{sec:sgr_correlation}. 

\begin{figure}
\includegraphics[width=\linewidth]{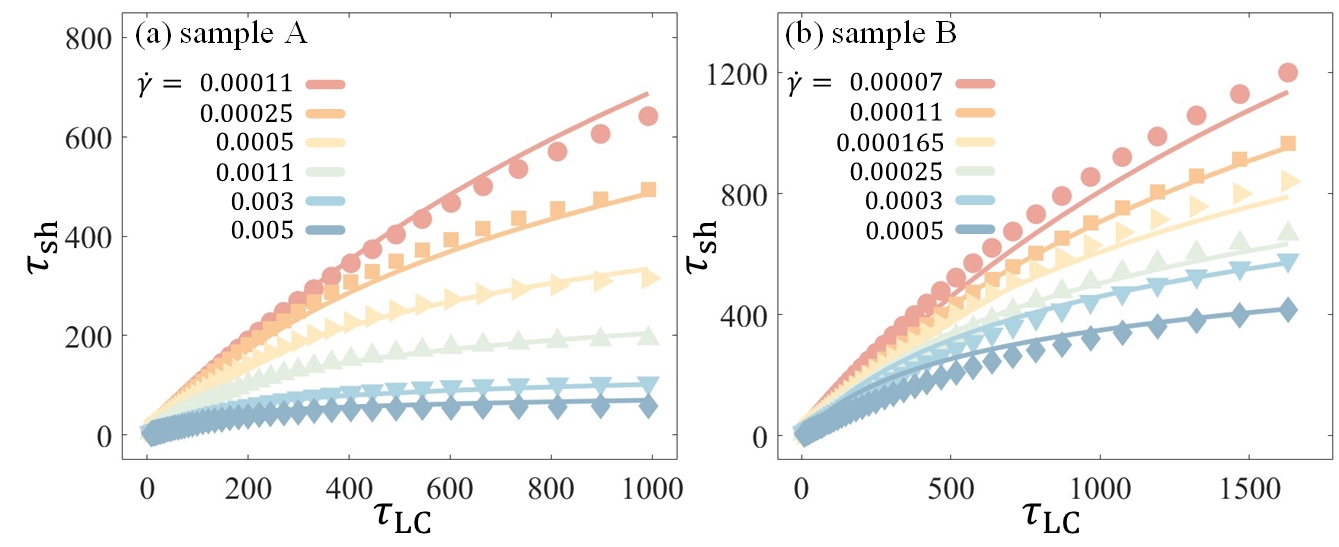}
    \caption{\label{fig:sgr_taucr_sh} 
    $\tau_\mathrm{sh}(\tau_\mathrm{LC})$ calculated from the model analysis (lines) for sample A (a) and sample B (b) at various $\dot{\gamma}$. 
    The results obtained from MD data (symbols) are also plotted. 
    In each panel, the results corresponding to the same $\dot{\gamma}$ are displayed in the same color. }
\end{figure}

Assuming that the basin starts deforming at $t=0$, 
the probability that a particle jumps at $t$ can be derived with equation~\ref{eq:sgr_outbarrier} and is expressed as:
\begin{eqnarray}
    f(t)&=&\Gamma \exp \left [ -\frac{E_\mathrm{b}-E(\dot{\gamma}t)}{k_\mathrm{B}T}  \right ]  \label{eq:sgr_outbarrier_dist}
    \\ \nonumber 
    &&\times \exp \left (-\Gamma \int_{0}^{t} \exp \left [ -\frac{E_\mathrm{b}-E(\dot{\gamma}t')}{k_\mathrm{B}T} \right ] \mathrm{d}t'  \right ).
\end{eqnarray}
Then, we can calculate the mean waiting time of jump by:
\begin{align}
\bar{t}(E_\mathrm{b},\dot{\gamma}) &= \int_{0}^{\infty} t \, f(t) \, \mathrm{d}t \label{eq:sgr_outbarrier_tbar} \\ 
&= \int_{0}^{\infty} \exp \left ( -\Gamma \int_{0}^{t} \exp \left[ -\frac{E_\mathrm{b}-E(\dot{\gamma}t')}{k_\mathrm{B}T} \right] \mathrm{d}t'  \right ) \mathrm{d}t. \notag
\end{align}
Setting $\dot{\gamma}=0$, equation~\ref{eq:sgr_outbarrier_tbar} reduces to equation~\ref{eq:sgr_eqtaucr} and gives the value of $\tau_\mathrm{LC}$ at equilibrium. 
For $\dot{\gamma}>0$, $\bar{t}(E_\mathrm{b},\dot{\gamma})$ gives $\tau_\mathrm{sh}$ according to the picture shown in figure~\ref{fig:pel_tilt}.  
Therefore, with equation~\ref{eq:sgr_outbarrier_tbar}, we can calculate the form of $\tau_\mathrm{sh}(\tau_\mathrm{LC})$. 
Figure~\ref{fig:sgr_taucr_sh} shows $\tau_\mathrm{sh}(\tau_\mathrm{LC})$ calculated by this approach at various $\dot{\gamma}$ for both samples. 
For comparison, we also plot the MD results of $\tau_\mathrm{sh}(\tau_\mathrm{LC})$. 
It is seen that our model captures the general characteristics of $\tau_\mathrm{sh}(\tau_\mathrm{LC})$. 
Particularly, its form exhibits a greater flattening with $\dot{\gamma}$. 
From the viewpoint of this model, this flattening is due to the increasingly dominant role of shear deformation compared to thermal activation as $\dot{\gamma}$ increases.

\begin{figure}
\includegraphics[width=\linewidth]{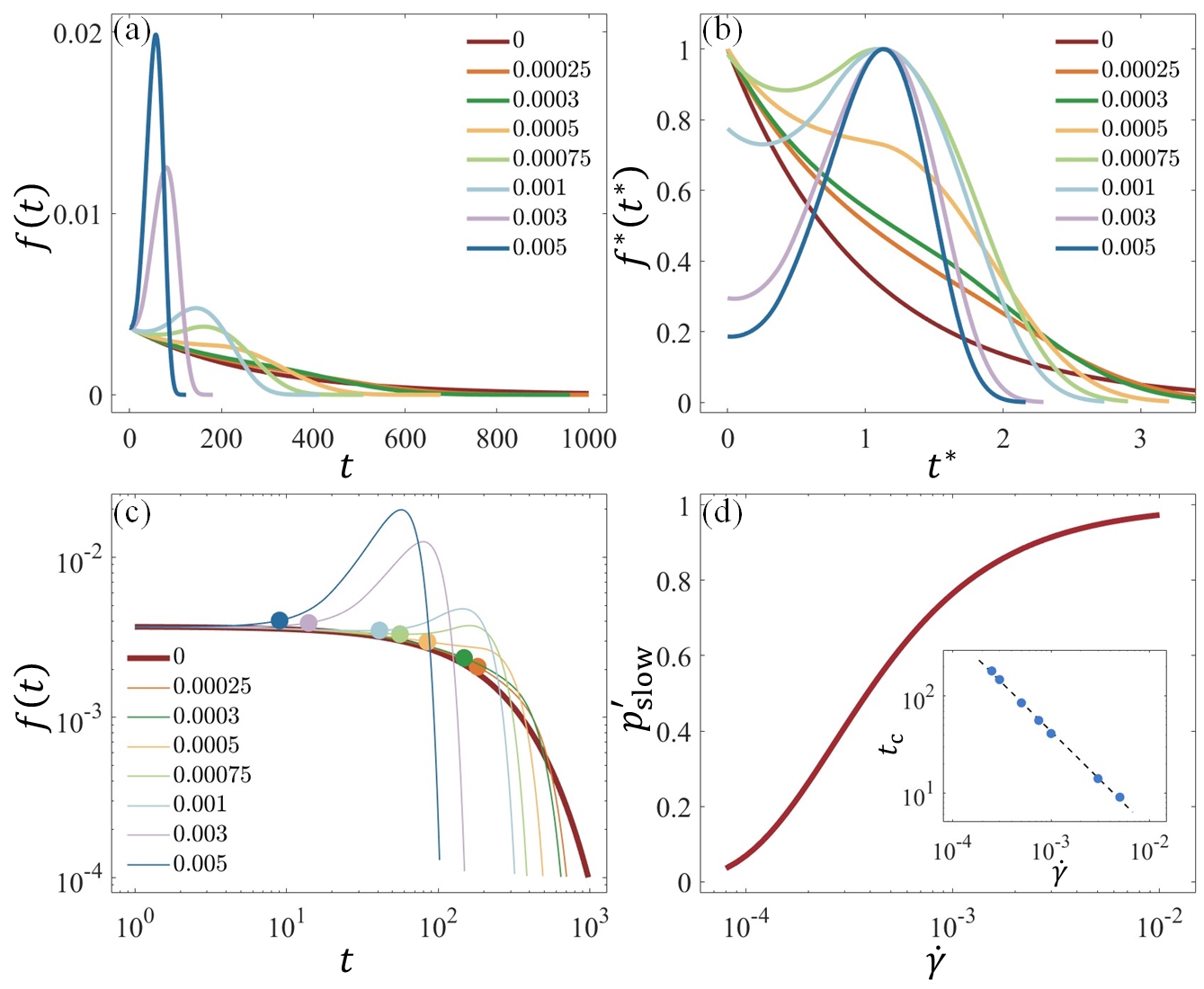}
    \caption{\label{fig:outbarrier_pdf} 
    Functional forms of $f(t)$ as various $\dot{\gamma}$. 
    (a) $f(t)$ in linear scale. 
    (b) $f^* (t^*)$, where $t^*=t/\bar{t}(E_\mathrm{b}', \dot{\gamma})$ and $f^*=f/f_\mathrm{max}$. 
    (c) $f(t)$ in log-log scale. The form at equilibrium $f_\mathrm{eq}(t)$ is highlighted by a thick line. 
    For each $\dot{\gamma}$, we mark the position $t_\mathrm{c}$ at which $f(t)$ deviates from $f_\mathrm{eq}(t)$ by a circle with its color the same to that of the corresponding $\dot{\gamma}$. 
    (d) The proportion of slow particles found from the model analysis $p_\mathrm{slow}'$ as a function of $\dot{\gamma}$. 
    Inset: $t_\mathrm{c}$ as a function of $\dot{\gamma}$ in log-log scale (symbols). 
    The dashed line denotes the relation $t_\mathrm{c} \sim \dot{\gamma}^{-1}$. }
\end{figure}

Figure~\ref{fig:outbarrier_pdf} (a) shows $f(t)$ at different $\dot{\gamma}$. 
To calculate $f(t)$, we set $E_\mathrm{b}$ in equation~\ref{eq:sgr_outbarrier_dist} with the most probable value of the equilibrium distribution of $E_\mathrm{b}$ of sample A, denoted as $E_\mathrm{b}'$. 
As $\dot{\gamma}$ increases, $f(t)$ as a whole shifts to smaller $t$. 
Moreover, the form of $f(t)$ also changes with $\dot{\gamma}$. 
In figure~\ref{fig:outbarrier_pdf} (b), we plot the scaled $f(t)$, denoted as $f^* (t^*)$, where $t^*=t/\bar{t}(E_\mathrm{b}', \dot{\gamma})$ and 
$f^*=f/f_\mathrm{max}$ ($f_\mathrm{max}$ is the maximum value of $f(t)$). 
It is seen that $f(t)$ gradually changes from an exponential form to a gaussian-like form as $\dot{\gamma}$ increases. 
Particularly, for intermediate shear rates, $f(t)$ exhibits a bimodal form composed of an exponential component and a gaussian component, 
with the gaussian component being enhanced by $\dot{\gamma}$. 

The alternation of the form of $f(t)$ corresponds to the observation in sections~\ref{sec:taucr} and \ref{sec:rheology} that more particles are facilitated by shear as $\dot{\gamma}$ increases. 
To clarify this point, we plot $f(t)$ in the log-log scale in figure~\ref{fig:outbarrier_pdf} (c). 
The one at equilibrium, $f_\mathrm{eq}(t)$, is highlighted by a thick solid line. 
For $f(t)$ with finite $\dot{\gamma}$, all of them align with $f_\mathrm{eq}(t)$ at small $t$, and deviate from $f_\mathrm{eq}(t)$ at large $t$. 
We denote the point at which $f(t)$ deviates from $f_\mathrm{eq}(t)$ as $t_\mathrm{c}$, and mark them in figure~\ref{fig:outbarrier_pdf} (c). 
Evidently, $t_\mathrm{c}$ can be viewed as the boundary between fast and slow groups: 
For $t<t_\mathrm{c}$, particles are not facilitated by external shear and should be classified as fast; 
while for $t>t_\mathrm{c}$, particles are facilitated and should be classified as slow. 
Then, we calculate the percentage of slow particles by $p_\mathrm{slow}'= \int_{t_\mathrm{c}}^{\infty} f(t) \, \mathrm{d}t$, and show the result in figure~\ref{fig:outbarrier_pdf} (d). 
The S-shaped dependence of $p_\mathrm{slow}'$ on $\lg{\dot{\gamma}}$ is similar to the results of $p_\mathrm{slow}$ shown in figure~\ref{fig:predict_viscosity_and_slowratio}, 
which are directly found from the MD results. 
The inset of figure~\ref{fig:outbarrier_pdf} (d) shows $t_\mathrm{c}$ as a function of $\dot{\gamma}$ in the log-log scale, 
which reveals a clear relation of $t_\mathrm{c} \propto \dot{\gamma}^{-1}$. 
This behavior is also consistent with the MD results shown in figure~\ref{fig:eq_taucr_sh} (c) and (d). 

\subsection{Steady shear}
In the preceding subsection, we model $\tau_\mathrm{LC}$, $\tau_\mathrm{sh}$ and their relation as the interplay among local configuration, 
convection and thermal activation. 
The discussion is primarily based on the start-up deformation represented by the behavior with respect to a single basin. 
In this subsection, we will further leverage the activation picture to rationalize our findings on the steady shear. 

By incorporating convective effects, Sollich \textit{et al}. generalized the trap model to the flow condition, known as the soft glass rheology (SGR) model \cite{cates1997prl, sollich1998pre}. 
For the condition of steady shear, the equation of motion of the SGR model is given by: 
\begin{equation}
    \frac{\partial P}{\partial t} = -\dot{\gamma}\frac{\partial P}{\partial \gamma}-\Gamma \exp \left [ -\frac{E_\mathrm{b} - E(\gamma)}{k_\mathrm{B}T} \right ]P + \Gamma(t)\rho(E_\mathrm{b})\delta(\gamma),
    \label{eq:sgr_eq}
\end{equation}
where $P=P(E_\mathrm{b},\gamma,t)$ is the probability that a coherent subsystem experiences a local energy barrier of $E_\mathrm{b}$ and a local strain of $\gamma$ at a given time $t$. 
In our picture, $P(E_\mathrm{b},\gamma,t)$ describes the dynamics at the particle level. 
The first term on the right-hand side of equation~\ref{eq:sgr_eq} represents the convective effect. 
The second term represents the activation from the distorted basin, where the energy barrier is given by $E_\mathrm{b} - E(\gamma)$. 
Here, the form of $E(\gamma)$ and the attempt frequency $\Gamma$ are set to those used in the preceding subsection. 
The third term represents the process that the yielded subsystems rearrange to new local equilibrium configurations. 
$\Gamma(t)$ is the total yielding rate, which is given by:
\begin{equation}
\Gamma(t)=\iint \Gamma \exp \left [ -\frac{E_\mathrm{b}-E(\gamma)}{k_\mathrm{B}T} \right ] P(E_\mathrm{b},\gamma, t) \, \mathrm{d}\gamma \, \mathrm{d}E_\mathrm{b}. 
\end{equation}
$\rho(E_\mathrm{b})$ is the density of states for new barriers. 
Here, it is assumed that $\rho(E_\mathrm{b})$ does not depend on $\dot{\gamma}$ \cite{cates1997prl}. 
Therefore, $\rho(E_\mathrm{b})$ can be obtained from the dynamics of equilibrium state as follows. 
First, we measure the distribution of $\tau_\mathrm{LC}$ value at equilibrium $H_\mathrm{eq}(\tau_\mathrm{LC})$. 
Second, assuming the activation picture, we combine $H_\mathrm{eq}(\tau_\mathrm{LC})$ and equation~\ref{eq:sgr_eqtaucr} to find the distribution of barrier height $P_\mathrm{b}(E_\mathrm{b})$. 
Then, $\rho(E_\mathrm{b})$ can be found by: 
\begin{equation}
    \rho(E_\mathrm{b}) \sim P_\mathrm{b}(E_\mathrm{b}) \exp \left (-E_\mathrm{b}/k_\mathrm{B}T \right ).
    \label{eq:sgr_rho_E}
\end{equation}
The model given by equations~\ref{eq:sgr_eq} -- \ref{eq:sgr_rho_E} provides a framework for quantifying the flow behaviors under shear.

As demonstrated in section~\ref{sec:structure}, the dynamics and rheology of supercooled liquids are strongly correlated to the local structure. 
It is interesting to understand this correlation from the shear-facilitated activation picture. 
We recall that local structural terms $g_{0,i}^0$ and $g_{2,i}^{-2}$ are most relevant to local dynamics compared to terms with other combinations of $l$ and $m$. 
In the present model, local dynamics is determined by the local energy barrier $E_\mathrm{b}$ and the local strain $\gamma$. 
As for $g_{0,i}^0$, it represents the local packing surrounding particle $i$. 
Notably, a recent work \cite{ajliu2016natphy} has established the quantitative connection between the energy barrier for particles and 
the local packing efficiency characterized by the particle-wise radial particle density, 
which is similar to $g_{0,i}^0$, for the equilibrium state. 
Following this result, it is clear that the importance of $g_{0,i}^0$ for local dynamics comes from its remarkable relation to the local energy barrier $E_\mathrm{b}$.  
Moreover, we find that the terms with $l=0$ and $m=0$ alter with $\dot{\gamma}$ only slightly. 
This result might justify the above approximation that $\rho(E_\mathrm{b})$ is $\dot{\gamma}$-independent. 

Our emphasis will be on local structural terms with $l=2$ and $m=-2$.  
These terms have received less attention in previous studies of the equilibrium state. 
However, they are very important for shear states as shown in section~\ref{sec:structure}.
To understand this observation, we first refer to the conclusion that the strain of an affine shear $\gamma_\mathrm{aff}$ is proportional to $g_2^{-2}(r)$ \cite{egami1987prb}:
\begin{equation}
    g_2^{-2}(r) \approx -\gamma_\mathrm{aff} \left [ \frac{1}{\sqrt{15}} r \frac{\mathrm{d}}{\mathrm{d}r} g_\mathrm{eq}(r)  \right ],
    \label{eq:g2m2}
\end{equation}
where $g_2^{-2}(r)$ is the ($l=2,m=-2$) coefficient of the spherical harmonic expansion of the global pair distribution function $g(\boldsymbol{r})$, 
and $g_\mathrm{eq}(r)$ is $g(\boldsymbol{r})$ at the undeformed state. 
Equation~\ref{eq:g2m2} suggests that $g_{2,i}^{-2}$ directly reflects the strain of the local configuration of particle $i$. 
Now, it is crucial to verify if the strain variable $\gamma$ in the activation picture, which can be calculated with the SGR model, 
can describe the configurational distortion reflected by $g_{2,i}^{-2}$ under steady shear. 
For this purpose, we perform the following calculations by employing the SGR model given by equations~\ref{eq:sgr_eq} -- \ref{eq:sgr_rho_E}. 

\begin{figure}
\includegraphics[width=\linewidth]{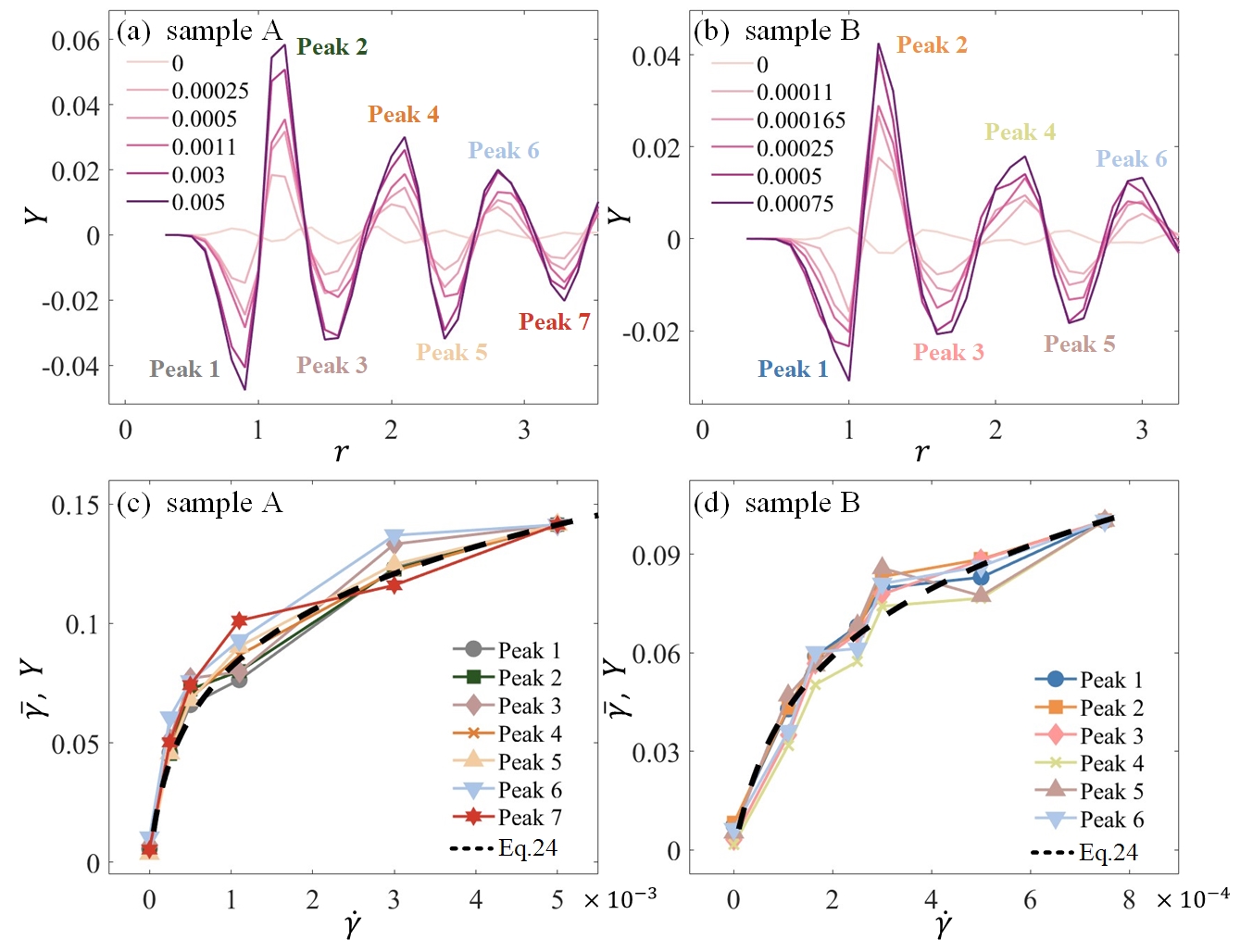}
    \caption{\label{fig:sgr_steady_strain} 
    (a) and (b) show the results of the structure feature $Y$, defined by equation~\ref{eq:sgr_tilt_vector}, as a function of the distance $r$ at various $\dot{\gamma}$ for sample A and sample B, respectively. 
    The first seven peaks of $Y(r)$ are marked. 
    (c) and (d) show the magnitudes of the first seven peaks of $Y(r)$ (symbols) at different $\dot{\gamma}$ for sample A and sample B, respectively. 
    The results of $\bar{\gamma}$ (dashed lines), defined by equation~\ref{eq:sgr_steady_strain_bar}, at different $\dot{\gamma}$ are also plotted. 
    In (c) and (d), the values of peak magnitudes of $Y(r)$ are scaled with a numerical factor to compare with $\bar{\gamma}$.  }
\end{figure}

According to equation~\ref{eq:sgr_eq}, it can be demonstrated that the distribution of local strain $\gamma$ under steady shear is given by the following equation \cite{cates1997prl, sollich1998pre}:
\begin{equation}
    P_\mathrm{s}(\gamma) \sim \int \rho(E_\mathrm{b}) \exp \left[ -\frac{1}{g(E_\mathrm{b})} \int_{0}^{\gamma} \mathrm{e}^{E(\gamma')/k_\mathrm{B}T} \mathrm{d}\gamma' \right] \mathrm{d}E_\mathrm{b},
\end{equation}
where $g(E_\mathrm{b})=\dot{\gamma}\mathrm{e}^{E_\mathrm{b}/k_\mathrm{B}T}/\Gamma$. 
With the preceding equation, we can find the average local strain $\bar{\gamma}$ under steady shear by:
\begin{equation}
    \bar{\gamma}=\frac{\int_0^\infty \gamma P_\mathrm{s}(\gamma) \, \mathrm{d}\gamma}{\int_0^\infty P_\mathrm{s}(\gamma) \, \mathrm{d}\gamma}.
    \label{eq:sgr_steady_strain_bar}
\end{equation} 
Here, we adopt the second generation of $g_{2,i}^{-2}$ to represent the local distortion, as suggested in section~\ref{sec:linear_regression},
and do not differentiate particle species when calculating $g_{2,i}^{-2}$ with equation~\ref{eq:glm_for_particle}. 
We construct the following vector for describing the distortion of the local configuration of particle $i$:
\begin{equation}
    Y_i=\left(g_{2,i}^{-2}(r_1), g_{2,i}^{-2}(r_2), \cdots, g_{2,i}^{-2}(r_n) \right)^{(2)}.
\end{equation}
Here, the set $\{r_1, r_2, \cdots, r_n\}$ gives a series of distances from the reference particle $i$. 
With $Y_i$, we find the average distortion by: 
\begin{equation}
Y(r_k)=\frac{1}{N} {\textstyle \sum_{j=1}^{N} Y_j(r_k)}, k=1,2,\cdots,n.
\label{eq:sgr_tilt_vector}
\end{equation}
We plot the results of $Y(r_k)$ with a series of $k$ values at different $\dot{\gamma}$ for sample A and sample B in figure~\ref{fig:sgr_steady_strain} (a) and (b), respectively. 
$Y(r_k)$ displays an oscillation as $r_k$ increases, which is similar to the $r$-dependence of $g_2^{-2}(r)$ \cite{wang2022prx}.
As expected, the amplitude of $Y$ grows with $\dot{\gamma}$. 
Figure~\ref{fig:sgr_steady_strain} (c) and (d) display the magnitudes of the first seven peaks of $Y(r_k)$, 
which are marked in figure~\ref{fig:sgr_steady_strain} (a) and (b), as a function of $\dot{\gamma}$ for sample A and sample B, respectively. 
Meanwhile, we also plot the results of $\bar{\gamma}$ found from equation~\ref{eq:sgr_steady_strain_bar} for both samples in figure~\ref{fig:sgr_steady_strain} (c) and (d). 
It is seen that the magnitudes of the peaks of $Y(r_k)$ and $\bar{\gamma}$ exhibit highly consistent dependences on $\dot{\gamma}$. 
This consistency holds for different peaks of $Y(r_k)$ within $r_k<3.5$. 
At larger $r_k$, this consistency weakens, implying the localization of such deformation coherency. 
The remarkable proportionality between $Y(r_k )$ and $\bar{\gamma}$ within $r_k<3.5$ implies a nice correspondence between $g_{2,i}^{-2}$ and the local strain $\gamma$ in the present model. 

\begin{figure}
\includegraphics[width=\linewidth]{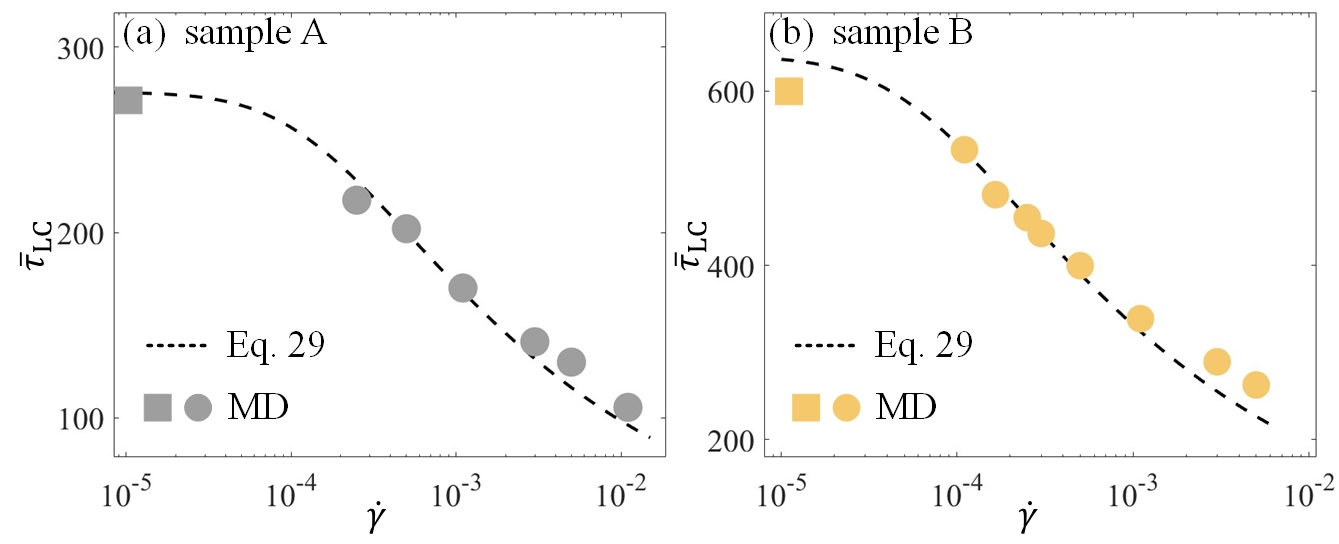}
    \caption{\label{fig:sgr_gamma_taucrave} 
    (a) and (b) show $\bar{\tau}_{\text{LC}}$ (dashed lines), defined by equation~\ref{eq:sgr_taucr_ave}, as a function of $\dot{\gamma}$ for sample A and sample B, respectively. 
    Meanwhile, the average $\tau_\mathrm{LC}$ measured from the MD data (symbols) are also shown. 
    For MD results, squares denote the results at the equilibrium state, and circles denote the results for finite shear rates.}
\end{figure}

In section~\ref{sec:linear_regression}, we show that local structural parameters constructed by $g_{0,i}^0$ and $g_{2,i}^{-2}$, 
which respectively reflect the energy barrier height $E_\mathrm{b}$ and the strain $\gamma$ of the particle-level basin according to the above discussion, 
well predict $\tau_\mathrm{LC}$ for all shear rates through a linear regression method. 
To explore the intrinsic mechanism of this correlation between local structure and dynamics, we will evaluate the dependence of $\tau_\mathrm{LC}$ on $\dot{\gamma}$ with the SGR model. 
Under steady shear, the distribution of local strain $\gamma$ and barrier height $E_\mathrm{b}$ in the SGR model is given by:
\begin{equation}
    P_\mathrm{s}(\gamma, E_\mathrm{b}) \sim \rho(E_\mathrm{b}) \exp \left [-\frac{1}{g(E_\mathrm{b})} \int_{0}^{\gamma} \mathrm{e}^{E(\gamma')/k_\mathrm{B}T} \mathrm{d}\gamma' \right].
\end{equation}
Within a basin characterized by $\gamma$ and $E_\mathrm{b}$, the average activation time is given by:
\begin{equation}
    \bar{\tau}(E_\mathrm{b},\gamma)=\frac{1}{\Gamma} \exp \left[ \frac{E_\mathrm{b}-E(\gamma)}{k_\mathrm{B}T}  \right].
\end{equation}
Combining the preceding two equations, the configuration-averaged $\tau_\mathrm{LC}$, denoted as $\bar{\tau}_\mathrm{LC}$, is written as
\begin{equation}
    \bar{\tau}_\mathrm{LC} = \frac{\iint P_\mathrm{s}(\gamma,E_\mathrm{b})\bar{\tau}(E_\mathrm{b},\gamma)\, \mathrm{d}E_\mathrm{b}\mathrm{d}\gamma}{\iint P_\mathrm{s}(\gamma, E_\mathrm{b}) \, \mathrm{d}E_\mathrm{b}\mathrm{d}\gamma}. 
    \label{eq:sgr_taucr_ave}
\end{equation}
Figure~\ref{fig:sgr_gamma_taucrave} gives $\bar{\tau}_\mathrm{LC}$ as a function of $\dot{\gamma}$ for both samples. 
We also plot the average $\bar{\tau}_\mathrm{LC}$ measured directly from MD data in figure~\ref{fig:sgr_gamma_taucrave}. 
It is seen that the results found by the SGR model are well consistent with the MD results.

The results shown in figures~\ref{fig:sgr_steady_strain} and \ref{fig:sgr_gamma_taucrave} suggest a clear picture for 
the universal correlation between local structure and $\tau_\mathrm{LC}$ at all shear rates shown in section~\ref{sec:linear_regression}. 
The local structural terms related to $g_{0,i}^0$ and $g_{2,i}^{-2}$ respectively associate with the local energy barrier height $E_\mathrm{b}$ and 
the local strain $\gamma$ of the particle-level basin. 
$\tau_\mathrm{LC}$ is the waiting time to jump out from the basin characterized by $\gamma$ and $E_\mathrm{b}$. 
The convective effect is reflected by the tilt of the basin, as illustrated in figure~\ref{fig:pel_tilt}. 
Under steady shear, the collective effects of thermal activation and convection are described by the SGR model. 
With these considerations, we can unify the values of $\tau_\mathrm{LC}$ from configurations at different shear rates. 

\begin{figure}
\includegraphics[width=\linewidth]{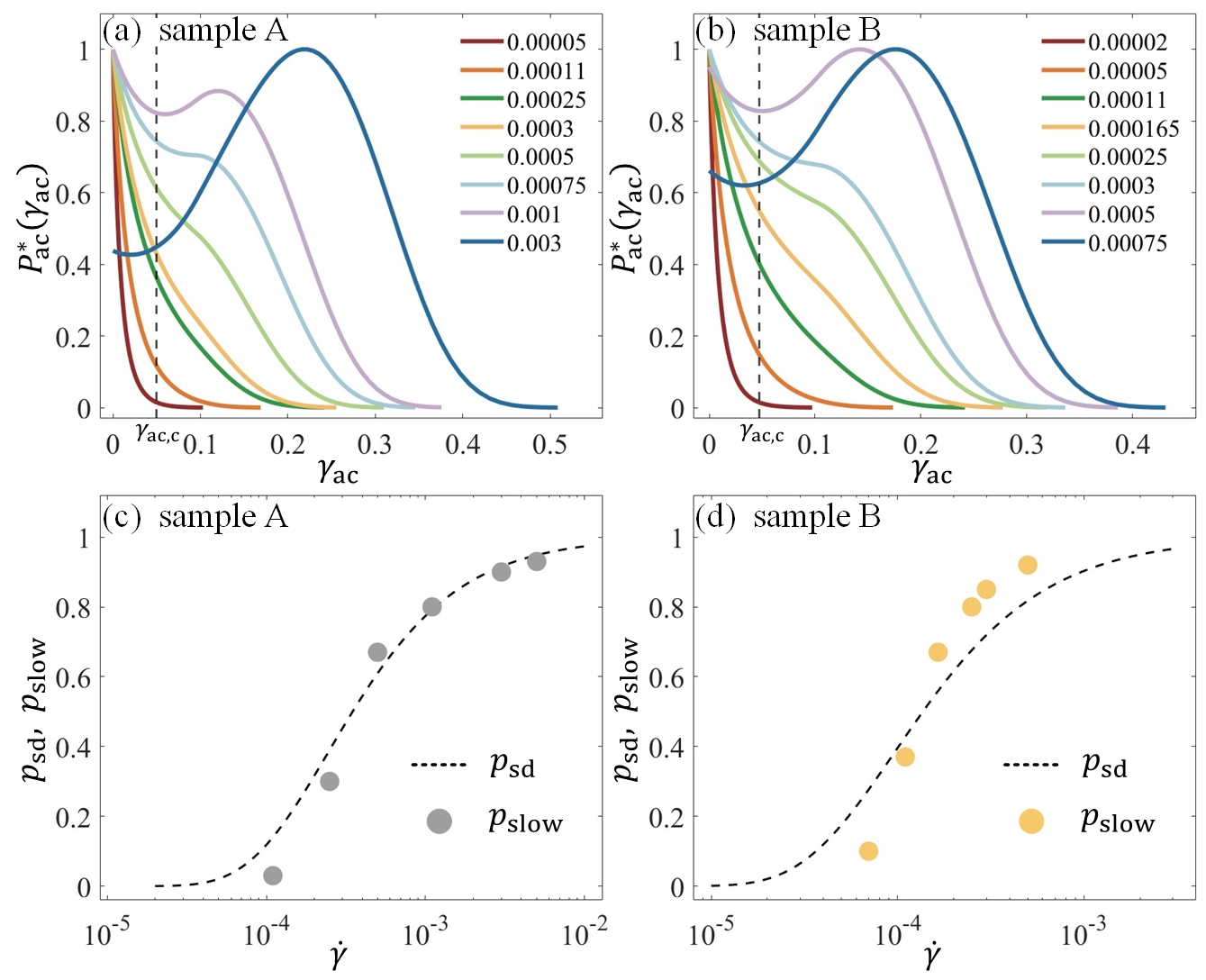}
    \caption{\label{fig:sgr_outstrain} 
    (a) and (b) show $P_\mathrm{ac}(\gamma_\mathrm{ac})$, given by equation~\ref{eq:sgr_outstrain}, at various $\dot{\gamma}$ for sample A and sample B, respectively. 
    Here, the values of $P_\mathrm{ac}$ are scaled as $P_\mathrm{ac}^*(\gamma_\mathrm{ac})=P_\mathrm{ac}(\gamma_\mathrm{ac})/P_\mathrm{ac,max}(\gamma_\mathrm{ac})$. 
    The vertical dashed line in each panel denotes the effective range of $P_\mathrm{ac}(\gamma_\mathrm{ac})$ at the lowest $\dot{\gamma}$. 
    (c) and (d) show the results of the percentage of shear-driven particles $p_\mathrm{sd}$ (dashed lines), defined by equation~\ref{eq:sgr_shear_driven_ratio}, at different $\dot{\gamma}$ for sample A and sample B, respectively. 
    The results of the proportion of slow particles $p_\mathrm{slow}$, which has been shown in figure~\ref{fig:predict_viscosity_and_slowratio}, are replotted (symbols). }
\end{figure}

In our framework, the solid-liquid duality, which is based on the grouping of slow/fast particles, is the key to connecting $\tau_\mathrm{LC}$ to rheological behaviors. 
Especially, this grouping is also strongly correlated to the local structure featured by terms associated with $g_{0,i}^0$ and $g_{2,i}^{-2}$, as demonstrated in section~\ref{sec:classification}. 
Thus, it is important to examine if the present model can give such grouping for conditions of steady shear. 
To explore this problem, we introduce the concept of activation strain $\gamma_\mathrm{ac}$, which represents the strain at which particles jump out from the basin. 
Particular attention is paid to the distribution of $\gamma_\mathrm{ac}$, denoted as $P_\mathrm{ac}(\gamma_\mathrm{ac})$. 
Under steady shear, $P_\mathrm{ac}(\gamma_\mathrm{ac})$ is given by:
\begin{equation}
    P_\mathrm{ac}(\gamma_\mathrm{ac})=\int \frac{1}{g(E_\mathrm{b})} \mathrm{e}^{E(\gamma_\mathrm{ac})/k_\mathrm{B}T} P_\mathrm{s}(\gamma_\mathrm{ac}, E_\mathrm{b}) \, \mathrm{d}E_\mathrm{b}.
    \label{eq:sgr_outstrain}
\end{equation}
Figure~\ref{fig:sgr_outstrain} (a) and (b) give the $P_\mathrm{ac}(\gamma_\mathrm{ac})$ of sample A and sample B, respectively. 
For the sake of clarity, we plot the scaled results $P_\mathrm{ac}^*(\gamma_\mathrm{ac})=P_\mathrm{ac}(\gamma_\mathrm{ac})/P_\mathrm{ac,max}(\gamma_\mathrm{ac})$, 
where $P_\mathrm{ac,max}(\gamma_\mathrm{ac})$ is the maximum value of $P_\mathrm{ac}(\gamma_\mathrm{ac})$.  
At the lowest $\dot{\gamma}$ shown here (in the Newtonian regime), $P_\mathrm{ac}(\gamma_\mathrm{ac})$ is exponential-like. 
In figure~\ref{fig:sgr_outstrain} (a) and (b), we use vertical dashed lines to mark the effective range of $P_\mathrm{ac}(\gamma_\mathrm{ac})$ at the lowest $\dot{\gamma}$, denoted as $\gamma_\mathrm{ac,c}$. 
As $\dot{\gamma}$ increases, $P_\mathrm{ac}(\gamma_\mathrm{ac})$ shifts to the regime of larger $\gamma_\mathrm{ac}$. 
More importantly, $P_\mathrm{ac}(\gamma_\mathrm{ac})$ gradually exhibits a bimodal feature composed of an exponential-like component 
at small $\gamma_\mathrm{ac}$ regime and a gaussian-like component at large $\gamma_\mathrm{ac}$ regime, with the latter being enhanced by shear. 
The emergence of the gaussian-like component at large $\gamma_\mathrm{ac}$ manifests the change in the way of relaxation. 
For particles corresponding to the gaussian-like component, their local configurations are significantly strained by external shear before relaxation. 
Qualitatively, these particles are likely slow particles: 
Their dynamics are slower than external shear, so that there is enough time for their local configurations to be strongly strained by external shear. 
On the contrary, remaining particles can be considered as likely fast particles, 
since there is no enough time for external shear to drive their local configurations to larger strains.

As seen in figure~\ref{fig:sgr_outstrain} (a) and (b), $\gamma_\mathrm{ac,c}$ gives a rough boundary between the exponential-like component and the gaussian-like component. 
Thus, we can evaluate the percentage of the gaussian-like component as follows:
\begin{equation}
    p_\mathrm{sd}=\int_{\gamma_\mathrm{ac,c}}^{\infty} P_\mathrm{ac}(\gamma_\mathrm{ac}) \, \mathrm{d}\gamma_\mathrm{ac}.
    \label{eq:sgr_shear_driven_ratio}
\end{equation}
Here, the subscript ``sd" denotes ``shear-driven", meaning that the local configuration and relaxation of particles corresponding to the gaussian-like component are strongly driven by shear. 
We plot the results of $p_\mathrm{sd}$ as a function of $\dot{\gamma}$ for two samples in figure~\ref{fig:sgr_outstrain} (c) and (d).  
In figure~\ref{fig:sgr_outstrain} (c) and (d), we replot the results of $p_\mathrm{slow}$ shown in figure~\ref{fig:predict_viscosity_and_slowratio}, which is obtained from the MD data, for comparison. 
Their dependences on $\dot{\gamma}$ are seen to be similar. 
Especially, $p_\mathrm{sd}$ gradually saturates as the system enters the shear-thinning regime. 
Figure~\ref{fig:sgr_outstrain} suggests that the SGR model exhibits two groups of responses to shear, 
which captures the major feature of the grouping of slow/fast particles. 

\begin{figure}
\includegraphics[width=\linewidth]{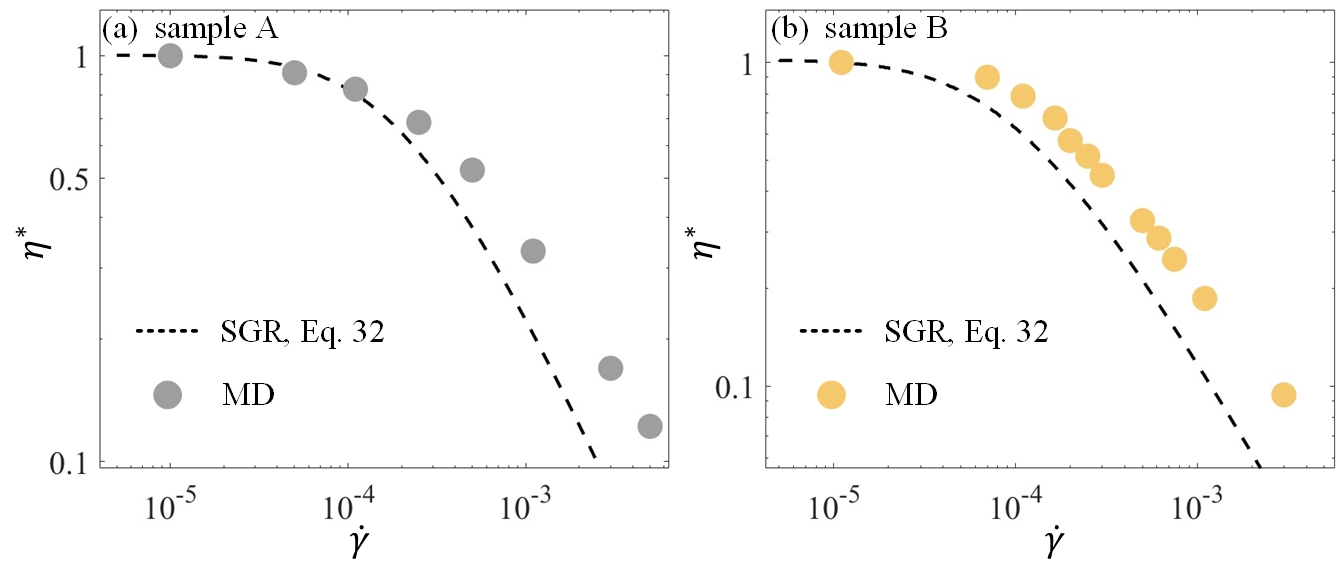}
    \caption{\label{fig:sgr_predict_viscosity} 
    Normalized shear viscosity $\eta^*$ as a function of $\dot{\gamma}$. 
    (a) and (b) show the results of sample A and sample B, respectively. 
    Dashed lines denote the SGR results calculated with equation~\ref{eq:sgr_sigmabar}. 
    Symbols denote the results found from MD data. }
\end{figure}

Knowing that the macroscopic stress is given by the average of the atomic level stress \cite{egami2011pms}, 
we evaluate the shear stress of the system with the stress experienced by a particle in its basin. 
For simplicity, this atomic level stress can be calculated by $\mathrm{d}E(\gamma)/\mathrm{d}\gamma$, where $E(\gamma)$ is given by equation~\ref{eq:pel_function}. 
Therefore, under steady state, the shear stress derived from the SGR model is given by:  
\begin{equation}
    \sigma_\mathrm{SGR} \sim \iint \frac{\mathrm{d}E(\gamma)}{\mathrm{d}\gamma} P_\mathrm{s}(\gamma, E_\mathrm{b}) \, \mathrm{d}E_\mathrm{b}\mathrm{d}\gamma.
    \label{eq:sgr_sigmabar}
\end{equation}
Then, the SGR viscosity can be found by $\eta_\mathrm{SGR}=\sigma_\mathrm{SGR}/\dot{\gamma}$. 
In figure~\ref{fig:sgr_predict_viscosity}, we show the results of $\eta_\mathrm{SGR}$, normalized as $\eta_\mathrm{SGR}^*=\eta_\mathrm{SGR}/\eta_\mathrm{eq}$, for two samples. 
$\eta_\mathrm{SGR}(\dot{\gamma})$ captures the general features of shear thinning: 
As $\dot{\gamma}$ increases, it is able to delineate three rheological regimes, i.e., Newtonian regime, crossover regime, and shear-thinning regime. 
Compared with figure~\ref{fig:sgr_outstrain} (c) and (d), it can be seen that the crossover regime appears when $p_\mathrm{sd}$ becomes finite, and the shear-thinning regime appears when $p_\mathrm{sd}$ starts to saturate. 
This result suggests that the emergence of shear thinning is due to the increasingly dominant role of shear-driven particles in the system. 
The picture given by the model analysis is consistent with our MD results shown in section~\ref{sec:rheology}.

We also plot the MD results of the normalized viscosity in figure~\ref{fig:sgr_predict_viscosity}. 
The SGR results and the MD results exhibit similar behaviors at the qualitatively level. 
Nevertheless, the quantitative deviation appears and grows as $\dot{\gamma}$ increases. 
We will discuss about this point in the next subsection.

To briefly summarize these two subsections, we try to model the MD results found in sections~\ref{sec:taucr} -- \ref{sec:structure}, 
which connect microscopic structure to macroscopic flow behaviors of supercooled liquids, by considering the shear-facilitated activation from basins.
This picture condenses the structural information characterized by $g_{0,i}^0$ and $g_{2,i}^{-2}$ into the height and strain characteristics of particle-level basins, 
enabling us to quantify the connection between local configuration and dynamics. 
The physical significance of $\tau_\mathrm{LC}$ and $\tau_\mathrm{sh}$, and the effects of shear flow and thermal activation, are clearly illustrated in this picture. 
With the aid of the SGR model, the relation between structure and rheology in flow state is clarified. 
Particularly, the model exhibits two distinct shear responses as evidenced in figure~\ref{fig:sgr_outstrain}. 
The relative proportion of these two responses dictates the variation of viscosity with respect to $\dot{\gamma}$, as illustrated in figure~\ref{fig:sgr_predict_viscosity}.

\subsection{Correlation}
\label{sec:sgr_correlation}
The deviation between model results and MD results shown in figure~\ref{fig:sgr_predict_viscosity} could be due to many reasons. 
For example, the basin shape $E(\gamma)$ we adopt could be oversimplified \cite{crocker2016natmat}; 
the restructuring process is ignored in the SGR model, etc. 
In our opinion, one of the key reasons is that the model given by equations~\ref{eq:sgr_eq} -- \ref{eq:sgr_rho_E} ignores the correlation among the cage-jump events of different particles. 
It has been demonstrated that there exists an elasticity-mediated long-ranged correlation among local flow events in glasses \cite{barrat2018rmp, bocquet2004epje}. 
In our recent work \cite{wang2024arxiv}, we also find the evidence of dynamic facilitation mediated by elasticity in solid-like regions in flowing supercooled liquids. 
In principle, as $\dot{\gamma}$ increases, such long-ranged correlation in a supercooled liquid should become more significant as the solid-like component grows. 
To account for this correlation while maintaining the simplicity of the model, one can replace the thermal term $k_\mathrm{B}T$ in 
equation~\ref{eq:sgr_eq} by the effective activation temperature $x$ \cite{cates1997prl}. 
Although the use of an effective temperature to describe the effect of mechanical noise has been questioned \cite{barrat2015epje}, 
it can qualitatively capture the phenomenon that the correlation among particles can facilitate the relaxation of a region.

Here, we tentatively adopt a more straightforward approach to phenomenologically incorporate this correlation. 
In the preceding two subsections, the activation volume $\Omega$ is set to the volume of a cage and is assumed to be constant with respective to $\dot{\gamma}$. 
This assumption is inconsistent with the long-ranged feature of the elasticity-mediated correlation in solid-like regions and the growth of the solid-like component with $\dot{\gamma}$. 
To explicitly account for this correlation, we assume that $\Omega$ increases correspondingly with $\dot{\gamma}$. 

\begin{figure}
\includegraphics[width=\linewidth]{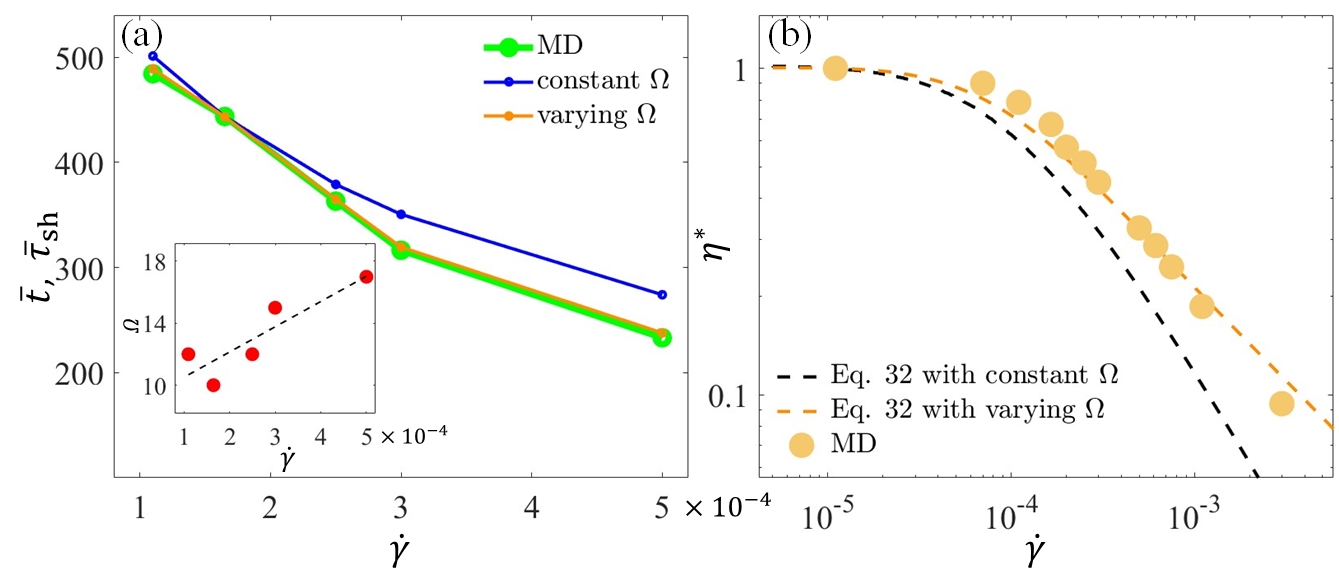}
    \caption{\label{fig:activate_volume} 
    (a) $\bar{\tau}_\mathrm{sh}(\dot{\gamma})$ (green circles, found from MD data) and $\bar{t}(\dot{\gamma})$ (defined in equation~\ref{eq:sgr_taushbar}) for sample B. 
    Here, $\bar{t}(\dot{\gamma})$ are calculated in two ways: with constant $\Omega$ (blue points) and with varying $\Omega$ determined by equation~\ref{eq:sgr_adjust_activate_volume} (orange points). 
    Inset: the varying $\Omega$ (red circles), determined by equation~\ref{eq:sgr_adjust_activate_volume}, as a function of $\dot{\gamma}$.  
    The dashed line denotes the linear fit of $\Omega$ with respect to $\dot{\gamma}$.  
    (b) Normalized viscosity of sample B. 
    Symbols denote the MD result. 
    Black dashed line denotes the result calculated with equation~\ref{eq:sgr_sigmabar} with constant $\Omega$. 
    This result has been shown in figure~\ref{fig:sgr_predict_viscosity} (b). 
    Orange dashed line denotes the result calculated with the varying $\Omega$ shown in the inset of (a). }
\end{figure}

We use the following strategy to determine $\Omega$. 
In section~\ref{sec:sgr_startup}, we introduce $\bar{t}(E_\mathrm{b},\dot{\gamma})$ in equation~\ref{eq:sgr_outbarrier_tbar} to represent $\tau_\mathrm{sh}$ in the start-up shear. 
The results are shown in figure~\ref{fig:sgr_taucr_sh}. 
It is seen that the values of $\tau_\mathrm{sh}$ evaluated by $\bar{t}(E_\mathrm{b},\dot{\gamma})$ and those measured from MD data have small but non-negligible discrepancy. 
Thus, we can determine a reasonable $\Omega$ by minimizing the difference between $\bar{t}(E_\mathrm{b},\dot{\gamma})$ and $\tau_\mathrm{sh}$ for each $\dot{\gamma}$. 
To be specific, we introduce the basin-averaged $\bar{t}(E_\mathrm{b},\dot{\gamma})$, denoted as $\bar{t}(\dot{\gamma})$:
\begin{equation}
    \bar{t}(\dot{\gamma})=\int P_\mathrm{b}(E_\mathrm{b}) \bar{t}(E_\mathrm{b},\dot{\gamma}) \, \mathrm{d}E_\mathrm{b}.
    \label{eq:sgr_taushbar}
\end{equation}
Conceptually, $\bar{t}(\dot{\gamma})$ corresponds to the globally averaged $\tau_\mathrm{sh}$, denoted as $\bar{\tau}_\mathrm{sh}(\dot{\gamma})$. 
The value of $\bar{t}(\dot{\gamma})$ depends on $\Omega$.  
Therefore, we can perform the following minimization by adjusting the value of $\Omega$ for each $\dot{\gamma}$:
\begin{equation}
    \min_{\Omega} \left | \bar{\tau}_\mathrm{sh}(\dot{\gamma}) - \bar{t}(\dot{\gamma}) \right |.
    \label{eq:sgr_adjust_activate_volume}
\end{equation} 

In figure~\ref{fig:activate_volume} (a), we present $\bar{\tau}_\mathrm{sh}(\dot{\gamma})$ and $\bar{t}(\dot{\gamma})$ using sample B as an illustrative example. 
Here, $\bar{t}(\dot{\gamma})$ is computed with two ways: the blue symbols give the result calculated with a constant $\Omega$; 
the orange symbols give the result calculated with the $\dot{\gamma}$-dependent $\Omega$ determined by equation~\ref{eq:sgr_adjust_activate_volume}.  
Generally speaking, incorporating a constant $\Omega$ leads to an over-estimation of $\tau_\mathrm{sh}$ at large $\dot{\gamma}$. 
In the inset of figure~\ref{fig:activate_volume} (a), we give the result of $\Omega(\dot{\gamma})$ determined by equation~\ref{eq:sgr_adjust_activate_volume} for sample B. 
$\Omega(\dot{\gamma})$ displays an increasing trend with $\dot{\gamma}$, 
corresponding to the enhancement of the correlation among flow events due to the increasingly dominant role of solid-like regions as $\dot{\gamma}$ increases. 
Within the mean-field framework of SGR, the increment of $\Omega(\dot{\gamma})$ can be understood by the enhancement of the rigidity \cite{biroli2021prl}. 
As the system becomes more rigid, the relaxation of the central particle will influence more surrounding particles, resulting in a larger $\Omega$ \cite{cavagna2009pr, dyre2006rmp}. 

By replacing the constant $\Omega$ in equation~\ref{eq:sgr_sigmabar} with $\Omega(\dot{\gamma})$, we recalculate the viscosity, and plot the result in figure~\ref{fig:activate_volume} (b). 
We also replot the MD result and the result calculated by equation~\ref{eq:sgr_sigmabar} with the constant $\Omega$, which have been shown in figure~\ref{fig:sgr_predict_viscosity} (b). 
It is seen that by using $\Omega(\dot{\gamma})$, the agreement between MD result and model result is significantly improved. 

\section{CONCLUDING REMARKS}
How do supercooled liquids flow? 
From the macroscopic viewpoint, Maxwell has laid the foundation for understanding the mechanical response of viscoelastic matter, 
including supercooled liquids, by introducing the Maxwell time $\tau_\mathrm{M}$ and the solid-liquid duality associated with $\tau_\mathrm{M}$. 
This approach, though extremely important in concept, lacks a clear molecular basis, which impedes the construction of a more general theory from microscopic structure and interaction. 
At the microscopic level, DH is one of the most important dynamical characteristics of supercooled liquids. 
Particularly, it has been proven that DH is closely correlated to the zero-shear viscosity \cite{cavagna2009pr}.
Knowing these facts, one may expect a microscopic approach, which embodies the conceptual solid-liquid duality by involving DH, for elucidating the flow behaviors of supercooled liquids. 

The key to bridging the gap between the microscopic and the macroscopic worlds is the local configurational relaxation time $\tau_\mathrm{LC}$. 
Defined at the particle level, $\tau_\mathrm{LC}$ well reflects the local mobility and the heterogeneous feature of dynamics. 
Moreover, $\tau_\mathrm{LC}$ plays a role similar to that of $\tau_\mathrm{M}$ in the Maxwell model: 
Depending on the comparison between $\tau_\mathrm{LC}$ and the external driving rate, the corresponding local region can exhibit solid-like or liquid-like response. 
Then, the macroscopic behavior is given by the average behavior of all local regions in the system. 
These results unify the concept of solid-liquid duality and DH into a micro-mechanical picture for the flow of supercooled liquids. 
$\tau_\mathrm{LC}$ has an unambiguous structural foundation. 
With the aid of statistical learning methods and a model considering the shear-facilitated activation from energy basins, 
we show that $\tau_\mathrm{LC}$ is determined by the combined effect of local packing and local configurational distortion, 
which are respectively represented by the barrier height and the tilted strain of the particle-level energy basin. 
With these efforts, we establish a chain relation of \textit{local configuration -- local dynamics -- local response -- flow behaviors} for supercooled liquids. 

As for the solid-liquid duality, extensive studies of amorphous matter, 
including but not limited to the glass transition and flow of supercooled liquids and the yielding of amorphous solids, 
are built on this intuition, with concepts such as soft modes \cite{harrowell2008natphy, lerner2021prl}, shear transformation zones \cite{langer1998pre}, 
cooperative rearranging regions \cite{yamamoto1998pre} and softness \cite{ajliu2015prl}. 
Most of these studies focus on the soft, liquid-like regions. 
On the other hand, our study highlights the increasing importance of the hard, 
solid-like regions in determining the flow behaviors of supercooled liquids as the flow is enhanced. 
This difference represents a shift of consideration \cite{wang2022prx}: 
For the yielding of solids, the key question is why a solid fails or even flows like a liquid. 
Liquid-like regions (for example, STZ) in the solid background naturally play as the precursor of the bulk yielding. 
For the flow of supercooled liquids, on the contrary, the key question is shifted to why a liquid exhibits strong viscoelasticity. 
By highlighting the solid-like regions in the liquid background, 
we are able to quantify the significant nonlinear behaviors of the flow.

Finally, we would like to point out that supercooled liquids are highly diverse, manifested by distinct kinds of microscopic interactions and dynamical characteristics \cite{dyre2012prx}. 
For example, water, interacting by hydrogen bonding, exhibits two liquid phases when entering the supercooled state \cite{stanley1992nat,soper2000prl, debenedetti2014nat, chen2015jpcl, gallo2016chemrev, nilsson2020science}. 
How these two phases respond to external shear, and how external shear affects these two phases, could be important for quantifying the nonlinear rheology of water \cite{patil2014pre, ribeiro2020prr}. 
Another example is the gaussian-core liquid \cite{stillinger1976jcp}. 
It belongs to ultrasoft liquids, which are featured by a very soft core in the interparticle interaction and represent a wide variety of soft materials \cite{likos2009rmp, likos2006sofmat}. 
The supercooled gaussian-core liquid behaves as a mean-field fluid \cite{hansen2000pre} and exhibits much weaker heterogeneity in dynamics compared with other model supercooled liquids \cite{miyazaki2011prl}, 
while maintaining an evident shear thinning \cite{mizuno2024comphy}. 
In this case, the role of DH, as well as the related structural features, in determining the rheology could be different. 
Note that, the method based on $\tau_\mathrm{LC}$ proposed here is also applicable to other systems. 
Such efforts could be helpful for achieving a more general framework for understanding the flow of supercooled liquids. 


\begin{acknowledgments}
This research was partially supported by the National Natural Science Foundation of China (no. 11975136). 
The authors acknowledge the Center of High Performance Computing, Tsinghua University for providing computational resources 
and Dr. Dejia Kong for his help in section~\ref{sec:structure}.
\end{acknowledgments}

\appendix
\section{ALGORITHMS FOR IDENTIFYING CAGE JUMPS AND CIRCLING CLUSTERS}
\subsection{Cage jump}
\label{sec:cage_jump_algorithm}
After obtaining the trajectory $S(t)$ of a particle from $0$ to $t_\text{end}$, 
(in shear condition, trajectory used is the non-affine trajectory, where the pure shear influence has been eliminated \cite{yamamoto1998pre}), 
an iterative process can be carried out to find cage jumps \cite{candelier2009prl, candelier2010prl}.
Specifically, the trajectory is divided at an intermediate time $t_\mathrm{m}$, resulting in two sub-trajectories: 
$d_1(t)$ from  $0$ to $t_\mathrm{m}$, and $d_2(t)$ from $t_\mathrm{m}$ to $t_\text{end}$.
To evaluate the degree of separation between $d_1$ and $d_2$ at $t_\mathrm{m}$, 
we calculate 
\begin{equation}
p(t_\mathrm{m}) = \zeta(t_\mathrm{m}) \sqrt{\big \langle d_1(t)- \overline{d}_2(t) \big \rangle ^2_{d_1} \cdot \big \langle d_2(t)- \overline{d}_1(t) \big \rangle ^2_{d_2}}, 
\end{equation}
where $\zeta(t_\mathrm{m})=\sqrt{t_\mathrm{m}/t_\text{end}\times(1-t_\mathrm{m}/t_\text{end})}$, 
$\overline{d}(t)$ represents the center of mass of the trajectory $d(t)$, 
and $\langle\cdot\cdot\cdot\rangle_{d(t)}$ denotes the average over the trajectory $d(t)$. 
This evaluation is performed for every time point along the trajectory to monitor when the $p(t_\mathrm{m})$ reaches its maximum value $p_{\text{max}}$,
and the corresponding time $t_\mathrm{m,max}$. 
Subsequently, the same process is applied to each sub-trajectory---from $0$ to $t_\mathrm{m,max}$ and $t_\mathrm{m,max}$ to $t_\text{end}$---to determine the critical times for these sub-trajectories. 
This division can be recursively applied to further sub-trajectories until $p_{\text{max}}$ falls below a threshold that, 
represents the cage size. 
Through this iterative procedure, cage jumps in the trajectory can be identified when $p(t)$ is maximal.

\subsection{Circling clusters}
\label{sec:circling_clusters}

In this part, we introduce the approach for partitioning slow and fast particles into clusters. 
As an exemplary illustration, we elaborate on the process of partitioning the slow particles. 
The similar procedure can be applied to fast particles. 

As seen from figure~\ref{fig:eq_spatial_taucr_dist}, the clusters of slow particles exhibit hierarchical characteristics. 
Thus, it can be imagined that slow particles ``grow" from different slow centers and form different slow clusters. 
Consequently, we can partition the particles based on the spatial positions of the slow centers. 
The identification of slow centers is carried out using the ``Grids" method. 
In brief, the simulation box is divided into small cubic grids. 
The side length of each grid is $L/20$, where $L$ is the length of the simulation box. 
Subsequently, with the following expression, each grid undergoes a convolution with particles' $\tau_\mathrm{LC}$, yielding a value denoted as $g$:
\begin{equation}
    g_i=\frac{ {\textstyle \sum_{j}} \tau_{\mathrm{LC},j}  \exp(-r_{ij}/l_\mathrm{cg} )}{ {\textstyle \sum_{j}} \exp(-r_{ij} / l_\mathrm{cg})},
    \label{eq:spatial_cg}
\end{equation}
where the subscript ``$i$" denotes the $i$th grid, $\tau_{\mathrm{LC},j}$ is the $\tau_\mathrm{LC}$ of particle $j$, 
$r_{ij}$ is the distance between particle $j$ and grid $i$, and $l_\mathrm{cg}$ is a characteristic length employed in the convolution and is set to $L/20$. 
The summations in equation~\ref{eq:spatial_cg} are performed over all particles. 

\begin{figure}
\includegraphics[width=0.8\linewidth]{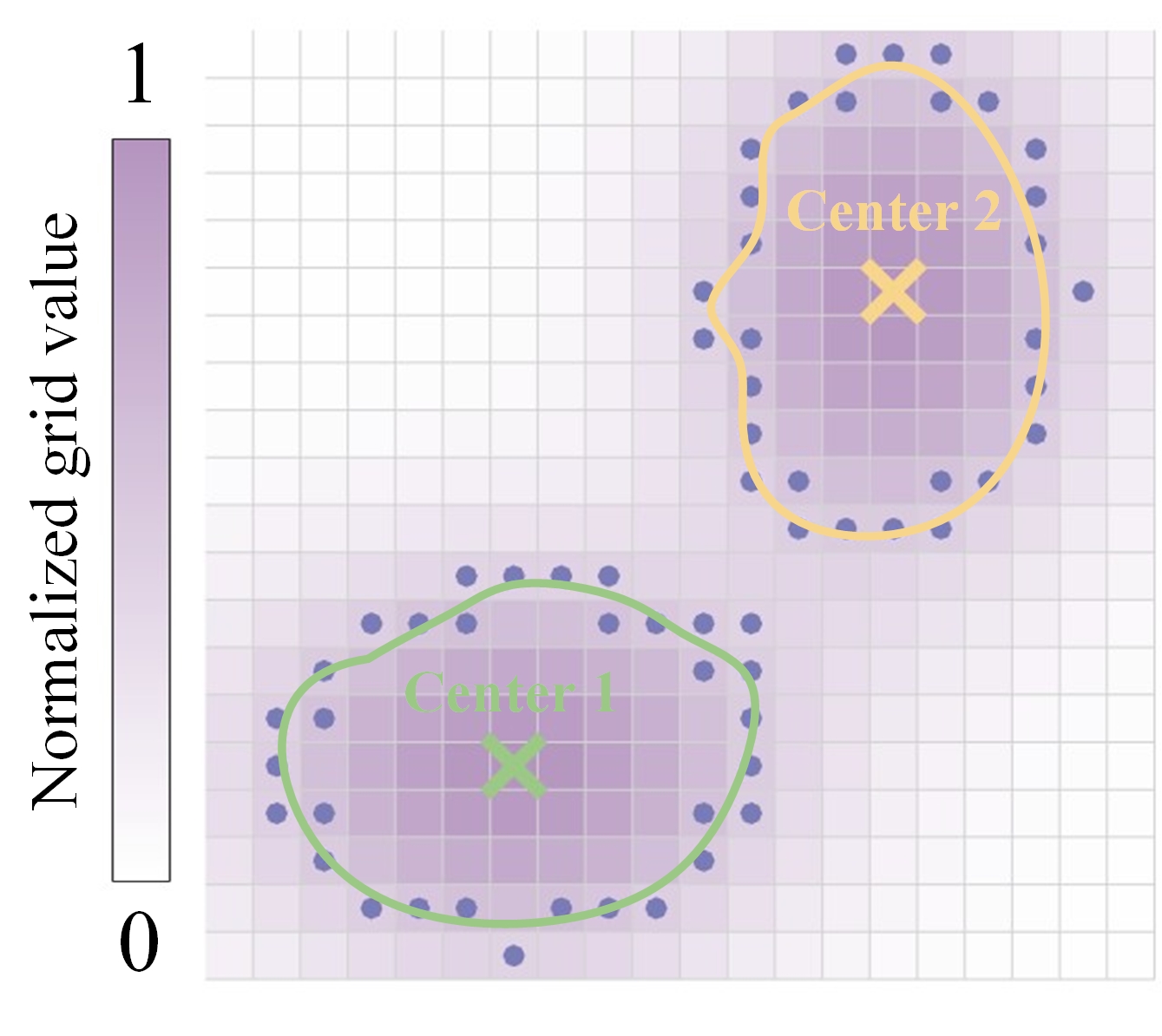}
    \caption{\label{fig:grids_method} 
    Illustration of the ``Grids" method for identification of slow/fast clusters. 
    Here, the system is divided into grids. 
    Each grid is assigned a value, which can be $\tau_\mathrm{LC}$ or other variables. 
    Without loss of generality, the values of the grids are normalized to the range of $0$ to $1$. 
    Giving a typical example, we define the regions of interest as those grids whose values are greater than $0.5$. 
    The boundaries of these regions can be identified. 
    Here, we use points to represent the boundaries. 
    Our objective is to divide the regions of interest into clusters. 
    First, we identify the center grids. 
    A center grid is defined as the grid whose value is greater than those of all its neighboring grids. 
    In this figure, cross symbols are used to denote the centers. 
    Subsequently, each grid within the region of interest is assigned to the nearest center grid. 
    As a result, each center grid has its ``jurisdiction area", which is represented by solid lines in the figure. }
\end{figure}

We define slow centers as the grids whose $g$ values are greater than those of their $26$ neighboring grids. 
This criterion guarantees that the slow centers are identified as local maxima. 
By utilizing these slow centers, each slow particle is assigned to the nearest center, thereby enabling the division of all slow particles into distinct clusters. 
The schematic diagram of this methods is presented in figure~\ref{fig:grids_method}. 

\begin{figure}
\includegraphics[width=\linewidth]{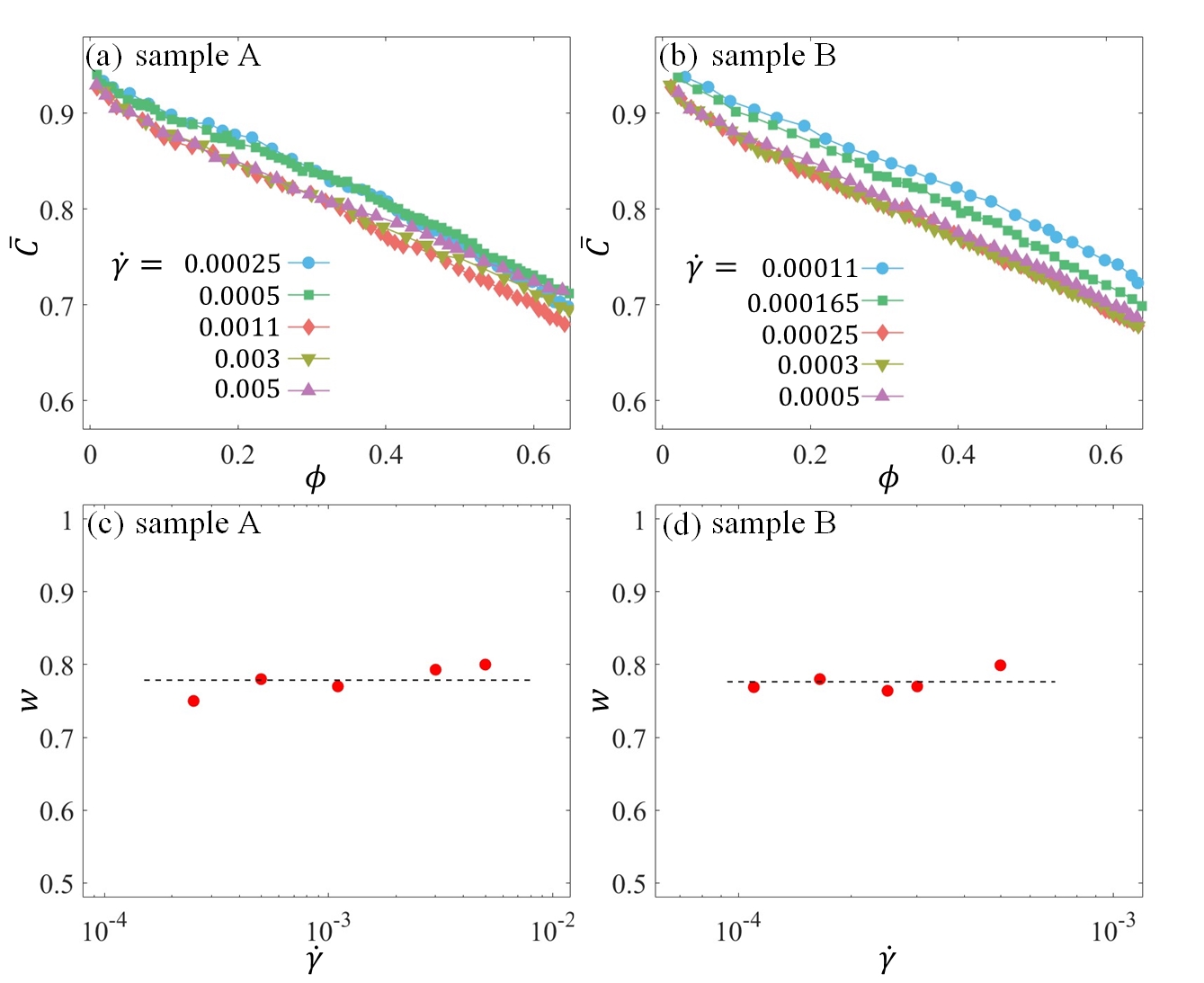}
    \caption{\label{fig:neighlost_with_jumpratio_plasticfactor} 
    (a) and (b) show the relation between the average neighbor-keeping ratio among slow particles, denoted as $\bar{C}$, 
    and the ratio of particles that have experienced cage jump in the system, denoted as $\phi$, for sample A and sample B, respectively, at various $\dot{\gamma}$. 
    (c) and (d) show the results of $w$ (symbols), found from equation~\ref{eq:neighlost_effective_modulus}, as a function of $\dot{\gamma}$ for sample A and sample B, respectively. 
    In (c) and (d), horizontal dashed lines denote the mean values of $w$ averaged over all $\dot{\gamma}$ for sample A and sample B. }
\end{figure}

\section{EFFECTIVE MODULUS OF LER}
\label{sec:effective_modulus}

In this part, we give the method for evaluating the effective modulus of LER from neighbor loss process. 
We first calculate the bond-breaking correlation function $C_i (t)$ for particle $i$ \cite{yamamoto1998pre, berthier2022prx} as $C_i (t)=n_i (t|0)/n_i (0)$, 
where $n_i (0)$ is the initial number of neighbors of particle $i$ at a given time origin, 
and $n_i (t|0)$ is the number of these initial neighbors that remain neighbors of particle $i$ after a time interval $t$. 
Here, a particle is considered a neighbor of a reference particle if the distance between them is within the first minimum of the pair distribution function $g(r)$. 
Inspired by Refs.~\cite{zaccone2017prl,zaccone2014prb} we posit that the modulus decays to 0 for a given region when, on average, 
half of the neighboring particles for particles within this region are lost. 
Furthermore, we assume that the change of modulus is approximately proportional to the average change in the number of neighbors. 
Then, we have:
\begin{equation}
    G(t) \approx G_\mathrm{ini}\frac{\bar{C}(t)-0.5}{0.5},
\end{equation}
where $\bar{C}(t)$ is the averaged $C_i(t)$ for a given region. 
$\bar{C}(t)$ is closely related to the percentage of jump $\phi(t)$, which represents the portion of particles that have undergone cage jump by time $t$ in a given region. 
Figure~\ref{fig:neighlost_with_jumpratio_plasticfactor} (a) and (b) exhibit the relation between $\bar{C}(t)$ of slow particles and 
$\phi(t)$ of the whole system for sample A and sample B, respectively, at several $\dot{\gamma}$. 
It is seen that $\bar{C}(t)$ and $\phi(t)$ are nearly linearly related, and their relation is not very sensitive to $\dot{\gamma}$ and sample. 
With these results, we can find $G(t)$, and the maximum stress accumulated within the elastic stage of LER is written as:
\begin{equation}
    \sigma_\mathrm{el} = \int_{0}^{\tau_\mathrm{el}} G(t) \, \dot{\gamma} \, \mathrm{d}t.
\end{equation}
The effective modulus $G_\mathrm{eff}$ is defined by $\sigma_\mathrm{el} = G_\mathrm{eff} \dot{\gamma} \tau_\mathrm{el}$. Thus, we have:
\begin{equation}
    G_\mathrm{eff} = w \, G_\mathrm{ini}=\frac{1}{\tau_\mathrm{el}}\int_{0}^{\tau_\mathrm{el}} G(t) \mathrm{d}t\approx\frac{G_\mathrm{ini}}{\tau_\mathrm{el}}\int_{0}^{\tau_\mathrm{el}} \frac{\bar{C}(t)-0.5}{0.5} \mathrm{d}t.
    \label{eq:neighlost_effective_modulus}
\end{equation}
The values of $w$ obtained by the preceding equation for sample A and sample B are given in figure~\ref{fig:neighlost_with_jumpratio_plasticfactor} (c) and (d), respectively. 

\section{IMPORTANCE OF STRUCTURAL FEATURES}
\label{sec:algorithm_select_features}

Once the $N_X$-dimensional feature vector $X$ for each particle has been computed, it is insightful to investigate the contribution of each element in $X$ to the final correlation. 
Thus, one can rank the $N_X$ elements in $X$ based on their importance.

Firstly, we directly compute the correlation coefficient between each element in $X$ and the target property. 
The element with the maximum correlation is denoted as $F_1$. 
Clearly, we should extend $F_1$ to include more elements to capture a more comprehensive relationship. 
To determine the second most important element, $F_2$, we adopt a method of exhaustion. 
Specifically, we combine $F_1$ with each of the remaining elements in $X$ to form $N_X-1$ two-element vectors. 
For each two-element vector, we use a linear regression model to establish the correlation between this vector and the target property, and record the correlation coefficient. 
This operation results in a list of correlation coefficients, each corresponding to a different two-element vector. 
The maximum correlation coefficient in this list is identified, and the corresponding two-element vector, which has the highest correlation, is noted. 
The second element in this vector is designated as $F_2$.

Based on $F_1$ and $F_2$, we can search for the third most important element, $F_3$. 
The method is similar to the one used to determine $F_2$. 
Specifically, we exhaustively combine each of the remaining $N_X-2$ elements with $F_1$ and $F_2$ to form three-element vectors. 
For each three-element vector, we compute the correlation coefficient and identify the best combination. 
The third element in the best combination is designated as $F_3$. 
By following this routine, we can subsequently determine $F_4$, $F_5$, and so on. 

Table~\ref{tab:rank_feature} lists the top $15$ most important features for two samples.

\begin{table}
\caption{\label{tab:rank_feature}The $15$ important features for two samples.}
\begin{ruledtabular}
\begin{tabular}{ccccccccc}
 &\multicolumn{4}{c}{sample A}&\multicolumn{4}{c}{sample B}\\
 Rank & gen\footnote{``gen" denotes the generation of the feature.}
 & $l$ & $m$ & $r$ & gen\footnotemark[1]
 & $l$ & $m$ & $r$ \\ \hline
 1  & $2$ & $0$ & $0$  & $1.3$ & $2$ & $0$ & $0$  & $1.3$ \\
 2  & $2$ & $2$ & $-2$ & $0.9$ & $2$ & $2$ & $-2$ & $1.2$ \\
 3  & $2$ & $0$ & $0$  & $1.1$ & $2$ & $0$ & $0$  & $1.4$ \\
 4  & $2$ & $0$ & $0$  & $1.2$ & $2$ & $0$ & $0$  & $0.9$ \\
 5  & $2$ & $0$ & $0$  & $3.0$ & $2$ & $0$ & $0$  & $1.2$ \\
 6  & $0$ & $0$ & $0$  & $1.0$ & $1$ & $2$ & $-2$ & $1.2$ \\
 7  & $1$ & $2$ & $-2$ & $0.9$ & $1$ & $0$ & $0$  & $1.4$ \\
 8  & $2$ & $0$ & $0$  & $0.8$ & $2$ & $0$ & $0$  & $1.1$ \\
 9  & $2$ & $0$ & $0$  & $1.9$ & $2$ & $0$ & $0$  & $2.3$ \\
 10 & $2$ & $0$ & $0$  & $1.4$ & $0$ & $0$ & $0$  & $1.3$ \\
 11 & $2$ & $0$ & $0$  & $0.7$ & $2$ & $0$ & $0$  & $2.1$ \\
 12 & $2$ & $0$ & $0$  & $1.1$ & $2$ & $0$ & $0$  & $3.4$ \\
 13 & $2$ & $2$ & $-2$ & $2.9$ & $2$ & $0$ & $0$  & $1.5$ \\
 14 & $2$ & $0$ & $0$  & $1.7$ & $2$ & $0$ & $0$  & $0.8$ \\
 15 & $2$ & $0$ & $0$  & $1.6$ & $0$ & $2$ & $-2$ & $5.0$ \\

\end{tabular}
\end{ruledtabular}
\end{table}

\bibliography{draft_maxwell.bib}

\end{document}